\documentclass[a4paper]{article}
\pdfoutput=1

\usepackage{arxiv}

\usepackage[utf8]{inputenc} 
\usepackage[T1]{fontenc}    
\usepackage{url}            
\usepackage{booktabs}       
\usepackage{amsfonts}       
\usepackage{nicefrac}       
\usepackage{microtype}      
\usepackage{lipsum}		
\usepackage{graphicx}
\usepackage{doi}

\usepackage{amsmath}
\usepackage{amsmath,bm}
\usepackage{mathtools}
\usepackage{slashed}
\usepackage{dsfont}
\usepackage{amssymb}
\usepackage{bbold}
\usepackage{diffcoeff}
\usepackage{float}
\usepackage[toc]{appendix}

\usepackage{caption}
\usepackage{subcaption}
\usepackage{multirow}

\title{Probing Parity Manifest Minimal Left-Right Symmetric Model through $CP$ Violation \& Anomalous Magnetic Moment of Charged Leptons Constrained by Lepton Flavor Violating $l_i \rightarrow l_j\gamma$ \& $l_i \rightarrow 3l_j$ Channels}

\author{{Rafid Buksh} \\
	Department of Physics\\
	University of Dhaka\\
	P.O. Box 1000, Dhaka, Bangladesh \\
	\texttt{rafubox123@gmail.com} \\
	\And
        {Samim Ul Islam} \\
	Department of Theoretical Physics\\
	University of Dhaka\\
	P.O. Box 1000, Dhaka, Bangladesh \\
	\texttt{mdsamimul-2018812477@tphy.du.ac.bd} \\
}

\date{}

\renewcommand{\headeright}{ }
\renewcommand{\undertitle}{ }
\renewcommand{\shorttitle}{Probing $P$ even MLRSM through EDM and MDM of Charged Leptons Constrained by CLFV}

\hypersetup{
pdftitle={Probing Parity even MLRSM through EDM and MDM of Charged Leptons Constrained by CLFV},
pdfsubject={hep-ph},
pdfauthor={Rafid Buksh, Samim Ul Islam},
pdfkeywords={First keyword, Second keyword, More},
}

\begin{document}
\maketitle

\begin{abstract}
While the Standard Model remains the prevailing description of natural phenomena, several observed phenomena continue to elude its explanation. To address these challenges, we investigate the Minimal Left-Right Symmetric Model, an immediate extension of the Standard Model. This model adeptly resolves issues related to parity violation and neutrino mass smallness by incorporating the seesaw mechanism. Our investigation focuses on assessing the Parity Manifest Minimal Left-Right Symmetric Model's contributions to specific phenomena, namely the anomalous magnetic moment (AMM) and CP violation of charged leptons. First, the model's coupling parameter space is derived from the experimental bounds of charged lepton flavor-violating $l_i \rightarrow l_j\gamma$ and $l_i \rightarrow 3l_j$ channels. These bounds emanate from the extended Higgs sector of the model. Subsequently, we evaluate the one-loop contributions of the model to magnetic dipole moment (MDM) and electric dipole moment (EDM), which provide insights into the model's estimations for AMM and CP violation of charged leptons based on the experimental bounds of the aforementioned processes. The outcomes of this analysis will shed light on whether the Parity Manifest Minimal Left-Right Symmetric Model stands as a highly promising candidate for physics beyond the Standard Model.
\newline
\newline
(The research of this article is partially derived from both author's master's theses.)
\end{abstract}


\section{Introduction}

The Standard Model (SM) of Particle Physics stands as one of the most accurate theories in the history of physics. It elegantly describes all known fundamental interactions except gravity, from precisely predicting the fine structure constant $\alpha$ to the monumental discovery of the Higgs boson by ATLAS and CMS in 2012. However, despite its successes, the SM leaves several fundamental questions unanswered and reveals theoretical gaps such as neutrino oscillations, muon anomalous magnetic moments, CP violations, and Lepton flavor violations. These unexplained aspects of the natural world signal the need for physics beyond the Standard Model (BSM). Consequently, we must extend or modify the SM to accommodate these phenomena. To facilitate this discussion, we will provide a brief overview of these puzzling phenomena below.
\newline
\newline
One prominent cosmological enigma is the baryon asymmetry, also known as the matter-antimatter asymmetry. It refers to the imbalance between baryonic matter and antimatter in the observable universe. Antimatter naturally arises due to the SM's gauge symmetry, where interactions are mediated by spin-1 particles. These interactions involve complex-valued fields, doubling the degrees of freedom and creating antimatter, identical to matter in kinematic properties but with opposite charges. Distinguishing between matter and antimatter would seem straightforward if interactions violated charge conjugation symmetry (C). However, this alone is insufficient due to additional internal degrees of freedom, like helicity. Therefore, the necessary condition for distinguishing matter from antimatter is the violation of both charge conjugation (C) and CP symmetry (the combination of C and spatial inversion, or parity, P). The weak interaction notably violates C, but introducing the CP violation proves more intricate.

In the SM, CP violation exists in the quark sector, contingent upon the presence of three generations of fermions. Unfortunately, the magnitude of this CP violation is insufficient to explain the observed Baryon asymmetry in the universe. However, promising hints of substantial CP violations have emerged in the lepton sector, linked to neutrino flavor mixing. This CP violation in the lepton sector can generate matter-antimatter asymmetry through a process called leptogenesis. It could potentially serve as the Standard Model's explanation for the universe's matter-antimatter asymmetry, providing experimental confirmation of lepton sector CP violation. Should this CP violation prove too small, additional CP-violating sources from physics beyond the Standard Model will become necessary. Novel BSM particles may provide these sources, as CP is not a ground state symmetry, as exemplified by the $K-\bar{K}$ mixing (see Section \ref{CP Violation}). In essence, the universe can generate baryon asymmetry if it satisfies three conditions, known as Sakharov's conditions: baryon number violation, C and CP violation, and departure from equilibrium.
\newline
\newline
Another intriguing phenomenon is Charged Lepton Flavor Violation (CLFV). In the SM, the existence of three fermion families is acknowledged, with distinctions based solely on particle masses, or flavors. For example, the charged leptons in the standard model - electron ($e$), muon ($\mu$), and tau ($\tau$)— represent three different flavors of charged leptons, while neutrinos correspond to the same flavor categories. Each flavor corresponds to a defined mass. SM interactions, for the most part, conserve flavor, ensuring that the same flavor number is maintained from the initial to the final state.

e.g.- For leptons, lepton number, $L_i$, is conserved in SM interactions, where $L_i$ is defined as:
$$L_i = \begin{cases} 
      1 , & \begin{pmatrix}
\nu_i \\ i
\end{pmatrix} \ ; \ i = e , \mu , \tau\\
      0, & otherwise
      \end{cases}$$ 
Therefore, SM interactions maintain flavor conservation; starting with one flavor, the final state retains the same flavor (e.g.- an electron or an electron-neutrino). However, lepton flavor-violating processes violate this conservation, resulting in a change in the number of particles with a particular flavor from the initial to the final state. The search for charged lepton flavor violation traces back to the early 1930s following the identification of the muon ($\mu$) as a distinct particle. Notably, flavor violation is well-established in the quark sector, but in the lepton sector, experimental evidence of flavor violation remains elusive.
\newline
\newline
Additionally, the Super Kamiokande Collaboration in 1998 provided definitive evidence for Neutrino oscillations, implying the existence of at least two massive neutrinos. This observation necessitates the introduction of new particles not predicted by the Standard Model. Just like all successful theories, the Standard Model not only offers new insights but also poses new challenges. The mystery of the three families of elementary particles, with their varying masses and mixing, gives rise to the flavor problem, which demands resolution. It is evident that if neutrinos were massive and exhibited mixing, this mixing would extend to charged leptons due to their shared coupling ($\nu-e-W$). Thus, massive neutrinos necessitate an extension of the Standard Model. The key question arises: how are these massive neutrinos introduced?

The most straightforward approach assumes that neutrinos possess Dirac-type masses, generated in the same manner as all other SM fermion masses via the Higgs mechanism. This entails the addition of neutrino Yukawa couplings to the theory. While this method is conceptually simple, it encounters a serious fine-tuning problem. Experimental data indicate that neutrino mass, $m_\nu$, is less than 1 eV \cite{abazajian2011}, implying that the neutrino Yukawa coupling, $Y_{\nu}$, must be on the order of $10^{-12}$. This value is precise and consistent with experimental data but appears unnatural when compared to the Yukawa couplings of charged fermions in the SM, typically ranging from $10^{-6}$ to 1. Furthermore, the SM extended with massive Dirac neutrinos predicts exceedingly small rates for charged lepton flavor-violating (CLFV) processes rendering such interactions nearly impossible to detect. Detecting such phenomena would provide evidence of contributions from a higher energy theory. Consequently, charged lepton flavor-violating transitions would serve as clear signals of physics beyond the Standard Model (BSM). Depending on the specific realization of neutrino mass, the rates for CLFV processes can vary. Numerous high-energy models typically predict CLFV at small branching ratios, rendering them challenging (if not impossible) to detect. In contrast, measurable CLFV rates are expected if the scale of new physics is close to the electroweak scale. These low-scale mechanisms for generating neutrino masses hold greater appeal from a phenomenological perspective, offering a window into new physics through promising CLFV perspectives. Moreover, these mechanisms can be directly tested at the LHC via the production of new particles, provided they are sufficiently light. While scientists have yet to make definitive discoveries, numerous experimental programs, such as MEGII, Mu3e, and COMET, are actively investigating this phenomenon, with searches conducted across various channels, with the most sensitive involving muons ($\mu$).
\newline
\newline
In light of the phenomena discussed thus far, it becomes evident that an extension of the Standard Model is imperative. While it can be extended more radically, we have primarily considered minimal extensions that introduce massive neutrinos and nothing more. One prominent method for generating neutrino masses within the SM avoids the fine-tuning problem associated with neutrino Yukawa couplings. This model, known as the Seesaw Mechanism, posits that neutrinos possess both Dirac and right-handed Majorana masses (see Section \ref{Neutrino Physics}). This arrangement allows the left-handed neutrino mass to remain tiny (on the order of eV) while permitting the right-handed mass to be extremely large (on the order of $10^{16}$ GeV). Additionally, it facilitates naturally-sized neutrino Yukawa couplings (on the order of 1). Consequently, in the Seesaw mechanism, neutrinos possess both Dirac and Majorana mass terms. This configuration permits neutrinos to exhibit Dirac masses comparable to those of charged leptons, while their Majorana nature keeps their mass low—a solution to the mass hierarchy challenge in the lepton sector via the Seesaw mechanism. However, given that Higgs doublets alone yield Dirac masses, solving the dark matter (DM), CP violation, and neutrino physics problems necessitates extending the Higgs sector to generate Majorana masses. In the Type-II Seesaw mechanism, a Higgs triplet ($\Delta$) resolves these issues (DM, CP Violation, and Seesaw) while providing a natural explanation for the missing piece of the Standard Model: the right-handed neutrino. One such Seesaw-compatible extension of the SM is the ``Left-Right Symmetric Model," wherein the Higgs sector is extended to accommodate right-handed neutrinos, forming a weak doublet that decouples following a new Higgs transition. This natural extension results in $P$ violation, which we will explore explicitly, along with increased CP violation and CLFV stemming from the extended Higgs sector. These, in turn, contribute to Electric Dipole Moments (EDM) and Magnetic Dipole Moments (MDM).
\newline
\newline
In this article, we will evaluate couplings parameter space of the Left-Right Symmetric Model from Charged Lepton Flavor Violation (CLFV) processes like $l_i \rightarrow l_j \gamma$, $l_i \rightarrow 3l_j$ (for instance, $\mu \rightarrow e \gamma$, $\mu \rightarrow 3e$ etc.) and calculate the values of Magnetic Dipole Moment (MDM) and Electric Dipole Moment (EDM) of charged leptons, thereby providing insights into the model's estimation for AMM and EDM based on experimental bounds posed by $l_i \rightarrow l_j \gamma$, $l_i \rightarrow 3l_j$ channels. Such analysis will shed light on whether the Minimal Left-Right Symmetric Model stands as a highly plausible candidate for physics beyond the Standard Model. 
\newline
\newline
The article is organized as follows: In Section \ref{Neutrino Physics}, we will review the neutrino mass smallness problem and its possible extension leading beyond the Standard Model (BSM) to solve this neutrino puzzle, which will take us to the see-saw mechanism. In Section \ref{Left-Right Symmetric Model (LRSM)}, we will study the model itself in detail. In Sections \ref{CP Violation}, \ref{Anomalous Magnetic Moment (AMM)} and \ref{Charged Lepton Flavor Violation (CLFV)}, we will discuss why SM can't explain phenomena such as CPV, AMM, and CLFV respectively, and how the LR model will explain them. In Section \ref{Couplings vs Right handed Neutrino Mass}, we will find the phenomenologically interesting region of couplings for right-handed Neutrino mass at $1 TeV$ to maintain the parity manifest condition of the model. In Section \ref{Phenomenology}, we present and interpret the numerical results of discussed phenomena. Finally, we conclude in Section \ref{Conclusion}.

\section{Neutrino Physics}
\label{Neutrino Physics}
\subsection{Neutrinos in Standard Model}

The neutrinos are spin-$\frac{1}{2}$ particles and have no electric charge. They interact only via weak interaction. The neutrinos are therefore leptons. There are also electrically charged leptons in the Standard Model, called electron, $e$; muon, $\mu$, and tau, $\tau$; of which the electron is the lightest and the tau the heaviest. There are also three neutrinos in the Standard Model, called electron neutrino,$\nu_e$; muon neutrino, $\nu_\mu$, and tau neutrino, $\nu_\tau$. The neutrinos are named like that, because in the SM when the $W^-$ boson decays into a charged lepton, it is always accompanied by an antineutrino of the same flavor: $$W^- \rightarrow l_i + \overline{\nu_i} \; \  i = e,\ \mu,\ \tau$$
In the Standard Model, the neutrinos are assumed to be massless. They are also assumed to be Dirac particles, which means that they have distinct antiparticles. The zero-mass forces the neutrino flavor to be conserved for a trivial flavor mixing matrix. So according to the Standard Model, there can be no neutrino mixing.
$$ \mathcal{L}_{Dirac} = m_D \bar{\psi}_L \psi_R + H.c. $$

So, they could have Majorana mass. However, due to gauge symmetry, it is forbidden. Because Majorana mass term in the Lagrangian is 
$$ \mathcal{L}_{Majorana} = M_M \psi^T_L C^{-1} \psi_L $$

which is clearly breaking the $SU(2)_L$ gauge symmetry and also Lepton number $U(1)_{L}$ symmetry. Thus, for gauge symmetry and the absence of right-handed neutrinos, neutrinos in the SM can not acquire mass. However, it is experimentally known that different neutrino flavors can mix. In neutrino mixing, a neutrino that started with a definite flavor ($\nu_e, \nu_\mu, \nu_\tau$) can be found with a different flavor after some propagation. This is possible, because the neutrinos taking part in the interactions are not the mass eigenstates, i.e. the states with a definite mass, but linear combinations of them.

Neutrino mixing has undeniably been verified in ``disappearance" experiments with atmospheric and accelerator neutrinos, solar neutrinos, and reactor neutrinos and in ``appearance" experiments with solar neutrinos. This kind of mixing is only possible when neutrinos are massive and their masses are different.

Therefore, the neutrino oscillations clearly show that the neutrinos have nonzero masses, and therefore that the Standard Model is not the complete description of nature. The SM has to be extended to include the massive neutrinos. The method of their introduction affects the magnitude of charged lepton flavor violation and leptonic CP violation.

\subsection{Neutrinos in Beyond Standard Model}

To solve the mass neutrino term, there is no other option but to add a right-handed neutrino. Then neutrinos can get Dirac mass terms. Majorana mass term violates lepton number conservation. But $U(1)_L$ is an accidental symmetry that occurs due to the fermionic structure and gauge interactions. There is no fundamental reason why it should be preserved. So, we can introduce either exotic massive fermions or scalars to the theory, after SSB (acquiring vev) it will produce Majorana mass or introduce gauge invariant non-renormalizable term ($D > 4$). Both lead to the same physics as these terms break the lepton number. 
\newline
\newline
If we consider neutrinos to be Majorana particles then this Majorana mass term is generated once the SM is considered a low energy Effective Field Theory (EFT) (see Appendix \ref{Effective Field Theory}) via Dim 5 Weinberg operator \cite{weinberg1979baryon} which is the only lowest dimensional nonrenormalizable operator and violates the conservation of lepton number by two units:
$$\mathcal{L}_5 = c^{ij} \frac{( \overline{L^c_{Li}} \cdot \widetilde{H^*}) (\widetilde{H^\dagger} \cdot L_{Lj} )}{\Lambda} + H.c.$$
After SSB, we get the major mass term - 
$$\mathcal{L}^\nu_{Yuk} = c_\nu \frac{v_H^2}{\Lambda} \nu_L^T C^{-1} \nu_L $$
where the Majorana mass is identified as $M_M = c_\nu \frac{v_H^2}{\Lambda}$
\begin{figure}[H]
\begin{center}
\includegraphics[width=60mm]{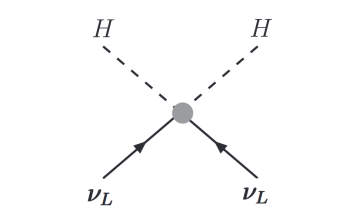}
\end{center}
\caption{Representation of Dim 5 Weinberg Operator }
\label{Weibg}
\end{figure}
To accommodate neutrino masses $m_\nu < 1$ eV \cite{abazajian2011}, $c_\nu \sim 1$ and $\Lambda$ must come from a BSM scale of $\mathcal{O}(10^{14})$ GeV which is close to the unification scale of electroweak and strong interactions i.e. GUT scale. 
\newline
\newline
Although the Majorana mass term for neutrinos generated through the Weinberg operator conserves electric charge, it still breaks the total lepton number conservation. This might be interpreted as a consequence of the SM being an effective low-energy theory of a much more general theory at higher energies.

There are 3 ways to generate this Dim 5 Weinberg operator (Figure \ref{Weibg}) at tree level. These are known as type-I, type-II, and type-III seesaw mechanisms, with an $SU(2)$ singlet fermion, triplet scalar, and triplet fermion respectively, as mediator given below (Figure \ref{seesaw1}, \ref{seesaw2}, \ref{seesaw3}) :
\begin{enumerate}
\item \textbf{Type-I Seesaw \cite{Minkowski} \cite{Yanagida}} : It assumes the existence of right-handed neutrinos $\nu_R : (1 , 1 , 0)$.
\graphicspath{{./seesaw1/}}
\begin{figure}[H]
\begin{center}
\includegraphics[width=80mm]{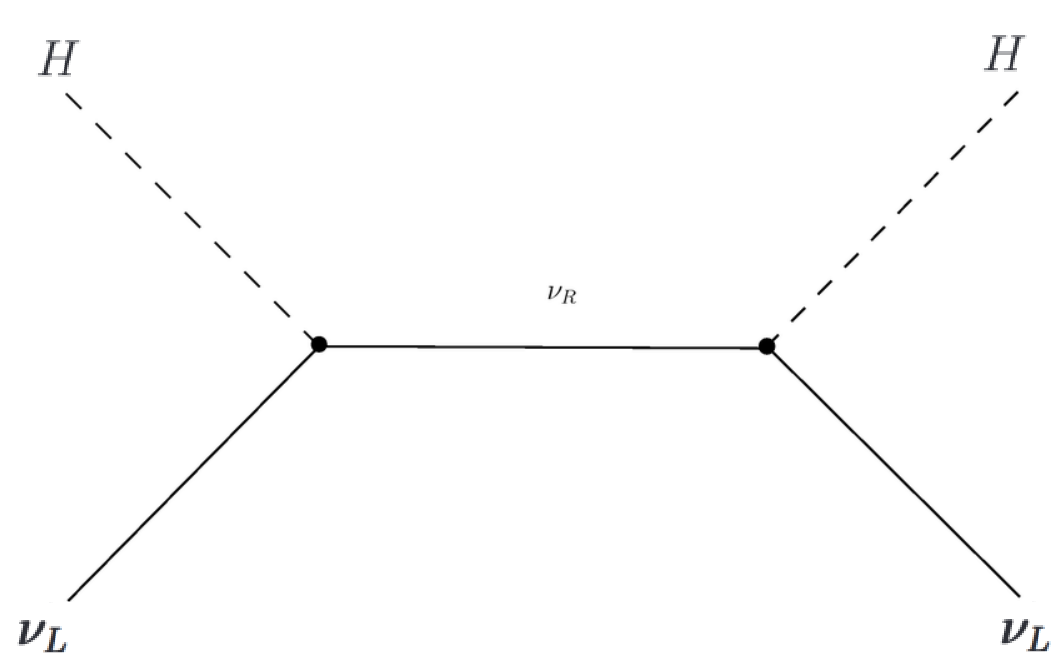}
\end{center}
\caption{Type-I Seesaw($SU(2)$ singlet fermion)}
\label{seesaw1}
\end{figure}
 These neutrinos are thus completely sterile under the SM gauge group. They are allowed then to have a nonzero mass term $M_R$ at tree level, and a Dirac Yukawa coupling with the Higgs boson $y_D \overline{\nu_L} H \nu_R + H.c$ then the scale of neutrino mass generation will be $\frac{M_R}{y^2_D}$.
\item \textbf{Type-II Seesaw \cite{Magg}} : It assumes the existence of a scalar triplet $\Delta = (\Delta^{++},\Delta^{+},\Delta^0) : (1 , 3 , 1)$.
\graphicspath{{./seesaw2/}}
\begin{figure}[H]
\begin{center}
\includegraphics[width=30mm]{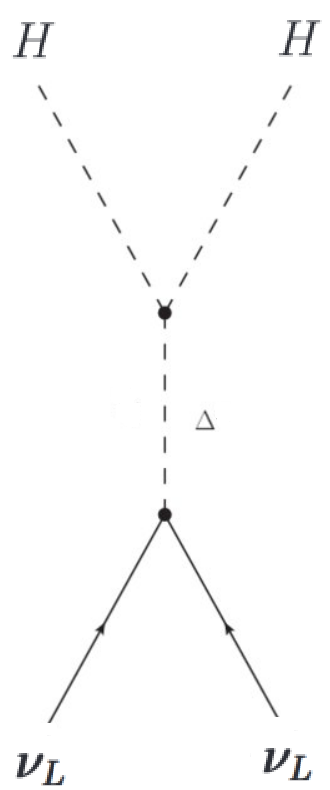}
\end{center}
\caption{Type-II Seesaw($SU(2)$ triplet scalar)}
\label{seesaw2}
\end{figure}
 They are allowed then to have a nonzero mass term $M_\Delta$ at tree level, Yukawa coupling with the lepton doublet $y_\Delta \overline{L_L^c}\Delta_L L_L + H.c. $ and a coupling with the Higgs doublet $\mu_\Delta \widetilde{H^\dagger} \Delta^\dagger_L H$. The scale of neutrino mass generation will then be $\frac{M_\Delta}{\mu_\Delta y_\Delta}$.
\item \textbf{Type-III Seesaw \cite{Foot}} : It assumes the existence of a fermion triplet $\Sigma = (\Sigma^{+},\Sigma^{0},\Sigma^-) : (1 , 3 , 0)$ whose middle component mixes with the left-handed neutrino. 
\begin{figure}[H]
\begin{center}
\includegraphics[width=80mm]{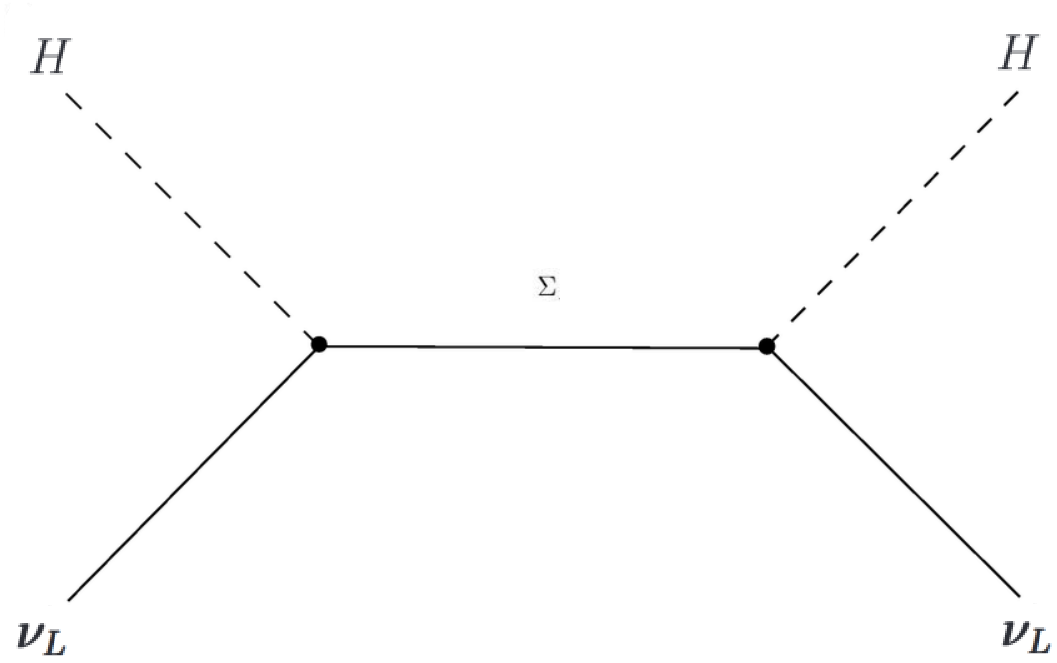}
\end{center}
\caption{Type-III Seesaw($SU(2)$ triplet fermion)}
\label{seesaw3}
\end{figure}
Analogous to type-I, they are allowed then to have a nonzero mass term $M_\Sigma$ at tree level, and a Yukawa coupling with the Higgs boson $y_\Sigma \overline{L_L} \Sigma \widetilde{H} + H.c$ then the scale of neutrino mass generation will be $\frac{M_\Sigma}{y^2_\Sigma}$. 

\end{enumerate}

\subsubsection{Neutrino Mass and Mixing}

Now, we have both left and right-handed neutrinos with both Dirac and Majorana mass. The Yukawa Lagrangian (mass term) is given by
$$-\mathcal{L}_{Yuk} = y_\nu\bar{L}_L\cdot H \nu_R+\frac{1}{2} M_M \nu_R^T C^{-1} \nu_R $$

If we want to avoid the Majorana term (2nd term), then we just need to impose Lepton number conservation. The neutrino in this Majorana basis is $\begin{pmatrix} 
\nu_L \\ \nu_R^c
\end{pmatrix}$ where, $\nu_L$ and $ \nu_R^c$, each are $3\times1$ matrix. The total multiplet is $6\times1$. On this basis, the mass matrix is (proved in the LR model)
$$ M_\nu = \begin{pmatrix} 
0 & m_D \\ 
m_D^T & M_M
\end{pmatrix} $$
Consider the eigenvalues of the matrix are $\lambda$, then
$$ det\begin{pmatrix} 
-\lambda &m_D \\ 
m_D^T & M_M-\lambda
\end{pmatrix}=0 $$

$$ \implies \lambda^2-M_M\lambda-m_D^2=0$$

Assuming $m_D \ll M_M$, then we get the masses 
$$ \lambda = m_{1,2} = \frac{M_M\pm\sqrt{M_M^2+4m_D^2}}{2} = M_M  ,\  \frac{-m_D^2}{M_M} $$
One massive state remains very heavy $M_M$ and the light neutrino states acquire a tiny mass $\frac{-m_D^2}{M_M}$. This light neutrino state is our standard model neutrino($\nu_L$). The heavy one($\nu_R$) is decoupled due to the symmetry-breaking pattern ($v_R \gg v_L$). 

The mixing between active $\nu$ and heavy $\nu$ will emerge but it will be very small. From diagonalization, 

$$ tan(2\theta)=\frac{2M_{12}}{M_{22}-M_{11}}=\frac{2m_D}{M_M} $$

where $\theta$ is the angle between the interaction (flavor) eigenstate and mass (physical) eigenstate which is very small. It means light neutrino which is active and heavy which is inert and very small. Thus, this mechanism can explain the smallness of neutrino mass and mixing naturally only if we can anyhow introduce the Majorana mass term for neutrino. 
\newline
\newline
We will see in the LR model (Section \ref{Left-Right Symmetric Model (LRSM)}) that with $\Delta$ Higgs coupling to neutrinos, will give Majorana mass term for neutrinos. It also has several phenomenological consequences (leptogenesis, LFV, etc.). There are some disadvantages of this mechanism. New particles are typically too heavy to be produced. As mixing with a heavy partner is very small, it is hard to detect. In general, this theory is hard to test. 

\subsubsection{Relation between Neutrino Mixing Matrix and Dirac Mass}

From Yukawa Lagrangian,
$$ -\mathcal{L}_{Yuk} = \frac{1}{2}
\begin{pmatrix} 
\bar{\nu^0_L} & \bar{(\nu^0_R)^c}
\end{pmatrix}
\begin{pmatrix} 
0 &m_D \\ 
m_D^T & M_M
\end{pmatrix}
\begin{pmatrix} 
(\nu^0_L)^c  \\ \nu^0_R
\end{pmatrix} + H.c$$
Here, the mass matrix $M_\nu$ is a $6\times6$ complex symmetric matrix that can be diagonalized by a $6\times6$ unitary matrix known as Pontecorvo-Maki-Nakagawa-Sakata matrix (PMNS) in the following way: 
$$ U_{PMNS}^T \ M_\nu \ U_{PMNS}  = diag (m_{\nu_1} , m_{\nu_2}, m_{\nu_3} , m_{M_1} , m_{M_2}, m_{M_3})$$
The unitary matrix, $U_{PMNS}$ relates flavour to mass eigen states
$$\begin{pmatrix} 
\nu_e \\ 
\nu_\mu \\
\nu_\tau \\
\end{pmatrix} = U_{PMNS} \begin{pmatrix} 
\nu_1 \\ 
\nu_2 \\
\nu_3 \\
\end{pmatrix} $$ 
where $$U_{PMNS} = \begin{pmatrix} 
c_{12}c_{13} & s_{12}c_{13} & s_{13}e^{-i\delta}\\ 
-s_{12}c_{23}-c_{12}s_{23}s_{13}e^{i\delta} & c_{12}c_{23}-s_{12}s_{23}s_{13}e^{i\delta} & s_{23}c_{13} \\
s_{12}s_{23}-c_{12}c_{23}s_{13}e^{i\delta} & -c_{12}s_{23}s_{13}e^{i \delta} & c_{23}c_{13}
\end{pmatrix} .\  diag (e^{-i\frac{\phi_1}{2}} , e^{-i\frac{\phi_2}{2}}, 1) $$
Here, $c_{ij} = cos \theta_{ij}$ \ , $s_{ij} = sin \theta_{ij}$. $\theta_{ij}$ are the neutrino flavour mixing angles, $\delta$ is the Dirac phase and $\phi_{1,2}$ are the Majorana phases.
\newline
This gives $3$ light Majorana neutrino mass eigenstates $\nu_{L_i}$, with masses $m_{\nu_i} (i=1,2,3)$, and
three heavy ones $\nu_{R_i}$, with masses $m_{M_i} (i=1,2,3)$ such that
$$\begin{pmatrix} 
\nu^{0}_L \\ (\nu^0_R)^c
\end{pmatrix} = U^*_{PMNS}\begin{pmatrix} 
\nu_L \\ \nu_R^c
\end{pmatrix} \  
\begin{pmatrix} 
(\nu^0_L)^c  \\ \nu^0_R
\end{pmatrix} = U_{PMNS}\begin{pmatrix} 
\nu_L^c  \\ \nu_R
\end{pmatrix}$$
where $\nu_L$ and $\nu_R^c$, each are a $3\times1$ matrix.
Since we are working in a basis where $M_M$ is diagonal, the heavy eigen states are then given by $$D_{M_M} = diag (m_{M_1}, m_{M_2}, m_{M_3})$$ 
One can therefore write, from seesaw, the physical light neutrino masses as $$m_\nu = - m_D M_M^{-1} m_D^T$$
After electroweak (EW) symmetry breaking, Dirac neutrino mass matrices can be written as $m_D = \frac{Y v_H}{\sqrt{2}}$. Thus, it is convenient to define a matrix $\kappa$ as
$$\kappa = \frac{2 m_\nu}{v_H^2} = Y_\nu  M_M^{-1}  Y_\nu^T $$
Since the light neutrino states are physically observable, the light neutrino mass matrix can be diagonalized using the PMNS matrix
$$D_{\kappa}= U_{PMNS}^T \ \kappa \ U_{PMNS}  = diag (m_{\nu_1} , m_{\nu_2}, m_{\nu_3})$$
$$=> D_{\kappa}= U_{PMNS}^T  Y_\nu  D_{M_M^{-1}}  Y_\nu^T \ U_{PMNS} =  U_{PMNS}^T  Y_\nu  D_{\sqrt {M_M^{-1}}}D_{\sqrt {M_M^{-1}}} Y_\nu^T \ U_{PMNS}$$
where, in obvious notation, $D_{\sqrt{A}} = \sqrt{D_A}$. Multiplying both members of the above equation by $D_{\sqrt {\kappa^{-1}}}$ from the left and from the right, we get
$$\mathbb{1} = [D_{\sqrt {M_M^{-1}}} Y_\nu^T U_{PMNS} D_{\sqrt {\kappa^{-1}}}]^T[D_{\sqrt {M_M^{-1}}} Y_\nu^T U_{PMNS} D_{\sqrt {\kappa^{-1}}}] $$
whose solution is $D_{\sqrt {M_M^{-1}}} Y_\nu^T U_{PMNS} D_{\sqrt {\kappa^{-1}}} = R$, where $R$ is a $3 \times 3$ complex orthogonal matrix ($R^{T}R = \mathbb{1}$), which represents the possible mixing in the right-handed neutrino sector. $R$ can be parameterized in terms of three complex angles,$\theta_i (i=1,2,3)$ as 
$$R = \begin{pmatrix} 
1 & 0 & 0\\ 
0 & c_{23} & s_{23} \\
0 & -s_{23} & c_{23}
\end{pmatrix}
\begin{pmatrix} 
c_{31} & 0 & s_{31}\\ 
0 & 1 & 0 \\
-s_{31} &0 & c_{31}
\end{pmatrix} 
\begin{pmatrix} 
c_{12} & s_{12} &0\\ 
-s_{12} & c_{12} & 0 \\
0 &0 & 1
\end{pmatrix}$$
We will set R to the unit matrix and use the best-fit values for the neutrino oscillation parameters as given in \cite{Forero}
$$\Delta m_{12}^2 = 7.60 \times 10^{-5} eV^2 , \Delta m_{13}^2 = 2.48 \times 10^{-3} eV^2,  $$
$$ sin^2 \theta_{12} = 0.323, sin^2 \theta_{23} = 0.467, sin^2 \theta_{31} = 0.0234. $$
Hence, to reproduce the physical, low-energy, parameters i.e. the light neutrino masses (contained in $D_\kappa$) and mixing angles and CP phases (contained in $U$), the most general $Y_\nu$ matrix is given by $$Y_\nu^T = D_{\sqrt{M_M}}R D_{\sqrt{\kappa}} U^\dagger_{PMNS}$$
$$=> m_D^T = \frac{\sqrt{2}}{v_H} \sqrt{diag(M_M)} R \sqrt{diag(m_\nu)}U^\dagger_{PMNS}$$
This is called the Casas-Ibarra parametrization of the Dirac mass matrix \cite{Casas:2001sr}. The advantage of this parameterization is that instead of using as input parameters the seesaw mass matrices $m_D$ and $M_M$, it uses the three physical light neutrino masses, the three physical heavy neutrino masses, the $U_{PMNS}$ matrix and a general complex $3\times3$ orthogonal matrix $R$. Thus, instead of proposing directly possible textures for $m_D$, or $Y_\nu$, one proposes possible values for $m_{M_1} , m_{M_2}, m_{M_3}$ and $R$ and $m_{\nu_1} , m_{\nu_2}, m_{\nu_3}$ also $U_{PMNS}$ to their suggested values from the experimental data. 
\newline
\newline
A particular choice for $Y_\nu$ can be taken, which corresponds to take $R = \mathbb{1}$ in the above equation. This is equivalent to assuming that there exists a basis of $L_i$ and $\nu_R$ in which $Y_\nu$ and $M_\nu$ are simultaneously diagonal (though not $Y_e$). To see this notice that if $R = \mathbb{1}$ then $Y_\nu$ in the above equation can be made diagonal by rotating $L_i$ with $U_{PMNS}$. This hypothesis may be consistent with certain models, namely when all the lepton flavor violations can be attributed to the sector of charged leptons, but it is not general, and it is inconsistent with other scenarios. So, any hypothesis for $R$ different from the unit matrix will lead to an additional lepton flavor mixing, besides the one introduced by the $U_{PMNS}$.
\newline
\newline
From the above equation, we can see that the mixing matrix $U_{PMNS}$ now has functional dependence on $M_M$. As a result, $U_{PMNS}$ is now dependent on the right-handed neutrino scale.
\section{Left-Right Symmetric Model (LRSM)}
\label{Left-Right Symmetric Model (LRSM)}
The Standard Model (SM) of particle physics exhibits immediate shortcomings, primarily in its treatment of parity violation, as it essentially gauges the $V-A$ theory. This gauge theory predicts massless neutrinos because only left-handed components are charged under the SM gauge group. Upon initial examination, there appears to be no direct link between the absence of $V+A$ charged currents and the absence of massive neutrinos. However, through the study of neutrino oscillations, we have established that neutrinos possess small but non-zero masses. This challenges the SM's predictions, and it is reasonable to assume that $V+A$ currents may exist.

In response to these shortcomings, physicists pursue an approach known as Beyond the Standard Model (BSM). Rather than constructing an entirely new theoretical framework from scratch, the BSM approach aims to enhance the existing SM by addressing its limitations and explaining phenomena it cannot account for. A notable advantage of this approach is that the SM describes processes occurring at energies attainable by current particle accelerators. Consequently, BSM theories can potentially be explored and tested in these accelerators. One such prominent extension of the SM that addresses the issue of massive neutrinos is the Left-Right Symmetric Model (LRSM) \cite{Mohapatra:1974gc}, \cite{Senjanovic:1975rk}, \cite{Grimus:1993fx}, \cite{Corrigan:2015kfu}, and \cite{Liu:1985zp}. The LRSM introduces several key features:\begin{itemize}
\item \textbf{Spontaneous Parity Violation}: The LRSM predicts the spontaneous breaking of parity, with the breaking scale corresponding to the mass of the right-handed $SU(2_R)$ gauge bosons, denoted as $W_R$. Importantly, the mass of the $W_R$ bosons ($M_{W_R}$) is expected to be significantly greater than that of the left-handed $W_L$ bosons, and the SM is recovered in the limit where $M_{W_R}$ tends towards infinity.
\item \textbf{Neutrino Masses}: Within the same process of spontaneous symmetry breaking (SSB), right-handed neutrinos also acquire large masses. These heavy right-handed neutrinos, in turn, induce small masses for their left-handed counterparts via the seesaw mechanism \cite{Mohapatra:1979ia}. The LR symmetry of the model enables the prediction of Dirac mass terms for neutrinos based on the light and heavy neutrino masses and their mixing, shedding light on the origin of neutrino masses through the Higgs mechanism.
\end{itemize}
This transformation results in right-handed neutrinos transitioning from sterile particles to fully interacting ones, akin to their left-handed counterparts. Furthermore, the Yukawa coupling ($y_D$) to the Higgs boson, responsible for the Type I seesaw contribution, depends on only a few observable parameters. This unique characteristic gives rise to a remarkable signature of lepton flavor violation in collider experiments. Additionally, the LRSM introduces novel gauge bosons and scalar particles, which are expected to have larger masses than their SM counterparts. This expanded particle spectrum creates opportunities for their potential detection in high-energy colliders.

\subsection{Gauge Boson Sector}
\subsubsection{Gauge Group Structure}

LRSM has been studied extensively since the 1970s \cite{Mohapatra:1974gc}. The most common and popular version of the model is called the Minimal LRSM and is defined by its scalar sector. The usual SM Higgs doublet is replaced by a bi-doublet and two complex triplets (left-handed and right-handed) are introduced into the theory. We also expect three extra gauge bosons, two charged $W^\pm_R$ analogous to the $W^\pm_L$ boson and one neutral $Z_R$ analogous to the $Z$ boson. The vev of the right-handed triplet is large, resulting in the gauge bosons $Z_R$ and $W^\pm_R$ having much larger masses than the SM gauge bosons. We take the transformation between L and R fields to be parity and impose parity invariance before spontaneously breaking down to $U(1)_{EM}$.  We also need to add a $U(1)_X$ factor so that we get mixing (Recall SM: we define hypercharge U(1) so that neutral gauge bosons can mix. The completeness of nonabelian group algebra implies this mixing). As the color part in the gauge structure is trivial and unbroken gauge group, we will not discuss that part. The simple extension is 
$$ SU(2)_L \times SU(2)_R \times U(1)_X $$
where we must find out the generator of this group. The structure is the same as the left part except that it acts on the right-handed doublets. Thus, fermion representations will change now. 
We know that this LR model gauge structure must break ultimately in the electromagnetism in the low energy. Thus,
$$Q_{EM}=\frac{\tau ^3}{2} + \frac{X}{2}$$

where, $\tau ^i $ are the $2 \times 2$ Pauli matrices. 

Now, first, the left-handed and right-handed doublets have the same electromagnetic charge. And $\tau ^3 $ being the same for them, "X" must be the same for both the left and right parts. 
\begin{itemize}
\item For electron: $Q_{EM}=-1$	, $\tau ^3 =-1$ , thus $X = -1$ 
\item For neutrino: $Q_{EM}=0$	, $\tau ^3 =+1$ , thus $X = -1$ 
\end{itemize}
Thus, for leptons $X = -1$ and similarly from the calculation, we find that for quarks $X = 1/3$. 

We can compactly write $X$: $X = B - L $ where B is the baryon number and L is the lepton number. (For electron B = 0 and L = 1; for quark B = 1/3 and L = 0)

So, we have found the gauge structure: 
$$ G_{LR} = SU(2)_L \times SU(2)_R \times U(1)_{B-L} $$
The irreducible representation of the gauge group: ($d_L$,$d_R$,$B-L$)
where, $D_{L,R}$ is the dimension of the representation of $SU(2)_{L,R}$.

The generators of the $SU(2)_{L,R}$ group is $T_L,R$. Thus, we can write the unbroken generator, (${T_3 = \frac{1}{2} \tau ^3}$)
$$Q_{EM}=T_{3L} +T_{3R}+ \frac{1}{2} \ (B-L)$$

\subsubsection{Gauge transformation of Multiplets}

Suppose $\psi$ is the fermions that are in the fundamental representation of the groups. $\phi$ is the Higgs bi-doublet. It is charged under $SU(2)$ and carries a $SU(2)$ index, thus, adjoint representation. 

$W$’s are the boson corresponding to $SU(2)$ group. It is in the adjoint representation (Non-abelian nature). $B$ is the boson corresponding to $U(1)$ group. It is in the adjoint representation (Abelian nature).
\newline
\newline
For $U_{L,R} \in SU(2)_{L,R}$ a $2 \times 2 $ unitary matrix,
$$\psi_{L,R} \rightarrow e^{-i(\frac{B-L}{2})\alpha} U_{L,R} \psi_{L,R}$$
$$\phi \rightarrow U_L \phi U_R^\dagger$$
$$\vec{T} \cdot \vec{W}_{L,R\mu} \rightarrow U_{L,R} \vec{T} \cdot \vec{W}_{L,R\mu} U_{L,R}^\dagger + \frac{i}{g} (\partial_\mu U_{L,R}) U_{L,R}^\dagger $$
$$B_\mu \rightarrow B_\mu + (\frac{1}{g'}) \partial_\mu \alpha$$

\subsubsection{Gauge Boson Lagrangian}

Kinetic terms for the gauge fields and interactions between them in the LRSM model are the same as the Standard model, but this time W bosons also couple to right-handed fermions due to symmetry extension $SU(2)_R$.
$$\mathcal{L} = -\frac{1}{4} F_{L\mu \nu}^a F^{\mu \nu}_{L a} - \frac{1}{4} F_{R\mu \nu}^a F^{\mu \nu}_{R a} - \frac{1}{4} B_{\mu \nu}B^{\mu \nu} $$ 
where, $SU(2)_{L/R}$ field strength tensor have the following form: $$F_{\mu \nu}^a = \partial_{\mu} W^a_{\nu}-\partial_{\nu} W^a_{\mu}+g \epsilon_{abc} W^b_{\mu} W^c_{\nu}$$ and $B_{\mu \nu}$ is the $U(1)_{B-L}$ field strength tensor $$B_{\mu \nu} = \partial_{\mu} B_{\nu}-\partial_{\nu} B_{\mu}$$

\subsection{Fermion Sector}

Fermions in this model form $SU(2)$ doublets in both left and right chiral fields. The fermion fields are
$$  L_{Li} = \begin{pmatrix}
\nu \\ e
\end{pmatrix}_{Li} , \  L_{Ri} = \begin{pmatrix}
\nu \\ e
\end{pmatrix}_{Ri} ; \ Q_{Li} = \begin{pmatrix}
u \\ d
\end{pmatrix}_{Li} , \ Q_{Ri} = \begin{pmatrix}
u \\ d
\end{pmatrix}_{Ri} $$
 Under the gauge group, their representation is given by 
$$\{Q_{L_i} (2,1,\frac{1}{3}), Q_{R_i} (1,2,\frac{1}{3}), L_{L_i} (2,1,-1),L_{R_i} (1,2,-1) \}$$

\subsubsection{Fermion-Gauge Lagrangian} 

The fermionic gauge Lagrangian includes kinetic terms for the fermions and interactions between fermion and gauge fields. By putting left- and right-handed fermions on the same footing, it is given by: 
$$\mathcal{L} = \sum_{\psi=Q,L}[\bar{\psi_L} i \gamma ^\mu (\partial_\mu+i g_L \vec{T} \cdot \vec{W}_{L \mu} +ig' \frac{B-L}{2} B_\mu) \psi_L +\bar{\psi_R} i \gamma ^\mu (\partial_\mu+i g_R \vec{T} \cdot \vec{W}_{R \mu} +ig' \frac{B-L}{2} B_\mu) \psi_R]$$
Here, $g_{L/R}$  is the gauge coupling of the group $SU(2)_{L/R}$ and $g'$ is the gauge coupling of the group $U(1)_{B-L}$. If $g_L = g_R$ is enforced, then it is called Minimally Parity Symmetric Theory.

\subsection{Higgs Sector}

We wish, as in the SM, to spontaneously break $SU(2)_L \times U(1)_Y \xRightarrow{\text{SSB}} U(1)_{EM}$. Thus, we must first break $SU(2)_R \times U(1)_{B-L} \xRightarrow{\text{SSB}} U(1)_{Y}$, requiring an extended Higgs sector. For phenomenological reasons, we require that this happens at a higher scale. This is to ensure
a high mass for the new (apparently hidden) $W_R$; $Z_R$ bosons; we discuss this further in Spontaneous Symmetry Breaking section.

\subsubsection{Standard Model like Yukawa Interaction}

To give fermions Dirac mass, we need to introduce Yukawa terms in the Lagrangian analogous to SM.
$$\mathcal{L}_{Yukawa}(\phi)= -\sum_{\psi=Q,L} [ \bar{\psi}_{Li} Y_{ij}^U \phi \psi_{Rj}+\bar{\psi}_{Li} Y_{ij}^D \tilde{\phi} \psi_{Rj} + h.c.]$$

where, $\phi$ is a $2 \times 2$ matrix of scalar fields and $\tilde{\phi}=\tau ^2 \phi^* \tau^2$. Here i and j are the flavor space index. 

$\psi$ and $\bar{\psi}$ are charged under $U(1)_{B-L}$. But $\bar{\psi}\psi$ is neutral under $U(1)_{B-L}$. Similarly, $\bar{\psi} \phi \psi$ is also neutral under $U(1)_{B-L}$. This implies that $\Phi$ transforms as (2,2,0) of the gauge group. So, $\phi$ cannot break the $U(1)_{B-L}$ group as it is neutral under that group. This bidoublet is useful only for breaking the $SU(2)$ group. Thus, we need to extend the Higgs sector.
\newline
\newline
\textbf{Charge of the $\phi$ fields}
\newline
\newline
The fields that are neutral under Electromagnetism can acquire vacuum expectation value (the vev) without violating $U(1)_{EM}$. This vev will do the Higgs transition and also give mass to the coupled fermions and bosons of the theory. 
As $\phi$ is in the adjoint representation, the charge operator on $\phi$ is given by ($\phi$ is neutral under $U(1)_{B-L}$) 
$$ Q\phi = [\frac{1}{2}\tau^3 , \phi] =\frac{1}{2}\tau^3  \phi -\frac{1}{2} \phi \tau^3 $$
$$=\frac{1}{2}\begin{pmatrix}
1 & 0 \\
0 & -1 
\end{pmatrix}
\begin{pmatrix}
\phi_{11} & \phi_{12} \\
\phi_{21} & \phi_{22} 
\end{pmatrix}
-
\frac{1}{2} \begin{pmatrix}
\phi_{11} & \phi_{12} \\
\phi_{21} & \phi_{22} 
\end{pmatrix}
\begin{pmatrix}
1 & 0 \\
0 & -1 
\end{pmatrix}
$$
$$= \begin{pmatrix}
0 \times \phi_{11} & +1 \times \phi_{12} \\
-1 \times \phi_{21} & 0 \times \phi_{22} 
\end{pmatrix}
$$
So, we can write the $\phi$ field as 
$$\phi=\begin{pmatrix}
\phi^0_1 & \phi^+_1 \\
\phi^-_2 & \phi^0_2 
\end{pmatrix} = \begin{pmatrix}
\phi_1 \\
\phi^c_2 
\end{pmatrix} : (2,2,0)$$

Thus, $\phi^0_1$ and $\phi^0_2$ are neutral under $Q_{EM}$ which can get vacuum expectation values.

\subsubsection{Extending the Higgs Sector}

We want to find a Higgs transition that takes us to the Standard model. But for that, the Higgs field must be charged under $U(1)_{B-L}$. As our $\Phi$ field is neutral under $U(1)_{B-L}$, the breaking of $\Phi$ can not break parity and thus Higgs transition does not lead to SM. 

To get a parity-violating Higgs field we must define a new field $\Delta$ such that it is singlet in $SU(3)$ color, but it must have $U(1)_{B-L}$ charge. For parity violation, we need two scalars $\Delta_L$ and $\Delta_R$ corresponding to charged under $SU(2)_L$ and $SU(2)_R$. By taking the two Higgs transitions corresponding to $\Delta_L$ and $\Delta_R$ at different energies we can get parity violated phase.  
\newline
\newline
\textbf{Representation of $\Delta$ Higgs}

Now we will determine the representation of $\Delta_L$ and $\Delta_R$ under the gauge group. $\Delta$ couple to fermions such that Lagrangian is gauge invariant. We will use this constraint to determine its representation. 

The representation of fermions and its conjugates under $$G_{LR} = SU(2)_L \times SU(2)_R  \times U(1)_{B-L}$$ 
is (For leptons) 
$$\psi_L : (2,1,-1) \ \& \ \hat{\psi}_R : (2,1,-1)$$ where the conjugate filed is defined by $\hat{\psi}_R = \gamma^0 C i\tau^2 \psi_L^*$ (Note: Under $SU(2)$, the conjugation is $\hat{\psi}\rightarrow i\tau^2 \psi$ and under spinor index, the conjugation is $\hat{\psi}(x)\rightarrow \gamma^0 C \psi^* (x)$. Our field fermion is spinor in spacetime and also $SU(2)$ fundamental representation).

The fermion bilinear is $$\bar{\psi}_L\hat{\psi}_R : (2,1,-1) \times (2,1,-1) = (1,1,-2) + (3,1,-2)$$
The first bilinear is neutral under both $SU(2)_{L, R}$. Thus, to couple $\Delta_L$ to the fermion bilinear, we must consider the second bilinear. The representation of $\Delta_L$ that can couple to fermion bilinear with Lagrangian gauge invariant is $\Delta_L : (3,1,2)$.

Similarly, for the right-handed part, we can show $\Delta_R : (1,3,2)$.
So, in general for fermions (leptons and quarks) with both left and right-handedness, we need the Higgs scalars
$$\Delta_L : (3,1,2) \ \& \ \Delta_R : (1,3,2)$$ 
to form gauge invariant Lagrangian and by breaking it, we will get the desired transition. 

Thus, the LRSM model contains no more than three scalar multiplets: bidoublet  $\phi$, and scalar triplets $ \Delta_L $, $ \Delta_R $.
Now, as $ \Delta_L $, $ \Delta_R $ are color neutral, due to $B-L = 2$ (will be discussed in the upcoming section) implies that $L = -2$ for $\Delta_L$, $\Delta_R$. ($B = 0$ for color neutral)

\subsubsection{Yukawa Interaction of Extended Higgs}

From the conjugate field definition 
$$\hat{\psi} = \gamma^0 C  \psi^*$$
$$\bar{\hat{\psi}}=\hat{\psi}^\dagger \gamma ^0 = (\gamma ^0 C \psi ^*)^\dagger \gamma ^0 = \psi ^T C^{-1}(\gamma ^0)^\dagger \gamma ^0 = \psi ^T C^{-1}$$
Thus, the Lagrangian is
$$\mathcal{L}_{Yukawa}(\Delta)=\sum_{\psi=L}[L_{Li}^T  G_{Lij} C^{-1}i \tau ^2 \Delta_L L_{Lj}+L_{Ri}^T  G_{Rij} C^{-1}i \tau ^2 \Delta_R L_{Rj}]+h.c.$$
where $G_{Lij}$ and $G_{Rij}$ are the Yukawa coupling of $\Delta$ corresponding to group $SU(2)_L$ and $SU(2)_R$ respectively. 
Their transformation under gauge group $G$ is (inserting $B-L = 2$)
$$ \Delta_L \rightarrow e^{-i \alpha}U_L \Delta_L U_L^\dagger$$ and $$\Delta_R \rightarrow e^{i \alpha}U_R \Delta_R U_R^\dagger$$
Thus, $ \Delta_L $ and $ \Delta_R $  are complex $SU(2)$ triplets with lepton number 2
$$ \Delta \equiv \frac{i}{\sqrt{2}}
\begin{pmatrix}
\delta_3 & \delta_1 - i \delta_2 \\
\delta_1 + i \delta_2 &  -\delta_3 
\end{pmatrix}
$$
\textbf{Charge of the $\Delta$ fields}
$$Q \Delta = [\frac{1}{2} \tau ^3 , \Delta]+\frac {(B-L)}{2} \Delta$$
For the vacuum to be electrically neutral, the field that takes on a non-zero vev must have an electric charge of zero. This means either $B-L = 0$ and $\delta^3$ breaks the symmetry, or $B-L = \pm 2$ and $(\delta^1 \pm i\delta^2)$ breaks the symmetry. The solution $B-L = 0$ does not completely break the $SU(2)_R$ symmetry as required. This will be shown in the Goldstone mode section by letting each generator act on the vacuum state and observing that the third generator still annihilates the vacuum, i.e. isn't broken. Therefore, we must choose $B-L = \pm 2$. It was shown in the Yukawa interaction of Extended Higgs that to form Majorana mass terms for the right-handed neutrinos we need a right-handed scalar field with $B-L = 2$. Now,  
$$Q \Delta =[\frac{1}{2} \tau ^3 , \Delta]+ 1 \Delta$$
$$ = \begin{pmatrix}
1 \cdot \Delta_{11} & 2 \cdot \Delta_{12} \\
 0 \cdot \Delta_{21}& 1 \cdot \Delta_{22}
\end{pmatrix}
$$
So, we can write the $\Delta$ field as, 
$$\Delta = 
\begin{pmatrix}
\frac{1}{\sqrt{2}} \delta^+ &  \delta^{++} \\
 \delta^0 & -\frac{1}{\sqrt{2}}\delta^+
\end{pmatrix}
$$
Only the neutral part can get the vacuum expectation value i.e., $\delta^0$.
\newline
\newline
Therefore, Total Scalar and Yukawa Lagrangian can be written as
$$\mathcal{L}_{scalar}= Tr[(D_\mu \Phi)^\dagger(D^\mu \Phi)+(D_\mu \Delta_L)^\dagger(D^\mu \Delta_L)+(D_\mu \Delta_R)^\dagger(D^\mu \Delta_R)]$$
$$\mathcal{L}_{Yuk}(\Delta , \phi)=-\sum_{\psi=Q,L} [ \bar{\psi}_{Li} Y_{ij}^U \phi \psi_{Rj}+\bar{\psi}_{Li} Y_{ij}^D \tilde{\phi} \psi_{Rj} + h.c.] + $$
$$\sum_{\psi=L}[L_{Li}^T  G_{Lij} C^{-1}i \tau ^2 \Delta_L L_{Lj}+L_{Ri}^T  G_{Rij} C^{-1}i \tau ^2 \Delta_R L_{Rj}]$$
where, 
$$D_\mu \phi = \partial_\mu \phi +ig (\vec{\frac{\tau}{2}}\cdot\vec{W}_{L\mu}\phi - \phi \vec{\frac{\tau}{2}}\cdot\vec{W}_{R\mu}) $$

$$D_\mu \Delta_{L,R}= \partial_\mu \Delta{L,R} + ig[\vec{\frac{\tau}{2}}\cdot\vec{W}_{L,R\mu} , \Delta_{L,R}]+ig' B_\mu \Delta_{L,R} $$
are the covariant derivatives of the fields.

\subsection{Physical Consequences of Spontaneous Symmetry Breaking}
\subsubsection{The vacuum expectation value}

Since the electromagnetic gauge symmetry $U(1)_{EM}$ should not be broken after SSB in any extension of the SM, only the electric neutral components (namely $\phi^0_1, \phi^0_2, \delta_L^0, \delta_R^0$) can acquire nonzero VEVs. In general, the charge conserving VEVs of bi-doublet $\phi$ and two triplets $\Delta_{L, R}$ should be all the neutral components of Higgs get vevs, 
$$ \langle \phi \rangle_0 = \frac{1}{\sqrt{2}} \begin{pmatrix}
ke^{i\alpha_1} & 0 \\
0 & k' e^{i\alpha_2}
\end{pmatrix} $$

$$\langle \Delta_L \rangle_0 = \frac{1}{\sqrt{2}} \begin{pmatrix}
0 & 0 \\
v_L e^{i\theta_1} & 0
\end{pmatrix}$$ 

$$\langle \Delta_R \rangle_0 = \frac{1}{\sqrt{2}} \begin{pmatrix}
0 & 0 \\
v_R e^{i\theta_2} & 0
\end{pmatrix} $$
Here we have four neutral components and thus four phases. We have three generators $T_{3L}$ , $T_{3R}$ and $B-L$ commuting with $Q_{EM}$. So, we can eliminate phases. Consider $\theta_L$, $\theta_R$ and $\theta_{B-L} $ are the transformation parameters of the generators respectively. Then the Higgs fields transform like
$$ \langle \phi \rangle \rightarrow e^{iT_{3L}\theta_L} \langle \phi \rangle  e^{-iT_{3R}\theta_R} $$
$$ \langle \Delta_L \rangle \rightarrow e^{iT_{3L}\theta_L} \langle \Delta_L \rangle  e^{-iT_{3L}\theta_L} e^{i\theta_{B-L}} $$
$$ \langle \Delta_R \rangle \rightarrow e^{iT_{3R}\theta_R} \langle \Delta_R \rangle  e^{-iT_{3R}\theta_R} e^{i\theta_{B-L}} $$
Thus, the phase transforms like 
$$\alpha_1 \rightarrow \alpha_1 +\frac{1}{2} \theta_L-\frac{1}{2} \theta_R $$
$$\alpha_2 \rightarrow \alpha_2 -\frac{1}{2} \theta_L+\frac{1}{2} \theta_R $$
$$\theta_{1,2} \rightarrow \theta_{1,2} - \theta_{L,R}+\theta_{B-L} $$
Thus, it is clear that there are two independent combinations of parameters, allowing removal in two phases only. It is a convention to set $\alpha_1 = 0 = \theta_2$. Thus, the Higgs vev's are given by 
$$\langle \phi \rangle_0 = \frac{1}{\sqrt{2}} \begin{pmatrix}
k & 0 \\
0 & k' e^{i\alpha}
\end{pmatrix} \ , \ \langle \Delta_L \rangle_0 = \frac{1}{\sqrt{2}} \begin{pmatrix}
0 & 0 \\
v_L e^{i\theta_L} & 0
\end{pmatrix} \ \& \  \langle \Delta_R \rangle_0 = \frac{1}{\sqrt{2}} \begin{pmatrix}
0 & 0 \\
v_R  & 0
\end{pmatrix} $$ 
Now, considering manifest CP violating LRSM (i.e. CP violation originates from Yukawa couplings so $\alpha = 0 = \theta_L$ ) we have
$$\langle \phi\rangle _0 = \frac{1}{\sqrt{2}}
\begin{pmatrix}
k &  0\\
0 & k'
\end{pmatrix} \ 
\& \  \langle \Delta_{L,R} \rangle _0 = \frac{1}{\sqrt{2}}
\begin{pmatrix}
0 &  0\\
v_{L,R} & 0
\end{pmatrix}
$$
Assuming the order of magnitude relations: $\lvert v_L \rvert ^2 <<\lvert k \rvert ^2+\lvert k' \rvert ^2 << \lvert v_R \rvert ^2$. This relation ensures that with decreasing temperature, right-handed gauge group breaking took place much earlier than the left part. This gives us the desired spontaneous parity-breaking mechanism.
 
Now just considering the $U(1)_R$, the right handed gauge group breaking along with the $B–L$ abelian group, generates $U(1)_Y$, the Standard Model hyper-charge symmetry
$$SU(2)_R \times U(1)_{B-L} \rightarrow U(1)_Y$$
Here, with this breaking of $\Delta_R$, we obtain SM. Thus, the total breaking scheme: 
$$ SU(2)_R \times SU(2)_L \times U(1)_{B-L} \xRightarrow{\text{SSB-1}} SU(2)_L \times U(1)_Y \xRightarrow{\text{SSB-2}} U(1)_{EM}$$
In the intermediate stage, we have an extended standard model with two Higgs doublets $\phi$, right-handed neutrino $\nu_R$, $\delta$’s, and the effect of the right-handed scale. However by breaking of $\langle \Delta_{L, R} \rangle _0$ lepton number is violated. But this will not generate Goldstone modes as $B-L$ is associated with the gauge boson $B_\mu$. There will be a Goldstone boson in the theory if the breaking of the symmetry is global. But in our case it's local. 

The mass matrix and the mixings are obtained from the Lagrangian evaluating at vacuum expectation values of the three Higgs fields at tree level. The fermion cases are obtained from Yukawa Lagrangian, and the boson cases are obtained from scalar field Lagrangian i.e., 
$\mathcal{L}_{Yukawa}(\Delta=\langle \Delta \rangle_0, \phi = \langle \phi \rangle _0) $ and $\mathcal{L}_{Scalar}(\Delta=\langle \Delta \rangle_0, \phi = \langle \phi \rangle _0)$

\subsubsection{Gauge Boson Particle Spectrum}

\begin{enumerate}
\item \textbf{Gauge Boson Lagrangian after SSB-1}

Recall the covariant derivative for a field in the adjoint representation in $SU(2)_{L, R}$ and charged under $U(1)_{B-L}$ is given by 
$$D_\mu \Delta_{L,R}= \partial_\mu \Delta_{L,R} +ig [\vec{\frac{\tau}{2}} \cdot \vec{W}_{L,R \mu} ,\Delta_{L,R}] + i g' B_\mu \Delta_{L,R}$$
The Pauli matrices are given by 
$$ \tau^1 =
\begin{pmatrix}
0 & 1\\
1 & 0
\end{pmatrix}
 \
\tau^2 =
\begin{pmatrix}
0 &  -i\\
i & 0
\end{pmatrix}
 \
\tau^3 =
\begin{pmatrix}
1 & 0\\
0 & -1
\end{pmatrix}
$$
$$W_\mu ^a \tau^a = \begin{pmatrix}
W_3 ^\mu & \sqrt{2}W^{+\mu}\\
\sqrt{2}W^{-\mu} & -W_3 ^\mu
\end{pmatrix}
$$
Thus,
$$ D_\mu \langle \Delta_L \rangle = \frac{i v_L}{\sqrt{2}}
\begin{pmatrix}
\frac{g}{\sqrt{2}}W_{L\mu}^+ &  0\\
-gW_{L \mu}^3+g'B_\mu & \frac{-g}{\sqrt{2}}W_{L\mu}^+
\end{pmatrix}$$
where $$ W_{L,R}^\pm =\frac{1}{\sqrt{2}}(W_{L,R1} \mp i W_{L,R2}) $$
Here, $W^{\pm}$ states are the interaction eigenstates and $W_{1,2}$ states are the mass eigenstates.

Taking Hermitian conjugate of  $D_\mu \langle \Delta_L \rangle$, we get
$$ (D_\mu \langle \Delta_L \rangle)^\dagger=  \frac{-i v_L}{\sqrt{2}}
\begin{pmatrix}
\frac{g}{\sqrt{2}} W_{L \mu}^- & -gW_{L \mu}^3 +g'B_\mu\\
0 & \frac{-g}{\sqrt{2}}W_{L \mu}^-
\end{pmatrix}$$

$$ [(D_\mu \langle \Delta_L \rangle)^\dagger  D^\mu \langle \Delta_L \rangle] = \frac{v_L^2}{2} 
\begin{pmatrix}
\frac{g^2}{2}W_{L \mu}^- W_L^{+\mu}+(-gW_{L\mu ^3} + g' B_\mu)^2  & -   \\
- & \frac{g^2}{2} W_{L \mu}^- W_L^{+ \mu}
\end{pmatrix}
  $$
Off-diagonal parts are irrelevant as we will take the trace. 
$$ Tr [ (D_\mu \langle \Delta_L \rangle)^\dagger  D^\mu \langle \Delta_L \rangle ] = \frac{1}{4}v_L^2 (2g)^2 W_{L \mu}^- W_L^{+\mu}+\frac{v_L^2}{2}(-g W_{L\mu}^3+g'B_\mu)^2 $$
Similarly, for the right-handed part: 
$$ Tr [  (D_\mu \langle \Delta_R \rangle)^\dagger  D^\mu \langle \Delta_R \rangle ] = \frac{1}{4}v_R^2 (2g)^2 W_{R \mu}^- W_R^{+\mu}+\frac{v_R^2}{2}(-g W_{R\mu}^3+g'B_\mu)^2 $$
\item \textbf{Gauge boson Lagrangian after SSB-2}

From the definition of the covariant derivative of the field $\phi$ in the $(2,2,0)$ representation, 
 $$D_\mu \phi = \partial_\mu \phi -i \frac{g}{2} (W_{L\mu} ^i \tau^i \phi - \phi W_{R\mu} ^i \tau^i)$$

$$ W_{L\mu} ^a \tau^a \langle \phi \rangle = \frac{1}{\sqrt{2}}\begin{pmatrix}
k W_{3L} ^\mu & \sqrt{2} k' W_L ^{+\mu}\\
\sqrt{2} k W_L ^{-\mu} & -k' W_{3L} ^\mu
\end{pmatrix}
$$
$$ \langle \phi \rangle W_{R\mu} ^a \tau^a = \frac{1}{\sqrt{2}}\begin{pmatrix}
k W_{3R} ^\mu & \sqrt{2} k W_R ^{+\mu}\\
\sqrt{2} k' W_R ^{-\mu} & -k' W_{3R} ^\mu
\end{pmatrix} $$
Substituting these we get the covariant derivative of the vacuum expectation value of the field and its derivative is, 
$$D_\mu \langle \phi \rangle = -i \frac{g}{2} (W_{L\mu} ^i \tau^i \langle \phi \rangle - \langle \phi \rangle W_{R\mu} ^i \tau^i)$$
$$= -i \frac{g}{2}  \begin{pmatrix}
\frac{k}{\sqrt{2}}(W_{3L\mu}-W_{3R\mu}) & k' W_{L\mu} ^+ - k W_{R\mu} ^+\\
 k W_{L\mu}^- - k' W_{R\mu} ^-&  - \frac{k'}{\sqrt{2}}(W_{3L\mu} - W_{3R\mu})
\end{pmatrix}$$
$$(D_\mu \langle \phi \rangle)^\dagger = i \frac{g}{2}  \begin{pmatrix}
\frac{k}{\sqrt{2}}(W_{3L\mu}-W_{3R\mu}) & k W_{L\mu} ^+ - k' W_{R\mu} ^+\\
 k' W_{L\mu}^- - k W_{R\mu} ^- &  - \frac{k'}{\sqrt{2}}(W_{3L\mu} - W_{3R\mu})
\end{pmatrix}$$
Multiplying them and taking trace we get, 
$Tr[(D_\mu \langle \phi \rangle)^\dagger D_\mu \langle \phi \rangle] = \frac{g^2}{4} \frac{1}{2}(k^2+k'^2)(W_{3L\mu}-W_{3R\mu})^2 + \frac{g^2}{4}[(k^2+k'^2) (W_{L\mu} ^+ W_{L \mu} ^- +W_{R\mu} ^+ W_{R \mu} ^-)  - 2 kk' ( W_{R\mu} ^+ W_{L\mu} ^- +W_{L\mu} ^+ W_{R\mu} ^- )]$
\newline
\newline
\item \textbf{Mass Term and Mixing of Gauge Boson}
\begin{enumerate}
\item \textbf{For Charged Gauge Boson}

Mass term Lagrangian for the charged bosons $W^{\pm}$ are, 
$$\mathcal{L}_{\Delta} ^{Charged} = \frac{1}{4} v_L ^2 (2 g^2) W_L ^{-\mu} W_{L \mu} ^+ + \frac{1}{4} v_R ^2 (2 g^2) W_R ^{-\mu} W_{R \mu} ^+ $$
$$\mathcal{L}_{\phi} ^{Charged} = \frac{g^2}{4} [(k^2+k'^2) (W_{L\mu} ^+ W_{L \mu} ^- +W_{R\mu} ^+ W_{R \mu} ^-)  - 2 kk' ( W_{R\mu} ^+ W_{L\mu} ^- +W_{L\mu} ^+ W_{R\mu} ^- )]$$
both Lagrangian above can be written in terms of mass matrix as
$$\mathcal{L}_{mass} ^{(W^{\pm})} = \begin{pmatrix}
W_L^- & W_R^-\\
\end{pmatrix}\begin{pmatrix}
\frac{1}{4} g^2 (k^2 + k'^2 +2 v_L ^2) &  - \frac{1}{2} g^2 kk' \\
- \frac{1}{2} g^2 kk' & \frac{1}{4} g^2 (k^2 +k'^2 +2v_R ^2)
\end{pmatrix}\begin{pmatrix}
W_L ^+ \\
W_R ^+
\end{pmatrix}$$
Here, $W^{\pm}_{L, R}$ is the states in the interaction eigenstate, not the mass eigenstate because their mass matrix is not diagonal. That means these are mixtures of mass states. By diagonalizing this mass matrix we can get the mass eigenstate $W_{1,2}^{\pm}$. So, now we need to determine the masses of these states and the rotation angle which is the angle we need to rotate the interaction eigenstates to get the mass eigenstates. As $W^{\pm}_{L, R}$ and $W_{1,2}^{\pm}$ are separately orthogonal systems, there is an angle between them. 
\newline
\newline
Note: these index $1,2$ are not the lie algebra index. $W^a$ where $a =1 ,2 ,3$ are the lie algebra index. We used $W_{L,R} ^{a=1}$ and $W_{L,R} ^{a=2}$ to define interaction eigenstates $W_{L,R}^{\pm}$ : 
$$ W_{L,R}^{\pm} =\frac{1}{\sqrt{2}}(W_{L,R} ^1 \mp i W_{L,R} ^2) $$
But now as we have found that these interaction states are not mass eigenstates, we are rotating this orthogonal system such that the mass matrix is diagonal giving us the exact massive states. 
$$\begin{pmatrix}
W_L ^\pm\\
W_R ^\pm
\end{pmatrix}
= \begin{pmatrix}
cos(\zeta) &  -sin(\zeta) \\
sin(\zeta) & cos(\zeta)
\end{pmatrix}
\begin{pmatrix}
W_1 ^\pm\\
W_2 ^\pm
\end{pmatrix}$$
These $W_{1,2}$ are the mass basis. Thus, we need to determine the mass of the bosons $M_{W_{1,2}}$ and the angle $\zeta$. Diagonalizing the mass matrix: To diagonalize the matrix we need to find the eigenvalues of the matrix. Say $\lambda$ are the eigenvalues. 
$$det \begin{pmatrix}
A -\lambda & B \\
C & D - \lambda
\end{pmatrix}
=0 $$
Our case: $$ A= \frac{1}{4} g^2 (k^2 + k'^2 +2 v_L ^2)$$
$$ B =  - \frac{1}{2} g^2 kk'$$
$$ C = - \frac{1}{2} g^2 kk'$$
$$ D = \frac{1}{4} g^2 (k^2 +k'^2 +2v_R ^2)$$
The eigenvalues are $$\lambda = \frac{(A+D)\pm \sqrt{(A+D)^2 - 4 (AD-BC)} }{2}$$
Substituting $A,B,C$ and $D$ for our case, we get
$$AD-BC = \frac{1}{4} g^4 [(k^2 +k'^2)^2 +(v_L^2+v_R^2)^2 + 2(k^2+k'^2)(v_L ^2 +v_R ^2)] $$
$$ A+D =\frac{1}{2} g^2 (k^2+k'^2+v_L^2+v_R^2) $$
$$\lambda =\frac{g^2}{4} [(k^2+k'^2+v_L^2+v_R^2) \pm \sqrt{(v_L^2 - v_R^2)^2+4 k^2 k'^2}] $$
The masses of the W-bosons in the limit $v_R  \gg k , k' \gg v_L $ are 
$$ M_{W_1}^2 \approx \frac{g^2}{4}(k^2+k'^2) $$
$$ M_{W_2}^2 \approx \frac{g^2}{4}(2v_R^2 +k^2+k'^2) $$
where $M_{W_1}$ is the standard model W-boson mass. Thus, $M_{W_2}$ is the mass of the heavy-charged gauge boson determined by the breaking scale of $v_R$. It means that the mass eigenstates are approximately the interaction eigenstate and so the angle between the interaction (or flavor) eigenstate and the mass eigenstate must be very small.

The angle between the orthogonal states: The diagonal basis and the interaction (or flavor) basis are related by a unitary matrix with angle $\zeta$. Then, for a given matrix $H$ 
$$H = \begin{pmatrix}
A & B \\
B & D
\end{pmatrix}
$$
we get the diagonal (mass basis) matrix 
$$ H_D = U H U^\dagger$$
$$\begin{pmatrix}
cos(\zeta) & sin(\zeta)\\
-sin(\zeta) & cos(\zeta)
\end{pmatrix}
\begin{pmatrix}
A & B \\
B & D
\end{pmatrix}
\begin{pmatrix}
cos(\zeta) & -sin(\zeta)\\
sin(\zeta) & cos(\zeta)
\end{pmatrix}
$$
As $H_D$ is a diagonal matrix, the off-diagonal elements are zero. Thus ${(H_{D})_{12} = 0}$ implies

$$\frac{1}{2} sin(2 \zeta)(D-A)+B cos(2\zeta) = 0 $$
$$ \implies tan(2\zeta) = \frac{2B}{D-A}$$

In our case: $tan(2\zeta) = \frac{-kk'}{v_R^2}$ in the limit $v_R  \gg k , k' \gg v_L $. Thus, the mixing angle is too much suppressed. This is what we suspected from the analysis of the mass spectrum compared with the standard model above.
\item \textbf{For Neutral Gauge Boson}

Mass term Lagrangian for the neutral bosons $W^{3}$ and $B$ are
$$\mathcal{L}_{\Delta} ^{Neutral} = \frac{1}{2} v_L ^2 g^2 W_{3L} ^2  + \frac{v_L ^2}{2}g'^2 B^2 - 2gg' \frac{v_L ^2}{2} W_{3L} ^\mu B_\mu  +  \frac{1}{2} v_R ^2 g^2 W_{3R} ^2 + \frac{v_R ^2}{2}g'^2 B^2 - 2gg' \frac{v_R ^2}{2} W_{3R} ^\mu B_\mu$$
$$\mathcal{L}_{\phi} ^{Neutral} = \frac{g^2}{4} \frac{1}{2}(k^2 + k'^2) (W_{3L}-W_{3R})^2$$
Now for neutral bosons, from $\mathcal{L}^{Neutral}_{\Delta / \phi}$ we can construct the mass matrix and diagonalize ${3 \times 3}$ matrix. We can get the neutral boson spectrum. We will get one eigenvalue or one diagonal element of the diagonal matrix to be zero. This is the photon $A_\mu$. This reflects the fact that $U(1)_{EM}$ is the unbroken symmetry and so the mediator photon is massless. (In superconductivity, the photon gets massive. In that case, there must be another SSB (spontaneous symmetry breaking) or phase transition to go to a superconducting state where a scalar field is coupled to the photon, giving mass to photons via phase transition.) Now, from neutral boson mass Lagrangian, we can write the Lagrangian in terms of mass matrix
$$\mathcal{L}_{mass}^{W_{L,R}^3 , B} = \frac{1}{2}\begin{pmatrix}
W_L^3 & W_R^3 & B \\ 
\end{pmatrix}\begin{pmatrix}
\frac{g^2}{4}(k^2+k'^2+4v_L^2) & -\frac{g^2}{4}(k^2+k'^2) & -gg'v_L^2 \\
-\frac{g^2}{4}(k^2+k'^2) & \frac{g^2}{4}(k^2+k'^2+4v_R^2) & -gg'v_R^2\\
-gg'v_L^2  & -gg'v_R^2 &  g'^2 (v_L^2+v_R^2)
\end{pmatrix}\begin{pmatrix}
 W_L^3 \\
 W_R^3 \\
 B
\end{pmatrix}$$
Diagonalizing the mass matrix and taking the limit $v_R  \gg k , k' \gg v_L $ we get the eigenvalues
$$ M_A ^2 =0 $$
$$ M_{Z_1} ^2 = \frac{(k^2+k'^2)g^2}{4cos^2(\theta_w)} (1-\frac{k^2+k'^2}{4v_R^2 cos^4(\theta_y)}) $$
$$ M_{Z_2} ^2 = g^2 v_R^2 $$
where $W_L^3$, $W_R^3$ and $B$ are the gauge or flavor eigenstate and $A$,$Z_1$ and $Z_2$ are the mass eigenstate. As they are related by the $3 \times 3$ matrix, we can define 
$$ \begin{pmatrix}
A \\
Z_1 \\
Z_2
\end{pmatrix}
=
\begin{pmatrix}
sin(\theta_w) & cos(\theta_w) sin(\theta_y)  & cos(\theta_w) cos(\theta_y) \\
-cos(\theta_w) &  sin(\theta_w) sin(\theta_y)  & sin(\theta_w) cos(\theta_y)  \\
 0  &  -cos(\theta_y)  &  sin(\theta_y)
\end{pmatrix}
.
\begin{pmatrix}
W^3 _L \\
W^3 _R \\
B
\end{pmatrix}
$$
Once again in the limit $v_R  \gg k, k' \gg v_L $ the $Z_1$ mass is the standard model $Z_0$ mass. The $Z_2$ is the mass of the heavy neutral boson determined by the breaking scale $v_R$.

\end{enumerate}

\end{enumerate}

\subsubsection{Fermion Particle Spectrum}

For the fermion mass term, due to Higgs or scalar field structure, there is a difference between the mass terms for quarks and lepton sector. Quarks are coupled to only the $ \phi $ scalar field but leptons are coupled to both $ \phi $ and $ \Delta $ scalar fields because we saw that $\Delta $ is neutral under Baryon number (i.e. $ B=0 $ ) but has Lepton number( $ L = -2 $ ).
\begin{enumerate}
\item \textbf{Quark Sector} \begin{enumerate}
\item \textbf{Yukawa Coupling with $\phi$}
$$\mathcal{L}_{Yukawa} ^{Quark} (\phi) = - \sum_{\psi = Q}[ \bar{\psi}_{Li} Y^Q_{ij} \phi \psi_{Rj} + \bar{\psi}_{Li} \tilde{Y}^Q_{ij} \tilde{\phi} \psi_{Rj} + h.c. ]$$
When the scalar field $\phi$ gets vacuum expectation value,
$$\mathcal{L}_{Yuk} ^{quk} (\langle \phi \rangle _0) =  - \sum_{\psi = Q}[ \begin{pmatrix}
\bar{u} & \bar{d} \\
\end{pmatrix}_{Li} Y^Q _{ij} \frac{1}{\sqrt{2}}  \begin{pmatrix}
k & 0 \\
0 & k'
\end{pmatrix}  \begin{pmatrix}
u \\
d
\end{pmatrix}_{Rj} + \begin{pmatrix}
\bar{u} & \bar{d} \\
\end{pmatrix}_{Li} \tilde{Y}^Q _{ij} \frac{1}{\sqrt{2}}  \begin{pmatrix}
k'^* & 0 \\
0 & k^*
\end{pmatrix}  \begin{pmatrix}
u \\
d
\end{pmatrix}_{Rj}]+ h.c.$$ 
(where $\tilde{\phi}= \tau^2 \phi^* \tau^2$ )
$$= -\frac{1}{\sqrt{2}}\sum_{Q} Y^Q_{ij} (\bar{u}_{Li} k u_{Rj}+ \bar{d}_{Li} k' d_{Rj}) -\frac{1}{\sqrt{2}}\sum_{Q} \tilde{Y}^Q_{ij} (\bar{u}_{Li} k'^* u_{Rj}+ \bar{d}_{Li} k^* d_{Rj}) + h.c.$$
\item \textbf{Quark Mass and Mixing}

We get the up-and-down type quark mass matrix given by 
$$ (M_U)_{ij} = \frac{1}{\sqrt{2}}(kY^Q_{ij} +k'^* \tilde{Y}^Q_{ij}) \ \& \ (M_D)_{ij} = \frac{1}{\sqrt{2}}(k'Y^Q_{ij} +k^* \tilde{Y}^Q_{ij}) $$
Their mass matrix is written in the interaction or flavor basis. To find the spectrum, we need to go to a mass eigenstate. For this we need to use unitary matrices (say $V$) to rotate to mass eigenstates in which the basis mass matrix is diagonal. Define mass eigenstates $u_{L,R}^m$ and $d_{L,R}^m$ by 
$$ u_{L,R} = V^u_{L,R} u_{L,R}^m \ \& \ d_{L,R} = V^d_{L,R} d_{L,R}^m$$
Diagonalized mass matrices are then given by (Lagrangian is invariant under this rotation thus from $\bar{u}_L M_U d_R $ invariance)
$$ \bar{u}_{Li} M_{Uij} u_{Rj} = \bar{u}_{Li}^m V_L^{u\dagger} M_{Uij} V_R^u u_{Rj}^m = \bar{u}_{Li}^m \hat{M}_{Uij} u_{Rj}^m$$
$$ \bar{d}_{Li} M_{Dij} d_{Rj} = \bar{d}_{Li}^m V_L^{d\dagger} M_{Dij} V_R^d d_{Rj}^m = \bar{d}_{Li}^m \hat{M}_{Dij} d_{Rj}^m$$
implies that, 
$$ \hat{M}_U = diag(m_u , m_c , m_t) = V_L^{u\dagger} M_U V_R^u $$
$$ \hat{M}_D = diag(m_b , m_s , m_b) = V_L^{d\dagger} M_D V_R^d $$
where $m_i$ with $i=u,c,t,d,s,b,$ are the masses of the individual states in the massive eigenstates. These are the physical particles. 
\newline
\newline
The standard model has only the left-handed $SU(2)$ gauge structure where right-handed particles are singlets of the group. But in the LR model (extension of SM) we have both $SU(2)_L$ and $SU(2)_R$. Thus, there are two mixing matrices in the LR model. Consider the charged current interactions (in the LR model) in the quark sector:
$$-\mathcal{L}_{c.c.}^Q = \frac{g}{\sqrt{2}}(\bar{u}_L \gamma^\mu d_L W_{L\mu}^+ + \bar{u}_R \gamma^\mu d_R W_{R\mu}^+) +h.c.$$
This is on a flavor or interaction basis. By rotating to mass eigen basis, 
$$ -\mathcal{L}_{c.c.}^Q = \frac{g}{\sqrt{2}}(\bar{u}_L^m V_L^{u\dagger} \gamma^\mu V_L^d d_L^m W_{L\mu}^+ + \bar{u}_R^m V_R^{u\dagger} \gamma^\mu V_R^d d_R^m W_{R\mu}^+) +h.c.$$
$$-\mathcal{L}_{c.c.}^Q = \frac{g}{\sqrt{2}}(\bar{u}_L^m  \gamma^\mu K_L d_L^m W_{L\mu}^+ + \bar{u}_R^m  \gamma^\mu K_R d_R^m W_{R\mu}^+) +h.c.  $$
This $K_L$ and $K_R$ are the two mixing matrices. They are given by, 
$$ K_L=V_L^{u\dagger} V_L^d = U_{CKM} $$
$$ K_R=V_R^{u\dagger} V_R^d $$
In general, $K_L$ and $K_R$ are different. There is a special case called ``Manifest CP invariance" where $K_L$ and $K_R$ are the same. In this case, $P$ symmetry ($Y^U$ and $Y^D$ are hermitian) and $CP$ symmetry ($Y^U$ and $Y^D$ are real) implies that $Y^U$ and $Y^D$ are real and symmetric. Thus, 
$$ V_{R}^{u,d} = V_{L}^{u,d} \implies K_{Lij} = K_{Rij} $$
\item \textbf{Yukawa Coupling with $\Delta$}

Now, for the spectrum (mass and mixing) of the lepton sector, there is a slight difference. It is obvious from the definition that $\Delta$ is color-neutral. Thus, there is no term of quark coupled to $\Delta$ in the Lagrangian. So, leptons are coupled to $\Delta$ but in a specific way. This $\Delta$ phase transition or vacuum expectation value $\langle \Delta \rangle _0 $ gives rise to the Majorana mass term. But the vacuum expectation value of $\phi$ gives the Dirac mass term. We coupled this $\Delta$ scalar with the lepton such that only neutrinos get the Majorana mass term. As there is no corresponding $\tilde{\Delta}$ of the field $\Delta$, we will only get the mass term of the upper element of the doublet which is the neutrino. If we needed to give electron Majorana mass then we need to add a $\tilde{\Delta}$ term giving Majorana mass to the lower element of the doublet which is charged leptons(like in the case of Dirac mass via $\phi$ vev) but this is excluded by the Higgs representation under gauge structure(no $\tilde{\Delta}$ field compatible in the extended model due to representation under gauge group). 
$$\mathcal{L}_{Yukawa}(\Delta)=\sum_{\psi=L}[L_{Li}^T  G_{Lij} C^{-1}i \tau ^2 \Delta_L L_{Lj}+L_{Ri}^T  G_{Rij} C^{-1}i \tau ^2 \Delta_R L_{Rj}]+h.c.$$
\end{enumerate}

\item \textbf{Lepton Sector}\begin{enumerate}
\item \textbf{Yukawa Coupling with $\phi$}

Charged lepton only gets Dirac mass like quark sector as it is not coupled to $\Delta$ scalar field. 

$$ \mathcal{L}_{Yukawa}^{Lepton} (\phi) = -\sum_{L} [\bar{L}_{Li} Y^L_{ij}\phi L_{Rj} + \bar{L}_{Li} \tilde{Y^L_{ij}} \tilde{\phi} L_{Rj}+h.c. ] $$

Like the quark case, leptons get Dirac masses with the vacuum expectation value of $\phi$,  
$$ (M^{D}_{neutrino})_{ij} = \frac{1}{\sqrt{2}}(kY^L_{ij} +k'^* \tilde{Y}^L_{ij}) \ \& \ (M^{D}_{charged \ lepton})_{ij} = \frac{1}{\sqrt{2}}(k'Y^L_{ij} +k^* \tilde{Y}^L_{ij}) $$
\item \textbf{Yukawa Coupling with $\Delta$}

When the scalar field $\Delta$ gets vacuum expectation value, 
$$ \mathcal{L}_{Yukawa}(\Delta)= L_{Li}^T G_{Lij}C^{-1}i \tau^2 \Delta_L L_{Lj} +  L_{Ri}^T G_{Rij}C^{-1}i \tau^2 \Delta_R L_{Rj} + h.c.$$
$$ \langle \Delta_{L,R} \rangle _0 = \frac{1}{\sqrt{2}}\begin{pmatrix}
0 & 0 \\
v_{L,R} & 0
\end{pmatrix}  \implies i\tau^2 \langle \Delta_{L,R} \rangle _0 = \begin{pmatrix}
\frac{v_{L,R}}{\sqrt{2}} & 0 \\
0 & 0
\end{pmatrix} $$
$$ \mathcal{L}_{Yukawa}(\langle \Delta \rangle _0 )= L_{Li}^T G_{Lij}C^{-1} \begin{pmatrix}
\frac{v_{L}}{\sqrt{2}} & 0 \\
0 & 0
\end{pmatrix}L_{Lj} +  L_{Ri}^T G_{Rij}C^{-1}\begin{pmatrix}
\frac{v_{R}}{\sqrt{2}} & 0 \\
0 & 0
\end{pmatrix} L_{Rj}+ h.c.$$
It will only pick up the (Majorana) mass term for neutrinos. Here lepton doublet is $L_{L,R} = \begin{pmatrix}
\nu \\
 e 
\end{pmatrix}_{L,R}$. Thus, the Majorana mass term for neutrinos are
$$\mathcal{L}_{Yukawa}(\langle \Delta \rangle _0 ) = \nu_L^T C^{-1} G_L \frac{v_{L}}{\sqrt{2}} \nu_L + \frac{1}{2}\nu_R^T C^{-1} G_R \frac{v_{R}}{\sqrt{2}} \nu_R +h.c.$$
The mass term for neutrino(fermion) is defined by 
$$ \mathcal{L}_{\nu}^{(mass)}= \frac{1}{2} \Omega_L^T C^{-1} M_{\nu} \Omega_L +h.c.$$
where $M_\nu$ is the mass matrix of the neutrino(fermion). Thus, in our case, Neutrino having both Dirac and Majorana masses, Dirac mass term mixes $\nu_L$ and $\nu_R$ but Majorana mass mixes $\nu_L$ and $(\nu_R)^c$, enlarging the freedom of the neutrino. Thus, putting all the 6- L-handed freedom into the vector,
$$\Omega_L = \begin{pmatrix}
\nu_L \\
 (\nu_R)^c 
\end{pmatrix}$$
and the mass matrix for the  (both Dirac and Majorana) is given by 
$$ M_\nu = \begin{pmatrix}
\sqrt{2}v_L G_L & M_D \\
M_D^T & \sqrt{2}v_R G_R
\end{pmatrix} $$
where $M_D$ is the Dirac mass term of the neutrino.

\end{enumerate}

\end{enumerate}

\subsubsection{Unbroken Generators and Goldstone Modes}

Goldstone theorem (J. Goldstone, Phys. Rev. 127, 965 1962) states that, when a continuous symmetry of a theory is spontaneously broken, there will be a massless particle (spin will depend on operators) corresponding to each of the broken generators, known as the Goldstone modes. Thus, they are along the direction of the broken generators and so these modes are absorbed in the broken gauge fields resulting in their masses. So, to find the Goldstone directions, we just need to apply broken generators to the vacuum of the theory. 
\newline
\newline
For breaking of a continuous symmetry (globally), the broken generators correspond to Goldstone modes that give mass to corresponding broken gauge bosons and the unbroken generators correspond to the residual symmetry of the ground. 
\begin{enumerate}
\item \textbf{SSB-1: $SU(2)_R \times U(1)_{B-L} \rightarrow U(1)_Y$}
\begin{enumerate}
\item \textbf{Breaking pattern}

In this model, spontaneous symmetry breaking occurs in two steps. First consider the breaking $SU(2)_R \times U(1)_{B-L} \rightarrow U(1)_Y$ in the limit $v_R \neq 0 , v_L = k = k' =0 $. The $SU(2)_R$ gauge bosons are $W_R^1 ,W_R^2 ,W_R^3$ and $U(1)_{B-L}$ gauge boson is $B$. We are considering the symmetry breaking similar to the Standard model. Thus, $W_R^1$ and $ W_R^2 $ mix to give $W_R^+ ,W_R^-$ and $W_R^3$ mixes with $B$ to give $Z_R$ and $A_R$ where $A_R$ will be the generator correspond to unbroken $U(1)_Y$. 
$$\begin{pmatrix}
W_R^1 \\
W_R^2 \\
A_R \\
Z_R
\end{pmatrix} = \begin{pmatrix}
1 &0 &0 &0 \\
0 &1 &0 &0 \\
0 &0 & \frac{g'}{\sqrt{g^2+g'^2}} & \frac{g}{\sqrt{g^2+g'^2}} \\
0 &0 &-\frac{g}{\sqrt{g^2+g'^2}} & \frac{g'}{\sqrt{g^2+g'^2}}
\end{pmatrix} \begin{pmatrix}
W_R^1 \\
W_R^2 \\
W_R^3 \\
B
\end{pmatrix}  $$
Unlike the Standard model, here the Higgs field $\Delta_R$ transforms as adjoint under $SU(2)R$.
\item \textbf{Vacuum Expectation Values}

The vacuum triplet $(\delta_1,\delta_2,\delta_3)$ in the adjoint representation is given by
$$ \Delta_R = \delta_R^a t^a = \begin{pmatrix}
\frac{\delta_R^+}{\sqrt{2}} & \delta_R^{++} \\
\delta_R^0 & - \frac{\delta_R^+}{\sqrt{2}}
\end{pmatrix}$$
where $t^a = \frac{1}{2} \sigma^a$ , $\sigma^a$ are the Pauli matrices. 

Thus, the components are related by 
$$\delta_R^1=\delta_R^{++}+\delta_R^0$$
$$\delta_R^2= i \delta_R^{++} - \delta_R^0$$
$$\delta_R^3=\sqrt{2}\delta_R^+$$
with the vacuum expectation values $\langle \delta_R^0 \rangle = \frac{v_R}{\sqrt{2}}$ , $\langle \delta_R^{++}\rangle = \langle \delta_R^{+}\rangle =0$. Thus, the vacuum is given by 
$$  \langle  \begin{pmatrix}
\delta_R^1\\
\delta_R^2 \\
\delta_R^3
\end{pmatrix} \rangle = \frac {v_R}{\sqrt{2}} \begin{pmatrix}
 1\\
 -i\\
0
\end{pmatrix} $$
\item \textbf{Determination of Unbroken Generators and Goldstone Modes}

Assume, $\delta$ is the infinitesimal transformation. Thus, for unbroken gauge group generators, $\delta$ acting on a vacuum must give us zero. For broken generators (which have eaten the Goldstone bosons after SSB), $\delta$ acting on vacuum gives us the desired Goldstone directions. 
$$\delta (vacuum) = i (Unbroken \ generator) (vacuum) = 0$$
$$\delta (vacuum) = i (Broken \ generator) (vacuum) = Goldstone \ directions$$
As $\Delta_R$ is adjoint under $SU(2)_R$, its transformation $\Delta_R \rightarrow U_R \Delta_R U_R^\dagger$ under $SU(2)_R$ implies in the infinitesimal form, the transformation under $SU(2)_R$ and $U(1)_{B-L}$ is 
$$ \Delta_R \rightarrow (1+i [w^a t^a,\Delta_R]+iw_{B-L}\Delta_R) $$
where $w_R^a$ and $w_{B-L}$ are the infinitesimal parameters of the gauge transformation. For determining the unbroken direction enforce $\delta \langle \Delta_R  \rangle = 0$ in the limit $v_R \neq 0 , v_L = k = k' =0 $.
$$\delta \langle \Delta_R  \rangle = i [w^a t^a,\langle \Delta_R \rangle]+iw_{B-L}\langle \Delta_R \rangle$$
$$ =  i v_L \begin{pmatrix}
\frac{1}{2} (w_R^1-iw_R^2) & 0 \\
-w_R^3+w_{B-L} & \frac{1}{2} (w_R^1-iw_R^2) \\
\end{pmatrix}$$
The condition $\delta \langle \Delta_R  \rangle = 0$ implies that $w_R^1 = w_R^2 = 0$ and $w_R^3 = w_{B-L}$

Thus, the generator $Y = T_R^3 + \frac{B-L}{2}$ generates the gauge transformation which leaves the vacuum invariant. So, it is the unbroken direction corresponding to the gauge boson $A_R$ and this gauge boson is massless. This $Y (= T_R^3 + \frac{B-L}{2})$ is the $U(1)_Y$ gauge group generator. 
\newline
\newline
Now, the broken generators, which correspond to $W_R^{\pm}, Z_R$ give the Goldstone direction, their generators (unbroken) are $T_R^1, T_R^2, (-g^2 T_R^3+g'^2 \frac{B-L}{2})$ respectively (determined from mixing matrix). Note, in the adjoint representation $(T_R^a)_{bc} = -i \epsilon^{abc}$. 

For Goldstone direction correspond to $Z_R$ boson, 
$$i.(generator)(vacuum) =  i (-g^2 T_R^3+g'^2 \frac{B-L}{2}) \langle \Delta_R \rangle $$
$$=  i (-g^2 \begin{pmatrix}
0&1&0\\
 -1&0&0\\
0&0&0
\end{pmatrix}+ g'  \begin{pmatrix}
1&0&0\\
 0&1&0\\
0&0&1
\end{pmatrix})\frac{v_R}{\sqrt{2}} \begin{pmatrix}
1\\
 -i\\
0
\end{pmatrix}= \frac{v_R}{\sqrt{2}}(g^2+g'^2) \begin{pmatrix}
i\\
1\\
0
\end{pmatrix}$$
Thus, Goldstone mode eaten by $Z_R$ is proportional to $[Im(\delta_R^1)+Re(\delta_R^2)]$ or using the vev conditions and rotate phase by 90 degrees (as still we have a $U(1)$ invariance), the Goldstone mode is proportional to $Im(\delta_R^0)$

The other two Goldstones correspond to $W_R^1$ and $W_R^2$ bosons are, 
$$i g \begin{pmatrix}
0&0&0\\
 0&0&-i\\
0&i&0
\end{pmatrix} \frac{v_R}{\sqrt{2}}\begin{pmatrix}
1\\
 -i\\
0
\end{pmatrix} = \frac{g v_R}{\sqrt{2}}\begin{pmatrix}
0\\
 0\\
i
\end{pmatrix}$$
$$i g \begin{pmatrix}
0&0&i\\
 0&0&0\\
i&0&0
\end{pmatrix} \frac{v_R}{\sqrt{2}}\begin{pmatrix}
1\\
 -i\\
0
\end{pmatrix} = \frac{g v_R}{\sqrt{2}}\begin{pmatrix}
0\\
0\\
1
\end{pmatrix}$$
The Goldstone mode eaten by $W_R^1$ and $W_R^2$ is proportional to $Im\delta_R^3=\sqrt{2}Im\delta_R^+$ and $Re\delta_R^3=\sqrt{2}Re\delta_R^+$ respectively. We know that gauge boson charged eigenstates are defined like $$ W_R^{\pm} =\frac{1}{\sqrt{2}}(W_R ^1 \mp i W_R ^2) $$ and so the Goldstone bosons corresponding to $W_R^{\pm}$, say $G_R^{\pm}$, is given by 
$$G_R^+ = Im (\delta_R^+) -i Re(\delta_R^+) = Re (\delta_R^+) + i Im (\delta_R^+) = \delta_R^+$$
$$G_R^- = Im (\delta_R^+) + i Re(\delta_R^+) = Re (\delta_R^+) - i Im (\delta_R^+) = \delta_R^-$$

\end{enumerate}

\item \textbf{SSB-2: $SU(2)_L \times U(1)_Y \rightarrow U(1)_{EM}$}
\begin{enumerate}
\item \textbf{Breaking pattern}
Now we consider the second SSB, $SU(2)_L \times U(1)_Y \rightarrow U(1)_{EM}$. The Higgs bidoublet responsible for this breaking is given by 
$$  \phi =\begin{pmatrix}
\phi_1^0 & \phi_1^+\\
\phi_2^- & \phi_2^0
\end{pmatrix}  $$ 
Like the first SSB, here we will consider the symmetry breaking is similar to what happens in the Standard model. Thus the mixing (similar to SM) is 
$$ \begin{pmatrix}
W_L^1 \\
W_L^2 \\
A \\
Z_L
\end{pmatrix} = \begin{pmatrix}
1 &0 &0 &0 \\
0 &1 &0 &0 \\
0 &0 & \frac{g'}{\sqrt{g^2+g'^2}} & \frac{g}{\sqrt{g^2+g'^2}} \\
0 &0 &-\frac{g}{\sqrt{g^2+g'^2}} & \frac{g'}{\sqrt{g^2+g'^2}}
\end{pmatrix} \begin{pmatrix}
W_L^1 \\
W_L^2 \\
W_L^3 \\
B_Y
\end{pmatrix}$$
\item \textbf{Vacuum Expectation Values}

Here, $B_Y$ is the $U(1)_Y$ gauge boson which corresponds to the unbroken direction in the SSB-1. 
Under $U(1)_Y$, where $Y=T_R^3+\frac{B-L}{2}$ , the $\Phi$ field transforms only under $SU(2)_R$ because $B-L=0$ for $\Phi$. 
$$ \phi \rightarrow  \begin{pmatrix}
\phi_1^0 & \phi_1^+\\
\phi_2^- & \phi_2^0
\end{pmatrix}   \begin{pmatrix}
e^{-i \frac{1}{2} w_Y } & 0\\
0 & e^{+i \frac{1}{2} w_Y }
\end{pmatrix} $$
So, we see that $\Phi$ is splitting into two doublets. 
$$  \phi_1 = \begin{pmatrix}
\phi_1^0 \\
\phi_2^-
\end{pmatrix} \ \& \ \phi_2 = \begin{pmatrix}
\phi_1^+ \\
\phi_2^0
\end{pmatrix} $$
with hyper charge $Y=-\frac{1}{2}$ and $Y=\frac{1}{2}$ respectively. 

In this case, the unbroken gauge group is $U(1)_{EM}$. Thus, the EM charge generator in this case(as the breaking pattern is the same) is given by 
$$Q = T_L^3 + Y = T_L^3 +T_R^3 + \frac{B-L}{2}$$. To verify this, we need to find the vacuum first. 
$$ \langle \phi_1 \rangle = \begin{pmatrix}
\frac{k}{\sqrt{2}} \\
0
\end{pmatrix} \ \& \ \langle \phi_2 \rangle = \begin{pmatrix}
0 \\
\frac{k'}{\sqrt{2}}
\end{pmatrix}$$
\item \textbf{Determination of Unbroken Generators and Goldstone Modes}

Now, applying $ I (generator) (vacuum)$, we must get zero to ensure it is unbroken. 
$$i(T_L^3+Y) \langle \phi_1 \rangle =  [\frac{1}{2}\begin{pmatrix}
1 &0 \\
0 &-1 
\end{pmatrix} - \frac{1}{2}\begin{pmatrix}
1 &0 \\
0 &1 
\end{pmatrix}] \begin{pmatrix}
\frac{k}{\sqrt{2}} \\
0 
\end{pmatrix} =\begin{pmatrix}
0 \\
0 
\end{pmatrix}$$
$$ i(T_L^3+Y) \langle \phi_2 \rangle =  [\frac{1}{2}\begin{pmatrix}
1 &0 \\
0 &-1 
\end{pmatrix} + \frac{1}{2}\begin{pmatrix}
1 &0 \\
0 &1 
\end{pmatrix}] \begin{pmatrix}
\frac{k'}{\sqrt{2}} \\
0 
\end{pmatrix} =\begin{pmatrix}
0 \\
0 
\end{pmatrix}  $$
Thus, this is the unbroken direction. This is the generator of $U(1)_{EM}$ and the corresponding boson is $A_\mu$ photon which is massless. Now, for the broken directions, the broken generators are $-g^2T_L^3+g'^2Y$, $gT_L^1$ and $gT_L^2$ corresponding to $Z$, $W_L^{1,2}$ respectively. Applying these generators to the vacuum, we get these corresponding Goldstone modes. 
$$ i(-g^2T_L^3+g'^2Y) \begin{pmatrix}
\frac{k}{\sqrt{2}} \\
0 
\end{pmatrix} = \frac{i}{2\sqrt{2}} [-g^2 \begin{pmatrix}
1 &0 \\
0 & -1 
\end{pmatrix}-g'^2 \begin{pmatrix}
1 &0 \\
0 &1 
\end{pmatrix}] \begin{pmatrix}
k \\
0 
\end{pmatrix} = -\frac{g^2+g'^2}{2\sqrt{2}} \begin{pmatrix}
i k \\
0 
\end{pmatrix} $$
and 
$$i(-g^2T_L^3+g'^2Y) \begin{pmatrix}
0 \\
\frac{k'}{\sqrt{2}} 
\end{pmatrix} = \frac{i}{2\sqrt{2}} [-g^2 \begin{pmatrix}
1 &0 \\
0 & -1 
\end{pmatrix}-g'^2 \begin{pmatrix}
1 &0 \\
0 &1 
\end{pmatrix}] \begin{pmatrix}
0 \\
k' 
\end{pmatrix} = -\frac{g^2+g'^2}{2\sqrt{2}} \begin{pmatrix}
0 \\
-i k 
\end{pmatrix}$$
Adding these Goldstone directions, we get that Goldstone mode is proportional to $k Im(\phi_1^0) -k' Im(\phi_2^0)$. Thus, normalizing we get the Goldston boson,
$$ G_Z = \frac{1}{\sqrt{(k^2+k'^2)}}[k Im(\phi_1^0) -k' Im(\phi_2^0)] $$
For Goldstones eaten by $W_L^{1,2}$,$G_L^{1,2}$, applying these broken generators on the vacuum, is given first for $G_L^1$
$$igT_L^1 \langle \phi\rangle =  \frac{ig}{2\sqrt{2}} \begin{pmatrix}
0 &1 \\
1 &0 
\end{pmatrix} \begin{pmatrix}
k &0 \\
0 & k'
\end{pmatrix} = \frac{g}{2\sqrt{2}} \begin{pmatrix}
0 & ik' \\
i k &0 
\end{pmatrix} $$
which is proportional to $k' Im(\phi_1^+) +k Im(\phi_2^-)$ and for $G_L^2$ correspond to $W_L^2$ is given by 
$$ igT_L^2 \langle \phi\rangle =  \frac{ig}{2\sqrt{2}} \begin{pmatrix}
0 &-i \\
i &0 
\end{pmatrix} \begin{pmatrix}
k &0 \\
0 & k'
\end{pmatrix} = \frac{g}{2\sqrt{2}} \begin{pmatrix}
0 & k' \\
-k &0 
\end{pmatrix} $$
which is proportional to  $k' Re(\phi_1^+) - k Re(\phi_2^-)$

As the gauge bosons charged eigenstates are given by $W^{\pm}=\frac{1}{\sqrt{2}}(W^1 \mp i W^2)$, the corresponding Goldstone modes $G_L^{\pm}$ are given by
$$G_L^+ =  \frac{1}{\sqrt{2}}(G_L^1 - i G_L^2) = \frac{1}{\sqrt{2}} [\{k' Im(\phi_1^+) +k Im(\phi_2^-)\} -i \{k' Re(\phi_1^+) - k Re(\phi_2^-)\}]$$
$$G_L^- =  \frac{1}{\sqrt{2}}(G_L^1 + i G_L^2) = \frac{1}{\sqrt{2}} [\{k' Im(\phi_1^+) +k Im(\phi_2^-)\} +i \{k' Re(\phi_1^+) - k Re(\phi_2^-)\}]$$
Now, with rotating phase by 90 degrees of the fields like the $\Delta$ case, we get the Goldstones 
$$G_L^+ = (\phi_1^+ - i \phi_2^+) \ \& \ G_L^- = (\phi_1^- - i \phi_2^-)$$
So, the normalized, charged eigenstate, Goldstone bosons are given by 
$$ G_L^{\pm} = \frac{1}{\sqrt{k^2+k'^2}}(\phi_1^{\pm}-i \phi_2^{\pm})$$

\end{enumerate}

\end{enumerate}

\subsubsection{Higgs Potential}
\label{Higgs Potential}
A complex scalar doublet $\phi$, the Standard Model Higgs field has the potential, 
$$V(\phi) = -\mu^2 \phi^\dagger \phi + \lambda (\phi^\dagger \phi)^2$$
Generalizing to bi-doublet, we determine the potential,
$$V(\phi) =-\mu_\phi^2 Tr(\phi^\dagger \phi) - \tilde{\mu}_\phi^2 [Tr(\tilde{\phi}^\dagger \phi)+h.c.] + \lambda_1 [Tr(\phi^\dagger \phi)]^2+\lambda_2 [ e^i\alpha_2 (Tr(\tilde{\phi}\phi^\dagger))^2 +h.c.]$$
$$+\lambda_3 [Tr(\tilde{\phi}\phi^\dagger)Tr(\tilde{\phi}^\dagger \phi)] +\lambda_4 Tr(\phi^\dagger \phi)[e^i\alpha_4 Tr(\tilde{\phi}\phi)+h.c.]$$
where $\tilde{\phi}=\epsilon \phi^*\epsilon$ is the conjugate field. As we need also the $\Delta$ Higgs to introduce Majorana mass for neutrino, there is also a triplet $\Delta$ field in the theory. Thus, assuming parity invariance and renormalizability, all possible combination of fields in the potential in the most general form is given below. 
\graphicspath{{./pot/}}
\begin{figure}[H]
\begin{center}
\includegraphics[width=125mm]{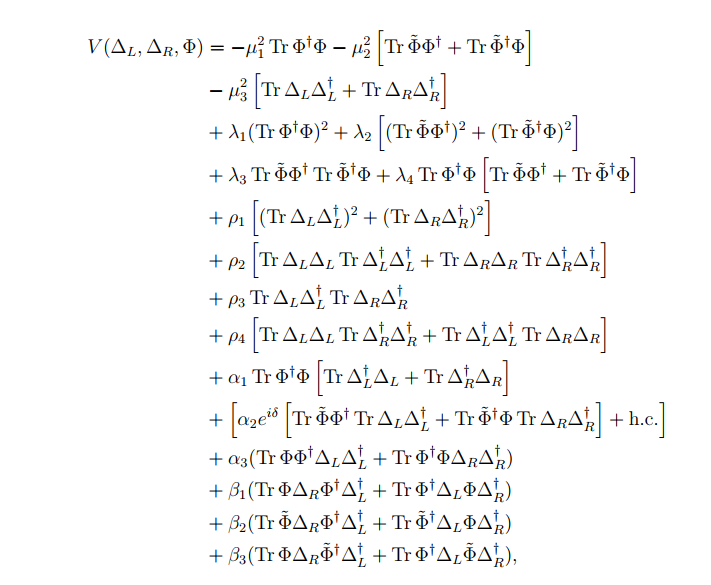}
\end{center}
\label{pot}
\end{figure}
The terms that are responsible for mixing between $\Delta_L$ and $\Delta_R$ are 
$$\beta_1[Tr(\phi \Delta_R\phi^\dagger \Delta_L^\dagger)+Tr(\phi^\dagger \Delta_L\phi \Delta_R^\dagger)] +\beta_2 [Tr(\tilde{\phi}\Delta_R\phi^\dagger\Delta_L^\dagger)+Tr(\tilde{\phi}^\dagger \Delta_L\phi \Delta_R^\dagger)] +\beta_3 [Tr(\phi \Delta_R\tilde{\phi}^\dagger\Delta_L^\dagger)+Tr(\phi^\dagger\Delta_L\tilde{\phi}\Delta_R^\dagger)]$$
The Higgs $\Delta_L$ and $\Delta_R$ mixing terms depend only on $\beta_i$ coefficient terms. In rest other terms there is no mixing or coupling between left and right-handed Higgs fields.
\newline
\newline
\textbf{Phenomenology of $\beta$ terms}
\newline
\newline
The Higgs potential with $\delta_2=0$ is the real Higgs potential (implying no explicit CP violation). In the potential, the only Higgs left and right-handed mixing terms are the $\beta$ coefficient terms. The only place in the Higgs potential where $v_L$ is generated from $\beta$ term. Else everywhere, $v_L^2=0$ due to phenomenology $v_L \ll k,k' \ll v_R$. 

The experimental bound from neutrino mass and gauge boson mass implies $\beta_i=0$. For neutrino smallness of mass, we need Majorana mass term, we need to couple $\Delta$ Higgs and set the limit $v_L \ll k,k' \ll v_R$. As $v_L$ is zero from neutrino phenomenology and $v_L$ is generated only from $\beta$ terms, the $\beta_i$'s must be zero to match neutrino phenomenology \cite{Senjanovic:1975rk}. That's why this theory is seesaw-compatible. In this general set up, we give vev to $$\phi_1^{0r},\phi_2^{0r},\delta_R^{0r},\delta_L^{0r}, \phi_2^{0i},\delta_L^{0i}$$
The potential is after spontaneous symmetry breaking, must be extreme at the vev 

$$ \frac{\partial V}{\partial \phi_1^{0r}}=\frac{\partial V}{\partial \phi_2^{0r}}=\frac{\partial V}{\partial \delta_R^{0r}}=\frac{\partial V}{\partial \delta_L^{0r}}=\frac{\partial V}{\partial \phi_2^{0i}}= \frac{\partial V}{\partial \delta_L^{0i}} $$ 

These equations evaluated at vev implies 
$$ \beta_2 = (-\beta_1 kk'-\beta_3k'^2+v_L v_R(2\rho_1-\rho_3))/k^2 $$
Note that, in the case of $\beta_i=0$, we find that 
$$ v_L v_R (2\rho_1-\rho_3) = 0 $$

This is known as the vev-seesaw relation. $v_R$ is non-zero to break $SU(2)_R$ and give large mass to $W_R$ and $Z_R$. Also $(2\rho_1-\rho_3)$ is non-zero due to phenomenology: As some Higgs have mass proportional to $(2\rho_1-\rho_3)$. If they are massless, they would open up new $Z$ decay channels with widths comparable to $Z \rightarrow \nu \overline{\nu}$ channels \cite{deshpande1991left}. Even with small mass contributions from loop corrections, such extra decays would be easily detectable.

Therefore, the only possibility to get $\beta_i=0$ is $v_L=0$. Consequently, we see that the mixing between $\Delta_L$ and $\Delta_R$ is very small.

\subsubsection{Physical Higgs}

We have calculated the Goldstone modes of the theory. These are the bosons created due to the breaking of Global symmetry. As we consider the local theory (interacting theory), these modes are absorbed by the broken gauge bosons. There are more modes in the Higgs field. These are the physical Higgs(not absorbed in any gauge bosons generator direction). When we study the Higgs loop sector, we only calculate the contributions from these physical Higgs, not the Goldstone directions. When we calculate the W-loops in Feynman 't Hooft gauge, as we discussed in the QFT structure chapter, we have to integrate not only W-loops where W's are now abelian bosons but also integrate the Goldstone modes in addition. Goldstone directions of the Higgs field are the direction of the broken gauge boson, not the direction of physical Higgs. We will not find the mass spectrum of the physical Higgs. We will classify the modes of the field. 

Expanding the potential around the vev, we can construct the bilinear terms. These will give us a mass matrix. From the mass matrix, we can read the Higgs basis, 

$$ \{ \phi_1^{0r},\phi_2^{0r},\delta_R^{0r},\delta_L^{0r} \} $$
$$ \{ \phi_1^{0i},\phi_2^{0i},\delta_R^{0i},\delta_L^{0i} \} $$
$$ \{ \phi_1^{+},\phi_2^{+},\delta_R^{+},\delta_L^{+} \} $$
$$ \delta_L^{++},\delta_R^{++} $$

We rotate this basis into flavor diagonal basis and excluding the Goldstone directions, we can read the physical modes. We also can explicitly find the mass matrix from the vev, diagonalize, and exclude the modes that have zero masses as these are the Goldstone modes to find the physical Higgs. 

A bi-doublet has four and a triplet has three complex numbers(we have two triplets). As one complex number has two degrees of freedom, there are $4\times2+3\times2 \times 2 = 20$ degrees of freedom. SSB produces 6 massive bosons and so 6 Goldstone modes are eaten by gauge bosons, leaving 14 degrees of freedom for physical Higgs bosons. 

These physical Higgs bosons are four real scalars $H_{0,1,2,3}^0$ and two real pseudo-scalars $A_{1,2}^0 $from the mixing of neutral fields, four complex scalars, the singly charged Higgs $H_{1,2}^{+}$ and two doubly charged Higgs $\delta_{1,2}^{++}$.

\subsection{Types of Left-Right Symmetric Model}

In the Higgs sector, There are 2 types of LRSM most studied based on the origin of CP violation (CPV) - \begin{enumerate} 
\item \textbf{Explicit CPV}

Here CPV contributions are from higher UV theories like GUT, and SUSY after SSB. $$ \mathcal{L}^{UV}_{D>4} \xRightarrow{\text{SSB}} \mathcal{L}^{IR}_{D=4} $$
That is, no contribution from Yukawa's coupling of IR theory.
\item \textbf{Implicit CPV}

Here, CPV is implicit in IR theory. Thus, in both models, the Higgs potential is real i.e. $\delta_2=0$ where $\delta_2$ is the explicit CP violating factor in the Higgs potential that we chose to be zero in such model. Based on Yukawa coupling implicit model can be further divided into - \begin{enumerate}
\item \textbf{Manifest CPV} - Here, CPV is not spontaneous. Thus, $\alpha = 0 = \theta_L$ where $\alpha$ and $\theta_L$ were complex phases in the vev of the field. The only source of CPV is from Yukawa couplings(therefore coupling is complex) and also $U_{CKM}^L=U_{CKM}^R$. 
\item \textbf{Pseudo-Manifest CPV} - Lagrangian in this model is considered $P$ and $CP$ invariant and also as there is no explicit CPV, $\delta_2=0$ which implies real Higgs potential. But in this case, vev could be complex and the Yukawa couplings are real and symmetric. Thus, the source of the imaginary factor is the vev of the field which is after the symmetry breaking. It turns out that this model can not produce enough CP asymmetry in $B \rightarrow \psi K$ decay even when maximum CPV is considered.

\end{enumerate}

\end{enumerate}

\section{CP Violation (CPV)}
\label{CP Violation}
\subsection{Introduction}

CP transformation combines charge conjugation, C with parity, P. Under C, particles and anti-particles are interchanged by conjugating all the internal quantum numbers, for instance: electric charge $Q_{EM} \rightarrow - Q_{EM}$. Under P, the handedness of space is reversed, $\vec{x}\rightarrow -\vec{x}$. Thus, a left-handed electron $e_L^-$ for instance, is transformed into a right-handed positron $e_R^+$ under CP.
\newline
\newline
Now, consider a scattering process $ab \rightarrow cd$ whose amplitude $\mathcal{M} \sim J^\mu_{ca} J^\dagger_{\mu bd}$ and the antiparticle reaction $\bar{a}\bar{b} \rightarrow \bar{c}\bar{d}$ whose amplitude $\mathcal{M}^\dagger \sim J^{\mu \dagger}_{ca} J_{\mu bd}$. So if we calculate $\mathcal{M}$ under the CP transformed process and get $\mathcal{M}_{CP} = \mathcal{M}^\dagger$ then the theory is CP invariant. 

We can describe this more formally by introducing 
$$ P |A\rangle = + |A\rangle$$ 
$$ P |B\rangle = - |B\rangle$$

$$ C |particle\rangle = + |antiparticle\rangle $$
$$ C |anti-particle\rangle = + |particle\rangle $$

where $|A\rangle$ and $|B\rangle$ are called even and odd under parity respectively. If $|A\rangle$ and $|B\rangle$ are physical eigenstates, then the theory is parity symmetric (similarly for $C$). Physical eigenstates are the mass eigenstates. On this basis, the mass matrix is real (or can be made real with symmetry transformation). So, consider a theory with state $|X\rangle$ which are its mass eigenstates (mass matrix diagonal and real). If  $|X\rangle$ is $CP$ eigenstate, the theory is $CP$ symmetric. 
$$ CP |X\rangle = \pm |X\rangle $$
but if $|X\rangle$ is not $CP$ eigen state, then $CP$ is violated in the theory.

\subsection{Experimental Evidence of CP Violation}

Deep down to the structure of QFT, Lorentz or Poincare invariance requires CPT symmetry of the theory even though the individual transformation may not be symmetry of the theory. Thus, CPT will hold for any QFT otherwise QFT will collapse as Special Relativity is violated so something is going very wrong. In this sense, CP violation implies T violation also.

Many experiments find that CP is violated. We will discuss the simplest example, the neutral kaon $K^0-\bar{K}^0$ mixing :
\newline
\newline
Pions are the lightest mesons and Kaons are the lightest strange mesons. Mesons are combinations of two quarks with opposite colors (observation must be color-neutral). With the 1st generation quark doublet $\begin{pmatrix}
u \\
d
\end{pmatrix}$ and a strange quark $s$, there are 4 possible combinations that can form kaons - 

$$ \bar{u}s (= K^-) \ ; \ \bar{s}u (= K^+) \ ; \ \bar{d}s  (=\bar{K}^0) \ ; \ \bar{s}d (= K^0) $$

These are strong eigenstates. These kaons are called pseudo scalar particles due to odd parity. William Chinowsky and Jack Steinberger showed that pions-like particles have negative parity. We can easily prove this by using the conservation of angular momentum and parity of a pion or kaon exchange system ($\pi^- + D \rightarrow n + n$). As they are pseudo scalars under parity
$$  P|K\rangle = -|K\rangle $$
And under charge conjugation operator(particle to antiparticle and vice versa), 
$$  C|K^\pm\rangle= +|K^\mp \rangle$$
Thus, these kaon states are parity and charge conjugation operator eigenstates. But neutral kaons carry strangeness so $K^0$ and $\overline{K^0}$ are oscillating in between and  we can see, under combined charge conjugation (C) and parity (P), the $K^0$ and $\overline{K^0}$ states (at rest with zero 3-momentum) may transform as
$$CP|K^0\rangle = -|\overline{K^0}\rangle$$
$$CP|\overline{K^0}\rangle = -|K^0\rangle$$
So, we define CP eigenstates of kaons as
$$ |K_1\rangle= \frac{1}{\sqrt{2}}[|K^0\rangle-|\bar{K}^0\rangle] \ \& \  |K_2\rangle= \frac{1}{\sqrt{2}}[|K^0\rangle+|\bar{K}^0\rangle] $$
where under CP transformation, 
$$ CP|K_1\rangle = +|K_1\rangle $$
$$ CP|K_2\rangle = -|K_2\rangle $$

Thus, $|K_1\rangle$ and $|K_2\rangle$ are the CP eigenstates with positive and negative eigenvalues respectively ($K^0$ and $\bar{K}^0$ are not CP eigenstates). 

We have defined kaons and also have found the CP eigenstates. Now, we just have to check if these CP eigenstates are the kaons physical or mass eigenstates to ensure the CP invariance of the theory. From observation, it was found that neutral kaons decay into two final states -
$$ K_S \rightarrow \pi^+ + \pi^- $$
$$ K_L  \rightarrow \pi^+ + \pi^- + \pi^0 $$
$ K_S \rightarrow  \pi^+ + \pi^-$ has more energy gap between initial and final states because there are two pions in the final state and has a mean lifetime $8.958\times10^{-11}$ s. Then, it has more phase space available, i.e. more kinetic energy. The probability of this decay must be higher. As it decay fast, that's why this kaon is called the `Short' Kaon ($K_S$). In the same way, as $ K_L  \rightarrow  \pi^+ + \pi^- + \pi^0$ and has a mean lifetime $5.18\times10^{-8}$ s and has less energy gap in between the initial and final states which implies less phase space available and so decay slow. That's why this kaon is called the `Long' Kaons ($K_L$). 
\newline
\newline
As pions are pseudo scalars, their CP eigenvalue is $(-1)$. The CP eigenvalue of two pions final state is $(-1)^2 = 1$. Thus, the initial state kaon ($K_S$) can be identified as $K_1$ with positive eigenvalue, and the CP eigenvalue of the three pions final state is $(-1)^3=-1$, where initial kaon ($K_L$) can be identified as $K_2$ with negative eigenvalue. Therefore, the theory is $CP$ symmetric iff $K_S = K_1$ and $K_L = K_2$ i.e.  
$$ K_1 \rightarrow \pi^+ + \pi^- $$
$$ K_2 \rightarrow \pi^+ + \pi^- + \pi^0 $$
Because by definition (as it is observed), $K_S$ and $K_L$ are the physical or mass eigenstates. The above condition implies the alignment of $CP$ eigenstates along the mass eigenstates (which gives $CP$ invariance). Thus, the $CP$ invariance condition(physical states are CP eigen states) is 
$$ CP |K_S\rangle = +|K_S\rangle $$
$$ CP |K_L\rangle = -|K_L\rangle $$
It was observed by studying phase space that $K_L$ decays into two pions final state with $BR \sim \mathcal{O}(10^{-3})$ which is the small effect but violates CP invariance explicitly. So, physical kaons $K_L$ and $K_S$ are expressed in terms of a linear combination of CP eigenstates, $K_1$ and $K_2$.

For small $\epsilon$ (and up to normalization), $K_L = K_2 + \epsilon K_1$ and similarly for $K_S$. Thus occasionally the $K_L$ decays as a  $K_1$ with $CP = +1$ and likewise the $K_S$ can decay with $CP = -1$. This is known as Indirect CP violation (CP violation due to mixing of $K^0$ and $\overline{K^0}$). There is also a Direct CP violation effect in which the CP violation occurs during the decay itself. Both are present because both mixing and decay arise from the same interaction with the W boson and thus have CP violation predicted by the $U_C$ matrix.
\newline
\newline
Therefore, we found a small CP violation as both eigenstates of $K_L$ and $K_S$ do not align. This is the reason we look for CP-violating models so that we get a natural explanation for CPV.

\subsection{Prospect of CP Violation in SM?}

As we introduced CP transformation, now we will analyze CP symmetry at the Lagrangian level. Consider a theory with a single scalar $\phi$ and two sets of $N$ fermions $\psi_L^i$ and $\psi_R^i$ where $i=1,2,\cdots N$. The Yukawa interaction term in the Lagrangian is 

$$ \mathcal{L}_{Yuk} = Y_{ij} \overline{\psi_{Li}} \phi \psi_{Rj} + Y^*_{ij} \overline{\psi_{Rj}} \phi^\dagger \psi_{Li} $$

The CP transformation of the fields is defined as 
$$\phi \leftrightarrow \phi^\dagger 	\ ; \ \psi_{Li}\leftrightarrow \overline{\psi_{Li}} \ ; \ \psi_{Ri}\leftrightarrow \overline{\psi_{Ri}} $$

Therefore, a CP transformation exchanges the operators, 

$$ \overline{\psi_{Li}} \phi \psi_{Rj} \xrightarrow{\text{CP}} \overline{\psi_{Rj}} \phi^\dagger \psi_{Li} $$

but leaves their coefficients $Y_{ij}$ and $Y^*_{ij}$ unchanged. This means CP is a symmetry of $\mathcal{L}$ if the couplings are real. 
$$  Y_{ij} = Y^*_{ij} $$

CP is violated if using all the freedom to redefine the phases of the fields, one can not find any basis where all couplings are real. If one can find such a basis, then CP is a good symmetry and the CP eigenstates are the physical or mass eigenstates. 
\newline
\newline
In the Standard model, we know that due to symmetry, the couplings of gluons, photons, and Z bosons are real. Because they are symmetries of the ground state and so the couplings must be real (Z boson is exceptional due to broken at ground). The fermion mass term or the Yukawa couplings and the charged weak gauge interactions, the relevant CP transformation laws are

$$ \overline{\psi_i} \psi_j \rightarrow \overline{\psi_j} \psi_i  $$

$$ \overline{\psi_i} \gamma^\mu W_\mu^+ (1-\gamma^5)\psi_j \rightarrow \overline{\psi_j} \gamma^\mu W_\mu^- (1-\gamma^5)\psi_i  $$

Thus, 3 mass terms and the charged current interaction terms are CP invariant if all the masses and the couplings are real. We can always choose masses to be real. Consider the coupling of $W_\mu^\pm$ to fermions.

$$ \frac{g}{\sqrt{2}} [G_{ij} \overline{u_i}\gamma^\mu W_\mu^+(1-\gamma^5)d_j + G_{ij}^* \overline{d_j}\gamma^\mu W_\mu^-(1-\gamma^5)u_i ]$$

The CP operation exchanges the two terms, except the couplings. Thus, CP would be a good symmetry if there were a mass basis where all masses and couplings are real.

\subsection{Experiment vs Standard Model on EDM}

At the Lagrangian level, an EDM violates parity (P) invariance and time reversal (T) invariance. Considering CPT invariance means that the EDM is a CP-violating (CPV) quantity. Thus, CP violation is directly connected with EDM via complex parameters in Lagrangian. Hence, EDM is necessarily the probe for the CP violation. 

It is well established that the flavor-conserving CP violation in the SM is very small, such that the induced particle electric dipole moments (EDMs) are vanishingly small. The non-zero SM contributions to lepton EDMs appear at the four-loop level and are thus strongly suppressed. For instance, the EDM of electron \cite{edfSM} and muon \cite{Babu} is estimated respectively  
$$|d_e^{SM}| \leq 10^{-38} \ e\cdot cm \ \& \ |d_\mu^{SM}| \leq 5 \times 10^{-23} \ e\cdot cm$$
Since it is far below the sensitivity of the current experimental techniques, any observation of a particle EDM will be an unambiguous sign of physics beyond the SM (BSM).
\newline
\newline
The current upper bound on the muon EDM is
$$|d_\mu^{Expt}| < 1.9 \times 10^{-19} \ e\cdot cm \ (95 \% C.L.)$$
set by the Muon $(g-2)$ Collaboration at Brookhaven National Laboratory (BNL) \cite{Muondf}, which is about ten orders of magnitude weaker than the one on the electron EDM \cite{edf} -
$$|d_e^{Expt}| < 4.1 \times 10^{-30} \ e\cdot cm \ (90 \% C.L.)$$
To improve the sensitivity of the muon EDM, there are several future experiments proposed to measure the muon EDM. For instance, J$-$PARC Muon $g-2 / EDM$ experiment \cite{MDMEDM} and the one using the frozen-spin technique at the Paul Scherrer Institute (PSI) \cite{frozenspin} will have sensitivities of $\sigma(d_\mu) \leq 1.5 \times 10^{-21} \ e\cdot cm$ and $\sigma(d_\mu) \leq 6 \times 10^{-23} \ e\cdot cm$, respectively.

\subsection{CP Violation in LRSM}

In the LR model, P and CP violations occur naturally due to its Higgs structure. This also gives enhanced contributes to EDM which will be evident from the loop calculation in Section \ref{Calculation of EDM and MDM Loop} where the QFT structure of EDM will pick only the imaginary part of the coupling ($Im(G_R^*G_L)$) eventually gives CP violation.

\section{Anomalous Magnetic Moment (AMM)}
\label{Anomalous Magnetic Moment (AMM)}
\subsection{Experiment vs Standard Model on AMM}

Land\'e g- factor for elementary charged fermions have value, $g = 2$ at tree level. Any deviation other than this is called Anomalous magnetic moment (AMM) defined by $a = \frac{(g-2)}{2}$. But quantum loop effects lead to small correction which deviates from 2 \cite{Schwinger}. Till now, The most precise theoretical value of AMM is given by the Standard Model. However, these theoretical values are in tension with experimental values. For instance, the recently reported combined value of Fermilab \cite{Fermilab} and BNL \cite{BNLg2} is
$$a_\mu^{Expt} = 116592059(22) \times 10^{-11}$$
while the SM prediction \cite{Muong2SM} is
$$a_\mu^{SM} = 116591810(43) \times 10^{-11} $$
which corresponds to a large $5.0\sigma$ disagreement
$$\Delta a_\mu = a_\mu^{Expt}-a_\mu^{SM} = 249(48) \times 10^{-11} $$ 
Besides, the recently reported precise value of the electron AMM ($a_e$) \cite{Fan:2022eto},
$$a_e^{Expt} = 115965218(12) \times 10^{-11}$$
The SM prediction for electron \cite{Electrong2} relies on the measurement of the fine-structure constant using the recoil velocity/frequency of atoms that absorb a photon, and currently, there is a $5.5\sigma$ discrepancy between the measurements using Rubidium-87 \cite{Morel} and Cesium-133 \cite{Parker} (We will use the Rb-measurement in this article).
$$\Delta a_e = \begin{cases} 
      48(30) \times 10^{-14} \ (Rb) \\
      -88(36) \times 10^{-14} \ (Cs)
   \end{cases}$$
These discrepancies of AMM of charged leptons are strong hints toward modification of the Standard Model.

\subsection{AMM in LRSM}

LR model introduces scalar $SU(2)$ triplet ($\Delta$) having hypercharge which handles the neutrino mass and mixing problem by introducing type-II Seesaw. This also gives enhanced contributions to AMM which will be evident from the loop calculation in Section \ref{Calculation of EDM and MDM Loop}. Thus, the LR model directly explains AMM discrepancy. 


\section{Charged Lepton Flavor Violation (CLFV)}
\label{Charged Lepton Flavor Violation (CLFV)}
\subsection{Prospect of Charged Lepton Flavor Mixing in SM?}

From the Standard Model, we can understand that flavor violation can only occur after the symmetry breaking, when the fermion masses are generated. Before the symmetry breaking the matter particles are massless so all matter particles are gauge eigenstates. After the symmetry breaking, mass splitting occurs (i.e. GIM mechanism) between generations of the fermion family as a result mass eigenstates originate which no longer are the same as the gauge eigenstates but are related by unitary transformation. This fact eventually implies that there are no flavor-changing neutral currents (FCNC) at the Lagrangian level i.e. the flavor cannot change in any process at the tree level where only neutral gauge bosons (photon,$\gamma$, and $Z$ boson) are exchanged. The flavor can only change in flavor changing charged currents (FCCC) at tree level where charged gauge bosons ($W^\pm$)are exchanged.

In general, fields can mix if they belong to the same representation under all the unbroken generators. That is, they must have the same spin, electric charge, and representation. If these fields also belong to the same representation under the broken generators, their coupling to the massive gauge boson is universal. However, if they belong to a different representation under the broken generators, their couplings in the interaction basis are diagonal but non-universal. These couplings become non-diagonal after rotation to the mass basis. Since the electroweak neutral currents retained their form when switched from the gauge eigenstates to the mass eigenstates, the flavor is not violated by them.
\newline
\newline
One usually wants to use the mass eigenstates instead of gauge eigenstates since the mass
eigenstates are the physical fields that are seen in the experiments. For this we need to use unitary matrices (say $V$) to rotate into mass eigenstates in which the basis mass matrix is diagonal. Define mass eigenstates $\psi_{L,R}^m$ such that 
$$\Psi_{L,R} = V_{L,R} \psi_{L,R}^m$$

Diagonalized mass matrices are then given by ( Lagrangian is invariant under this rotation thus from $\bar{\psi}_L M_U \psi_R $ invariance )
$$ \bar{\Psi}_{Li} M_{ij} \Psi_{Rj} = \bar{\psi}_{Li}^m V_L^{\dagger} M_{ij} V_R \psi_{Rj}^m = \bar{\psi}_{Li}^m \hat{M}_{ij} \psi_{Rj}^m$$
where, $$ \hat{M} = diag(m_1 , m_2 , m_3) = V^{\dagger}_L M V_R$$

where $m_i$ with $i=u,c,t,d,s,b,e,\mu,\tau$ are the masses of the individual states in the massive eigenstates. Hence, we have flavor-changing charged current (FCCC) interactions with the W bosons: 
$$\mathcal{L}_{cc} = \frac{g}{\sqrt{2}}[\overline{u}^m_L\gamma^\mu (V_u^\dagger V_d)d^m_L+\overline{\nu}^m_L\gamma^\mu (V_\nu^\dagger V_e)e^m_L]W^+_\mu + H.c.$$ where mixing matrix 
$$U_{CKM} = V_u^\dagger V_d$$ 
$$U_{PMNS}  = V_\nu^\dagger V_e$$
But in SM, neutrino having no mass, the neutrino matrix is trivial, i.e. $V_\nu = \mathds{1}$, leading to no mixing of neutrino. Then we can just go to charged lepton mass basis and mixing matrix of lepton sector $U_{PMNS} = V_e$ i.e. trivial. Thus, no possibility of the charged leptons mixing in the theory. 
\newline
\newline
From the above discussion, we can see that off diagonal term of mixing matrix or in general nontrivial mixing matrix acts as a source for the flavor violation. Mixing of neutrinos could suggest that the charged leptons- $e$, $\mu$, and $\tau$ could mix via FCCC. 

So, we can conclude from the Standard Model by saying that, there are no charged lepton flavor violating processes in the Standard Model. So any observation of a charged lepton flavor violating process via FCNC (if neutrino mass exists then FCCC occurs in both SM and BSM) would prove the existence of physics beyond the Standard Model (BSM).

\subsection{Effective Operators for CLFV Process}

The Standard Model Lagrangian is constrained to include only renormalizable terms, meaning that the coefficients of every term in the Lagrangian possess non-negative mass dimensions. Furthermore, all these terms must remain invariant under the Standard Model gauge group. However, when we relax the requirement of renormalizability, the Standard Model can be extended by incorporating gauge-invariant operators characterized by mass dimensions greater than 4 ($D > 4$). These operators emerge as remnants of a higher-level theory when we examine the low-energy limit. At energy scales lower than those associated with the higher-energy theory, it becomes improbable to produce the heavy particles from the higher theory during collisions or decays. Consequently, these heavy particles are effectively absent in the external states of the system. Nevertheless, these heavy fields persist as virtual particles, influencing the behavior of the system even though they remain undetected. This phenomenon is akin to how the effects of the electroweak theory were initially discovered, as virtual effects extending beyond the realm of Quantum Electrodynamics (QED), which was the prevailing Standard Model at the time.

\begin{figure}[H]
\begin{center}
\includegraphics[width=45mm]{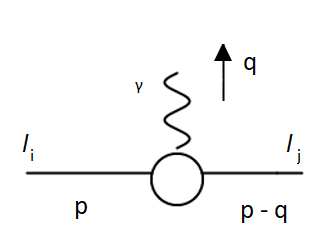}
\end{center}
\caption{Effective vertex of photonic flavor violating channels}
\label{LFV}
\end{figure}

The propagators of these heavy particles effectively collapse to single points, leading to a suppression factor approximately proportional to $\frac{1}{M^2_{NP}}$ in the low-momentum limit. This reduction in the propagators results in the predominance of contact interactions. The vertex function associated with these new vertices is referred to as an Effective Operator (see Appendix \ref{Effective Field Theory}). This is the fundamental idea of Effective field theories(EFTs)- 
\newline
\newline
``At energies much lower than the energy scale of the higher theory, the propagators of the heavy particles become factors which are suppressed by the energy scale of the higher theory."
\newline
\newline
Now, we will briefly study effective vertex originated from Feynman amplitude of photonic flavor violating channels where off-shell photon emission also contributes (Figure \ref{LFV}) 
\graphicspath{{./LFV/}}
written in a general form \cite{chengandli}
$$\mathcal{M} = \overline{u}_j(p-q)[\gamma_\lambda(A+B\gamma_5)+q_\lambda(C+D\gamma_5) + iq^\nu \sigma_{\lambda\nu}(E+F\gamma_5)]u_i(p)\epsilon^\lambda$$
where $A, B, C, D, E$ and $F$ are some constants or form factors. This is the most general form that can be simplified, however.

Using Ward identity($q^\mu \mathcal{M}_\mu(q) = 0$) since there is gauge invariance, we deduce that $A = B = 0$ when photon ($\gamma$) is on-shell ($q^2 = 0$). Also from transverse photon polarization i.e. $q^\mu \epsilon_\mu = 0$ for an on-shell photon ($\gamma$), the $C$ and $D$ term in the amplitude vanishes. So we are only left with electric and magnetic dipole transition term
$$\mathcal{M} = \overline{u}_j(p-q)[iq^\nu \sigma_{\lambda\nu}(E+F\gamma_5)]u_i(p) \epsilon^\lambda$$
We make further approximation if $ m_{l_i} >> m_{l_j} $ i.e. $m_{l_j} \approx 0$. This means that the outgoing lepton must be left-handed which is only possible if $E = F$. Then the Feynman
amplitude for the process is
$$\mathcal{M} = F\overline{u}_j(p-q)[iq^\nu \sigma_{\lambda\nu}(1+\gamma_5)]u_i(p) \epsilon^\lambda$$
Therefore, $\mathcal{M}$ correspond to a Dim 5 operator  
$$\mathcal{L}_5 = \frac{\mu^M_{ij}}{2}\overline{\psi}_i \sigma^{\mu\nu} \psi_j F_{\mu\nu}+ \frac{\mu^E_{ij}}{2}\overline{\psi}_i \sigma^{\mu\nu}\gamma_5 \psi_j F_{\mu\nu} + H.c.$$ 
Since mass dimension of $\psi$ and $F_{\mu \nu}$ are, 
$[\psi]=\frac{3}{2} , [F_{\mu \nu}]=2 $. Hence, the operator is nonrenormalizable as the term is 5 dimensional while space-time is 4 dimensional. So, it can not occur at the Lagrangian of the theory i.e. EDM and MDM contribution is zero at the tree level. This is 1 loop form (in IR limit) of Dim 5 Weinberg operator \cite{weinberg1979baryon} (see Appendix \ref{Effective Field Theory}).

From $\mathcal{L}_5$, the diagonal part of  $\mu^M_{ij}$ in the transition generates the anomalous magnetic moments (AMM) of the fundamental particles. Similarly, the diagonal part of $\mu^E_{ij}$ generates contributions to the electric dipole moments of the fundamental particles and indirect contribution to flavor violation. The off-diagonal elements contribute to CLFV processes. Since $ m_{l_i} >> m_{l_j} $, it is convenient to define the dipole form factor, $F_{MDM/EDM}$ such that $\mu^{M/E}_{ij} = \frac{em_{l_i} F_{MDM/EDM}}{2}$. So, naturally, any flavor non-diagonal coupling will activate CLFV processes and at the same time yield contributions to the anomalous magnetic moments.
\newline
\newline
So, we can conclude that CLFV can be thought of as remnants of the higher theory occurring via effective vertex in the low energy limit since we can not have CLFV processes at tree level in SM. Therefore, the reason for the least possibility of CLFV processes is - \begin{enumerate}
\item \textbf{Tree level or GIM Suppression}: EDM and MDM operators of $\mathcal{L}_5$ are nonrenormalizable and thus can not occur at Lagrangian of the theory. So, CLFV is suppressed at the tree level.
\item \textbf{Loop Suppression}: As EDM and MDM contributions are zero at the tree level, they will contribute via 1st-order quantum loop corrections ( $\mathcal{O}(\hbar)$ suppression) which is smaller than dominating tree-level contribution (no $\hbar$ suppression).
\item \textbf{Flavor Mixing Suppression}: Flavor mixing matrix is nontrivial in higher theories and both EDM and MDM operators of the CLFV process contribute to the off-diagonal terms of flavor mixing matrix which are small.

\end{enumerate}

\subsection{CLFV in LRSM}

So far, it is well understood that to make CLFV happen, we need  
\begin{enumerate}
\item an effective vertex that bypasses tree level or GIM suppression
\item and non trivial mixing matrix of lepton sector, $U_{PMNS} = V_\nu^\dagger V_e$ such that neutrino matrix, $V_\nu \neq \mathbb{1}$.
\end{enumerate}
Now, the Interacting Lagrangian from the LR model for the photonic CLFV process (Figure \ref{LFV}) is given by, 
$$ \mathcal{L}^{UV}_{int}= \bar{\psi} \gamma^{\mu}[G_L P_L + G_R P_R]\psi  W^a_\mu T_a + \bar{\psi}[G_L P_L + G_R P_R]\psi H $$
At the IR limit, this interacting Lagrangian becomes
$$ \mathcal{L}^{IR}_{int}= \frac{em_{l_i}}{2} \overline{\psi}_{l_i} (V^\dagger_{i \alpha}V_{\alpha j}) \sigma^{\mu\nu} [G_L P_L + G_R P_R] \psi_{l_j} F_{\mu\nu} + H.c.$$
Therefore, we can see that both requirements are naturally acquired in the Left-Right Symmetric Model because not only is there exists a nontrivial mixing matrix $V^\dagger_{i \alpha}V_{\alpha j}$  which is nothing but $U_{PMNS}$ (see Section \ref{Neutrino Physics}) due to seesaw compatibility but also allows Dim 5 effective vertex running BSM particles in 1 loop.
$$\delta \Gamma^\mu \sim \frac{1}{M^2_{NP}}(G_L^* G_L terms + G_R^* G_R terms) + \frac{m_0}{M^2_{NP}}(G_R^* G_L terms+G_L^* G_R terms)$$
\graphicspath{{./clfv/}}
\begin{figure}[H]
\begin{center}
\includegraphics[width=100mm]{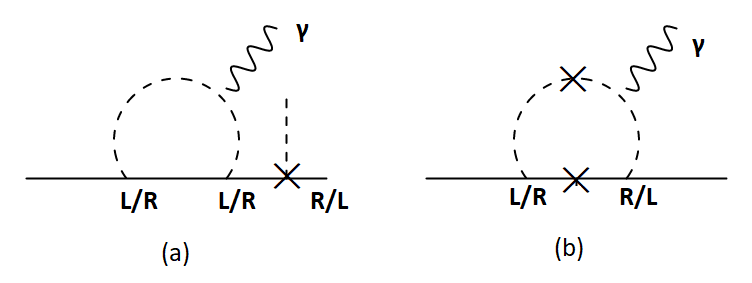}
\end{center}
\caption{All possible  photonic CLFV diagram where mass insertion (a) outside (b) inside}
\label{clfv}
\end{figure}
Therefore, the LR model manifests the CLFV phenomenon via FCNC where the only unknowns needed are energy scale and off-diagonal Yukawa couplings.


\section{Couplings vs Right-handed Neutrino Mass}
\label{Couplings vs Right handed Neutrino Mass}
\subsection{Yukawa Coupling in terms of Right-handed Neutrino Mass Matrix}

The right-handed neutrinos acquire mass from the $\Delta_R$ field getting vacuum expectation value. The Lagrangian term for right-handed neutrino mass (Section \ref{Left-Right Symmetric Model (LRSM)}) is 
$$\mathcal{L}_{\nu_R}^{Mass} =  \nu_{R}^T C^{-1} M_N \nu_R $$
The Yukawa Lagrangian corresponding to right-handed leptons coupling to $\Delta$ field is given by (Section \ref{Left-Right Symmetric Model (LRSM)}),
$$\mathcal{L}_{Yuk}(\Delta_R) = (Y_R)_{ij} L_{Ri}^T C^{-1} i \tau^2 \Delta_R L_{Rj} + h.c. $$
After the $\Delta_R$ field gets vacuum expectation value, the Lagrangian becomes,
$$\mathcal{L}_{Yuk}(\langle \Delta_R \rangle_0) = (Y_R)_{ij} \frac{v_R}{\sqrt{2}} \nu_{Ri}^T C^{-1}  \nu_{Rj} + h.c. $$
Thus, the right-handed neutrino mass matrix, from $\Delta_R$ acquiring vacuum expectation value, can be found comparing the right-handed neutrino mass Lagrangian with the above equation.
$$M_N = Y_R \frac{v_R}{\sqrt{2}}$$
This $M_N$ is a complex symmetric matrix. It can be diagonalized with a unitary matrix, say $U_R$. Thus,
$$M_N = U_R^\dagger \hat{M}_N U_R^* $$
Therefore, the right-handed Yukawa coupling can be written as:
$$Y_R = \frac{\sqrt{2}}{v_R} U_R^\dagger \hat{M}_N  U_R^* $$
Now, recall the right handed $W_R$ boson mass is given by (in the limit $v_R >> k,k'$),
$$M_{W_R} = g_R\frac{v_R}{\sqrt{2}}$$
Thus, $Y_R$ in terms of $M_{W_R}$ and $g_R$ can be written as:
$$Y_R = (\frac{g_R}{M_{W_R}}) U_R^\dagger \hat{M}_N  U_R^*  $$

\subsection{Plots and Discussion}
\begin{figure}[H]
     \centering
     \begin{subfigure}[b]{0.4\textwidth}
         \centering
         \includegraphics[width= \textwidth]{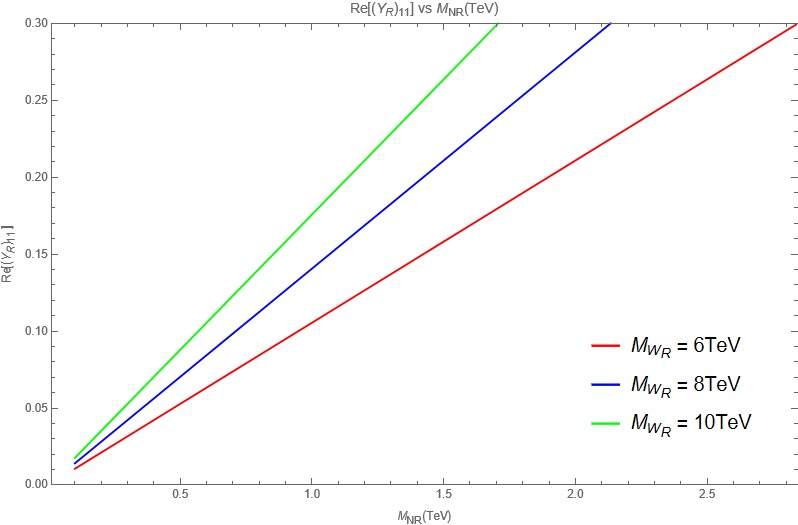}
         \caption{Re[$(Y_R)_{11}$] vs $M_{N_R}$}
         \label{dire}
     \end{subfigure}
     \hspace{1cm}
     \begin{subfigure}[b]{0.4\textwidth}
         \centering
         \includegraphics[width=\textwidth]{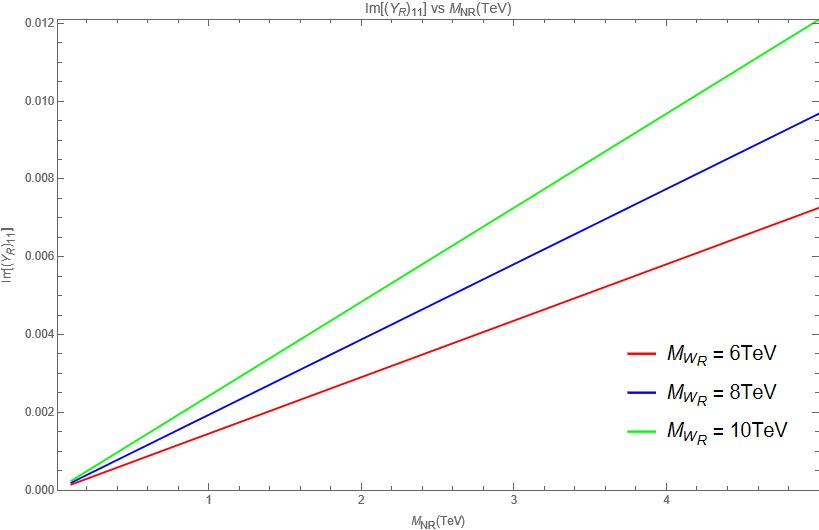}
         \caption{Im[$(Y_R)_{11}$] vs $M_{N_R}$}
         \label{diim}
     \end{subfigure}
     \vfill
     \begin{subfigure}[b]{0.4\textwidth}
         \centering
         \includegraphics[width=\textwidth]{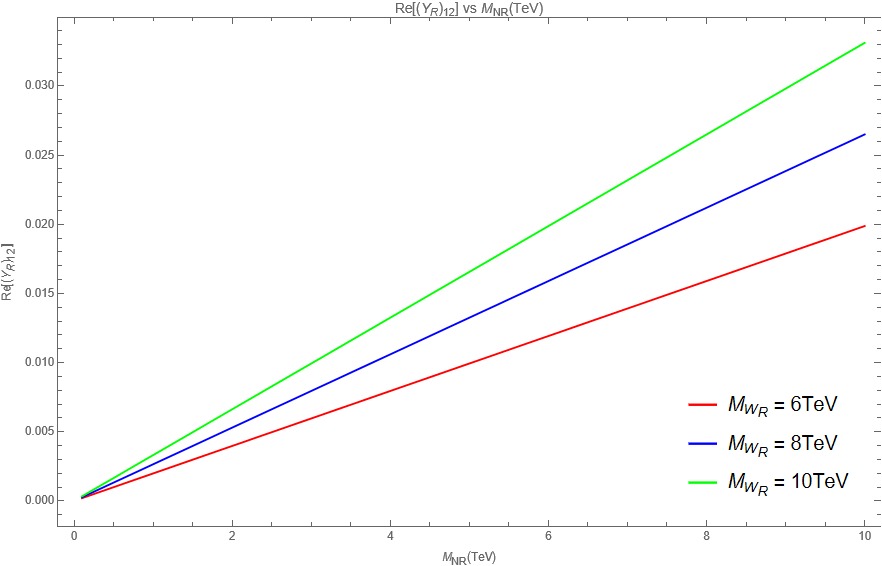}
         \caption{Re[$(Y_R)_{12}$] vs $M_{N_R}$}
         \label{odire}
         \end{subfigure}
         \hspace{1cm}
         \begin{subfigure}[b]{0.4\textwidth}
         \centering
         \includegraphics[width=\textwidth]{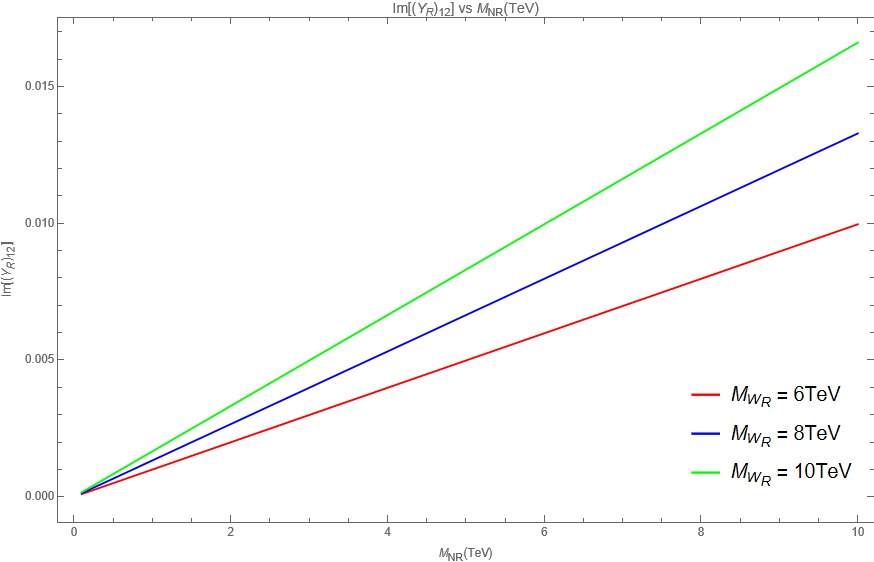}
         \caption{Im[$(Y_R)_{12}$] vs $M_{N_R}$}
         \label{odiim}
         \end{subfigure}
     \caption{Coupling parameters vs Right-handed Neutrino mass}
     \label{MNR}
\end{figure}

For phenomenological considerations, we seek to examine scenarios where the right-handed sector mirrors the characteristics of the left-handed sector. Explicitly, we want to analyze the case when 
\begin{itemize}
\item left-handed and right-handed weak couplings are same $g_L = g_R$ i.e. Parity manifest
\item the right-handed mixing matrix $U_R$ is same as left handed $U_{PMNS}$ matrix
\item Finally, all heavy neutrinos have the same masses.
\end{itemize}
Based on the above assumptions, we generate the $Y_R$ vs $M_{N_R}$ graphs for $M_{W_R}=$ 6 TeV, 8 TeV, and 10 TeV in Figure \ref{MNR}.

Therefore, adopting the aforementioned assumptions, it becomes evident that only the real component of the diagonal couplings attains magnitudes on the order of $0.1$ for neutrino masses at $1 TeV$. Conversely, all other plots evident significant suppressed couplings for a neutrino mass of $1 TeV$, reaching the order of $0.001$. This phenomenon holds only when we treat the left and right-handed sectors on an equal footing.

Thus, we will systematically vary the couplings within the phenomenologically interesting region i.e. the range of $0.001$ to $0.1$ when the neutrino mass is $1 TeV$ for a comprehensive analysis of parity manifest minimal LRSM in the subsequent section \ref{Phenomenology}.

\section{Phenomenological Consequences}
\label{Phenomenology}
In this section, we will generate parameter space of the Yukawa couplings originating from the extended Higgs sector of parity manifest minimal LRSM using experimental bounds of the following charged lepton flavor violating processes:
\begin{table}[H]
        \centering
      \begin{tabular}{ | c | c |} 
      \hline
Process  & $BR$ \\
\hline 
$\mu \rightarrow 3e$ & $ < 1.0 \times 10^{-12}$ \cite{SINDRUM} \\
\hline
$\tau \rightarrow 3e$ & $ < 2.7 \times 10^{-8}$ \cite{Belle}\\
\hline
$\tau \rightarrow 3\mu$ & $ < 3.3 \times 10^{-8}$ \cite{Belle}\\
\hline
\hline 
$\mu \rightarrow e\gamma$ & $ < 4.2 \times 10^{-13}$ \cite{MEG:2016leq}\\
\hline
$\tau \rightarrow e\gamma$ & $< 3.3 \times 10^{-8}$ \cite{BaBar} \\
\hline
$\tau \rightarrow \mu\gamma$ & $< 4.4 \times 10^{-8}$ \cite{BaBar}\\
\hline
     \end{tabular}
     \caption{Experiment Bounds on $l_i \rightarrow l_j\gamma$ and $l_i \rightarrow 3l_j$ channels}
     \label{clfvexpt}
\end{table}
where masses of internal fermions and bosons are -
\begin{table}[H]
        \centering
      \begin{tabular}{ | c | c |} 
      \hline
     Particle  & Mass\\
\hline 
$W_1$ & 80.377 GeV\\

$Z_1$ & 91.1876 GeV\\

$Z_2$ & $> 1.162$ TeV \cite{Aguila}\\
\hline
$H^+$ & $> 1.103$ TeV \cite{ATLAS:2018gfm}\\
$H^0$ & $> 1.613$ TeV \cite{CMS:2018rmh}\\
\hline
$N$ & 1 TeV (Section \ref{Couplings vs Right handed Neutrino Mass})\\
\hline
\end{tabular}
\caption{Masses of Internal fermions and bosons}
\label{mass}
\end{table}
Subsequently, we will evaluate the one-loop MDM and EDM contributions of the model (Appendix \ref{Calculation of EDM and MDM Loop}) to see whether they match with the recent experimental values of AMM and EDM or exceed:
\begin{table}[H]
        \centering
      \begin{tabular}{ | c | c | c |} 
      \hline
Particle  & $\Delta a = a^{Expt}-a^{SM}$ & $|d^{Expt}| \ (TeV^{-1}) $\\
\hline 
$e$ & $48(30) \times 10^{-14}$ \cite{Fan:2022eto},\cite{Morel} & $< 3.30645 \times 10^{-14}$ \cite{edf}\\
\hline
$\mu$ & $249(48) \times 10^{-11}$ \cite{Fermilab},\cite{Muong2SM} & $< 0.00153226$ \cite{Muondf}\\ 
\hline
     \end{tabular}
     \caption{Experiment values of AMM and EDM of Electron and Muon}
     \label{exptvalue}
\end{table}

\subsection{Coupling Parameter Space from Charged Lepton Flavor Violating Channels}
\label{Parameter Space of Couplings}
\subsubsection{Branching Ratio Formulas}

From Section \ref{Charged Lepton Flavor Violation (CLFV)}, we saw that minimal LRSM gives rise to CLFV channels due to the presence of non-zero off-diagonal couplings. Now, Yukawa Lagrangian for extended Higgs of minimal LRSM be \cite{Cirigliano},
$$\mathcal{L}_{Yuk}^{H^+} = \frac{g}{\sqrt{2}}[H^+ \overline{N}(\widetilde{h}P_L)l + H^- \overline{l}(\widetilde{h}^\dagger P_R)N]$$
$$\mathcal{L}_{Yuk}^{\Delta^{++}} = \frac{g}{2}[\Delta^{++}_{L/R} \overline{l^c}(h_{L/R}P_{L/R})l + \Delta^{--}_{L/R} \overline{l}(h^\dagger_{L/R}P_{R/L})l^c]$$
Here, $P_{L/R} = \frac{1 \mp \gamma_5}{2}$. For manifest LRSM, $h_L = h_R = h = K_R^T \frac{diag(M_\nu)}{M_{W_2}}K_R$ and $\widetilde{h} = K_L^* h_L$. Thus coupling relation with right-handed mixing matrix be
$$h_{ij} = \sum_{n = heavy} (K_R^\dagger)_{in}(K_R)_{nj}\sqrt{x_n} \ ; \ x_n = (\frac{M_n}{M_{W_2}})^2$$
$$(h^\dagger h)_{ij} = \sum_{n = heavy} x_n (K_R^\dagger)_{in}(K_R)_{nj} \ ; \ x_n = (\frac{M_n}{M_{W_2}})^2$$
Thus, Branching ratio formulas - \begin{itemize}
\item \textbf{For $l_i \rightarrow 3l_j$ Channels}
\graphicspath{{./l3l/}}
\begin{figure}[H]
\begin{center}
\includegraphics[width=50mm]{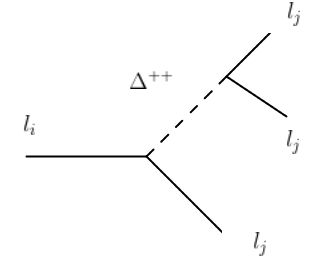}
\end{center}
\caption{Tree level diagram of $l_i \rightarrow 3l_j$ channel}
\label{ll}
\end{figure}
In minimal LRSM, $l_i \rightarrow 3l_j$ process at tree level exchange is mediated by doubly charged Higgs ($\Delta^{++}$)  so the branching ratio is \cite{Cirigliano},
$$BR(l_i \rightarrow 3l_j) = \frac{1}{2} |h_{ij} h^*_{jj}|^2 M_{W_1}^4 (\frac{1}{M^4_{\Delta_L^{++}}}+\frac{1}{M^4_{\Delta_R^{++}}})$$

\item \textbf{For $l_i \rightarrow l_j\gamma$ Channels} 

In minimal LRSM, $l_i \rightarrow l_j\gamma$ process at loop level exchange (Figure \ref{alldiagm} (a),(c),(d) ) is mediated by both charged gauge bosons and charged Higgses so branching ratio is \cite{Cirigliano},
$$BR(\mu \rightarrow e\gamma) = \frac{\Gamma(\mu \rightarrow e\gamma)}{\Gamma_\mu} = 384\pi^2 e^2 (|A_L|^2+ |A_R|^2)$$ 
Here, $\Gamma_\mu = \frac{G_F^2m^5_\mu}{192\pi^3}$. Coefficient $A_{L}$ has contributions from $W_2, \nu_R, \Delta^{++}_R$ and is given by
$$(A_{L})_{ij} = \frac{1}{16\pi^2}\sum_{n = heavy}(K^\dagger_R)_{in}(K_R)_{nj}M^2_{W_1}[S_3(x_n)\frac{1}{M^2_{W_2}}-\frac{x_n}{3}\frac{1}{M^2_{\Delta^{++}_R}}] \ ; \ x_n = (\frac{M_n}{M_{W_2}})^2$$
where,
$$S_3(x) = -\frac{x(1+2x)}{8(1-x)^2} + \frac{3x^2}{4(1-x)^2}[S_4(x)+1] \ \& \ S_4(x) = \frac{x(1-x + \log x)}{(1-x)^2}$$
and Coefficient $A_{R}$ has contributions from $H^+, \Delta^{++}_L$ and is given by
$$(A_{R})_{ij} = \frac{1}{16\pi^2}\sum_{n = heavy}(K^\dagger_R)_{in}(K_R)_{nj}x_n M^2_{W_1}[-\frac{1}{3M^2_{\Delta^{++}_L}}-\frac{1}{24M^2_{H^+}}] \ ; \ x_n = (\frac{M_n}{M_{W_2}})^2$$

\end{itemize}

\subsubsection{Plots \& Results}
We generated the data using Wolfram Mathematica's built-in function RandomReal to generate random numbers for off-diagonal couplings and appended the data into a data file. Finally, we plotted the data set using the ListPlot function with proper labels and legends utilizing experimental bound from Table \ref{clfvexpt} and masses from Table \ref{mass}.
\newpage
\subsubsection*{For $l_i \rightarrow 3 l_j$ process}

\textbf{Plots:} \begin{figure}[H]
     \centering
     \begin{subfigure}[b]{0.4\textwidth}
         \centering
         \includegraphics[width=\textwidth]{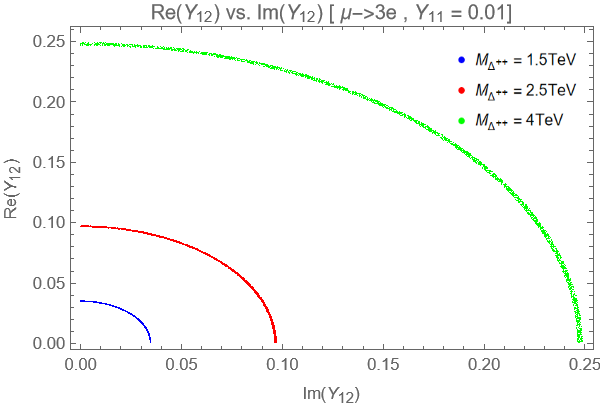}
         \caption{$Y_{12}$ coupling parameter space of $\mu \rightarrow 3e$ channel when $Y_{11} = 0.01$}
         \label{mu3e1}
     \end{subfigure}
     \hspace{1cm}
     \begin{subfigure}[b]{0.4\textwidth}
         \centering
         \includegraphics[width=\textwidth]{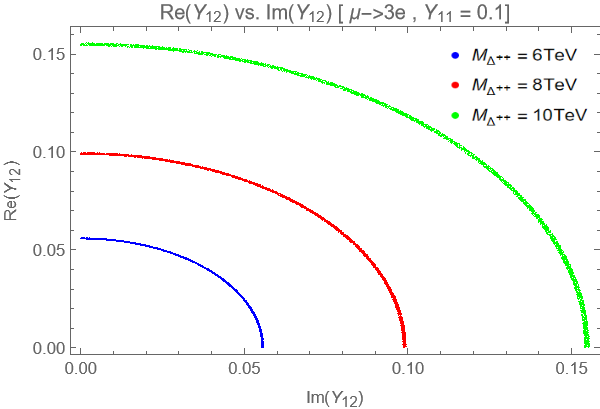}
         \caption{$Y_{12}$ coupling parameter space of $\mu \rightarrow 3e$ channel when $Y_{11} = 0.1$}
         \label{mu3e2}
     \end{subfigure}
     \vfill
     \begin{subfigure}[b]{0.4\textwidth}
         \centering
         \includegraphics[width=\textwidth]{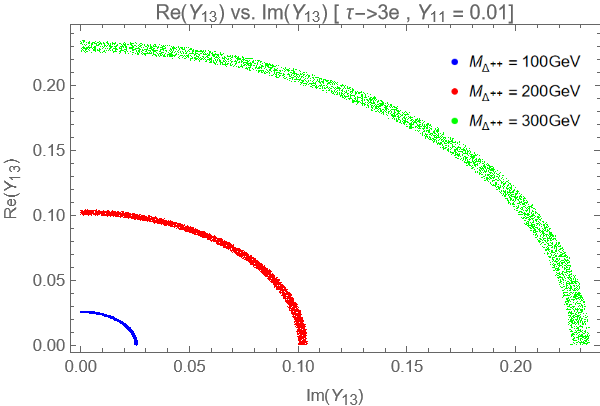}
         \caption{$Y_{13}$ coupling parameter space of $\tau \rightarrow 3e$ channel when $Y_{11} = 0.01$}
         \label{tau3e1}
         \end{subfigure}
         \hspace{1cm}
         \begin{subfigure}[b]{0.4\textwidth}
         \centering
         \includegraphics[width=\textwidth]{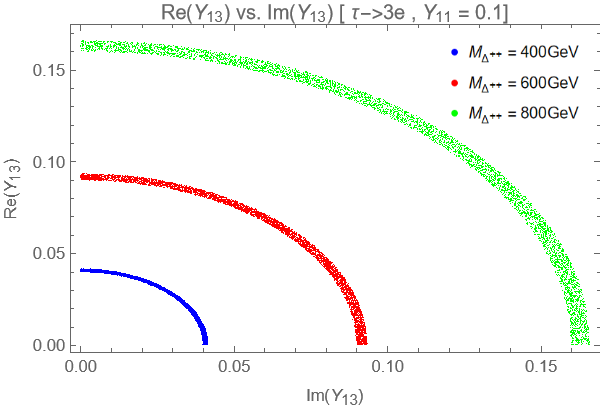}
         \caption{$Y_{13}$ coupling parameter space of $\tau \rightarrow 3e$ channel when $Y_{11} = 0.1$}
         \label{tau3e2}
         \end{subfigure}
         \vfill
         \begin{subfigure}[b]{0.4\textwidth}
         \centering
         \includegraphics[width=\textwidth]{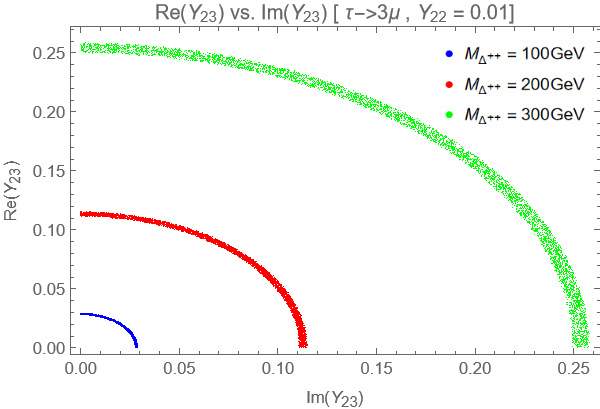}
         \caption{$Y_{23}$ coupling parameter space of $\tau \rightarrow 3\mu$ channel when $Y_{22} = 0.01$}
         \label{tau3mu1}
         \end{subfigure}
         \hspace{1cm}
         \begin{subfigure}[b]{0.4\textwidth}
         \centering
         \includegraphics[width=\textwidth]{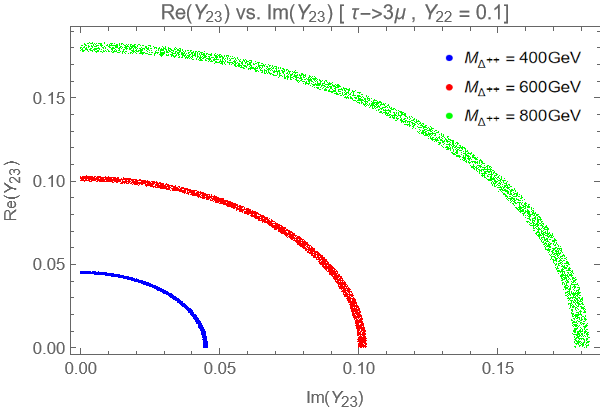}
         \caption{$Y_{23}$ coupling parameter space of $\tau \rightarrow 3\mu$ channel when $Y_{22} = 0.1$}
         \label{tau3mu2}
     \end{subfigure}
     \caption{Off diagonal coupling parameter space from $l_i \rightarrow 3l_j$ channels}
     \label{3l}
\end{figure}
Here, we plotted $Re(Y_{ij})$ vs $Im(Y_{ij})$ for diagonal couplings $Y_{ii}=0.1$ \& $Y_{ii}=0.01$ with doubly charged Higgs masses:
\begin{itemize}
\item $400 GeV$, $600 GeV$ and $800 GeV$ in $Y_{22}=0.1$ and $100 GeV$, $200 GeV$ and $300 GeV$ in $Y_{22}=0.01$ for $\tau \rightarrow 3\mu $ process.  
\item $400 GeV$, $600 GeV$ and $800 GeV$ in $Y_{11}=0.1$ and $100 GeV$, $200 GeV$ and $300 GeV$ in $Y_{11}=0.01$ for $\tau \rightarrow 3e $ process.
\item $6 TeV$, $8 TeV$ and $10 TeV$ in $Y_{11}=0.1$ and $1.5 TeV$, $2.5 TeV$ and $4 TeV$ in $Y_{11}=0.01$ for $\mu \rightarrow 3e $ process.
\end{itemize}
The reason for choosing these specific masses is, that we focused on the phenomenologically interesting region discussed in Section \ref{Couplings vs Right handed Neutrino Mass} (off-diagonal couplings lie in the region $0 < Y_{ij} \leq 0.1$).
\newline
\newline
\textbf{Results:} Off-diagonal Couplings $Y_{12}$, $Y_{23}$ and $Y_{13}$ are calculated from the plots of the processes $\mu \rightarrow 3 e$, $\tau \rightarrow 3 \mu$ and $\tau \rightarrow 3e$ respectively and summarized in the table below -
\begin{table}[H]
        \centering
\begin{tabular}{ | c | c | c | c | } 
      \hline
      \multicolumn{2}{| c |}{$Y_{11} = 0.01$} & \multicolumn{2}{c |}{$Y_{11} = 0.1$}\\     
\hline 
$M_{\Delta^{++}} = 1.5 TeV$ & $Y_{12} = 0.03452$ & $M_{\Delta^{++}} = 6 TeV$ & $Y_{12} = 0.05539$\\
\hline
$M_{\Delta^{++}} = 2.5 TeV$ & $Y_{12} = 0.09624$ & $M_{\Delta^{++}} = 8 TeV$ & $Y_{12} = 0.09859$\\
\hline
$M_{\Delta^{++}} = 4 TeV$ & $Y_{12} = 0.24579$ & $M_{\Delta^{++}} = 10 TeV$ & $Y_{12} = 0.15417$\\
\hline
\end{tabular}
\caption{$Y_{12}$ coupling values from $\mu \rightarrow 3e$ process}
\label{y12m3e}
\end{table}
\begin{table}[H]
        \centering
\begin{tabular}{ | c | c | c | c | } 
      \hline
      \multicolumn{2}{| c |}{$Y_{11} = 0.01$} & \multicolumn{2}{c |}{$Y_{11} = 0.1$}\\     
\hline 
$M_{\Delta^{++}} = 100 GeV$ & $Y_{13} = 0.02458$ & $M_{\Delta^{++}} = 400 GeV$ & $Y_{13} = 0.04095$\\
\hline
$M_{\Delta^{++}} = 200 GeV$ & $Y_{13} = 0.10213$ & $M_{\Delta^{++}} = 600 GeV$ & $Y_{13} = 0.09206$\\
\hline
$M_{\Delta^{++}} = 300 GeV$ & $Y_{13} = 0.22916$ & $M_{\Delta^{++}} = 800 GeV$ & $Y_{13} = 0.16378$\\
\hline
\end{tabular}
\caption{$Y_{13}$ coupling values from $\tau \rightarrow 3e$ process}
\label{y13t3e}
\end{table}
\begin{table}[H]
        \centering
\begin{tabular}{ | c | c | c | c | } 
      \hline
      \multicolumn{2}{| c |}{$Y_{22} = 0.01$} & \multicolumn{2}{c |}{$Y_{22} = 0.1$}\\     
\hline 
$M_{\Delta^{++}} = 100 GeV$ & $Y_{23} = 0.02738$ & $M_{\Delta^{++}} = 400 GeV$ & $Y_{23} = 0.04457$\\
\hline
$M_{\Delta^{++}} = 200 GeV$ & $Y_{23} = 0.11154$ & $M_{\Delta^{++}} = 600 GeV$ & $Y_{23} = 0.10041$\\
\hline
$M_{\Delta^{++}} = 300 GeV$ & $Y_{23} = 0.25429$ & $M_{\Delta^{++}} = 800 GeV$ & $Y_{23} = 0.18071$\\
\hline
\end{tabular}
\caption{$Y_{23}$ coupling values from $\tau \rightarrow 3\mu$ process}
\label{y23t3m}
\end{table}
\newpage
\subsubsection*{For $l_i \rightarrow l_j \gamma$ process}

\textbf{Plots:} \begin{figure}[H]
     \centering
     \begin{subfigure}[b]{0.3\textwidth}
         \centering
         \includegraphics[width=\textwidth]{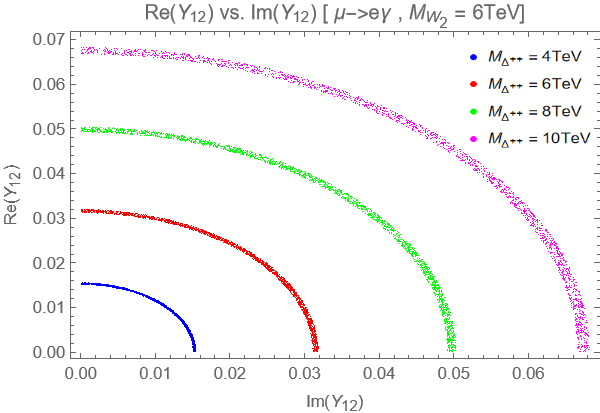}
         \caption{$Y_{12}$ coupling parameter space of $\mu \rightarrow e\gamma$ channel when $W_R = 6$ TeV}
         \label{mueg1}
     \end{subfigure}
     \hfill
     \begin{subfigure}[b]{0.3\textwidth}
         \centering
         \includegraphics[width=\textwidth]{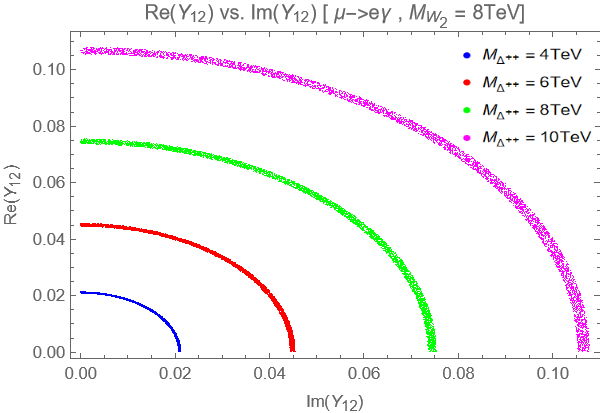}
         \caption{$Y_{12}$ coupling parameter space of $\mu \rightarrow e\gamma$ channel when $W_R = 8$ TeV}
         \label{mueg2}
     \end{subfigure}
     \hfill
     \begin{subfigure}[b]{0.3\textwidth}
         \centering
         \includegraphics[width=\textwidth]{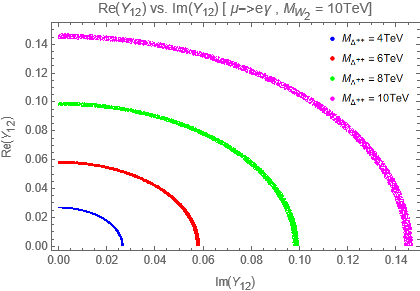}
         \caption{$Y_{12}$ coupling parameter space of $\mu \rightarrow e\gamma$ channel when $W_R = 10$ TeV}
         \label{mueg3}
     \end{subfigure}
     \vfill
     \begin{subfigure}[b]{0.3\textwidth}
         \centering
         \includegraphics[width=\textwidth]{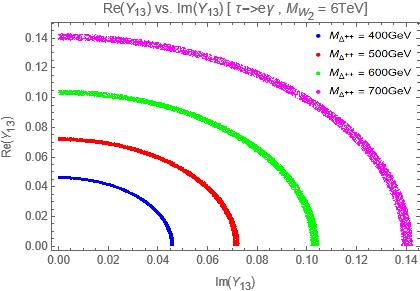}
         \caption{$Y_{13}$ coupling parameter space of $\tau \rightarrow e\gamma$ channel when $W_R = 6$ TeV}
         \label{taueg1}
     \end{subfigure}
     \hfill
     \begin{subfigure}[b]{0.3\textwidth}
         \centering
         \includegraphics[width=\textwidth]{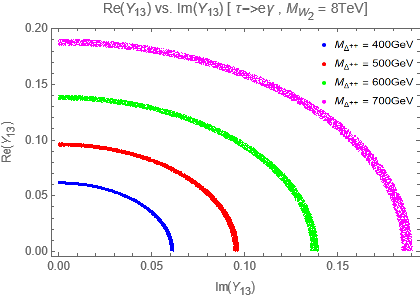}
         \caption{$Y_{13}$ coupling parameter space of $\tau \rightarrow e\gamma$ channel when $W_R = 8$ TeV}
         \label{taueg2}
     \end{subfigure}
     \hfill
     \begin{subfigure}[b]{0.3\textwidth}
         \centering
         \includegraphics[width=\textwidth]{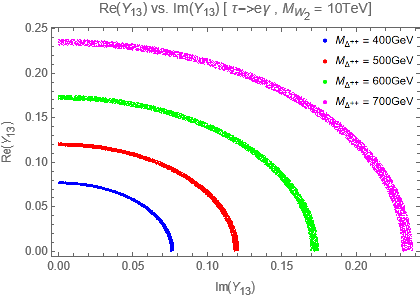}
         \caption{$Y_{13}$ coupling parameter space of $\tau \rightarrow e\gamma$ channel when $W_R = 10$ TeV}
         \label{taueg3}
     \end{subfigure}
     \vfill
     \begin{subfigure}[b]{0.3\textwidth}
         \centering
         \includegraphics[width=\textwidth]{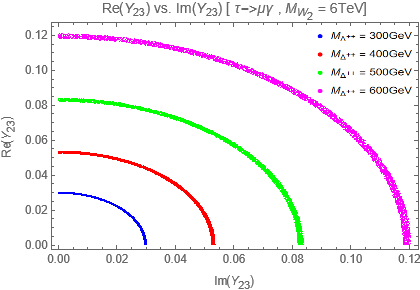}
         \caption{$Y_{23}$ coupling parameter space of $\tau \rightarrow \mu\gamma$ channel when $W_R = 6$ TeV}
         \label{taumug1}
     \end{subfigure}
     \hfill
     \begin{subfigure}[b]{0.3\textwidth}
         \centering
         \includegraphics[width=\textwidth]{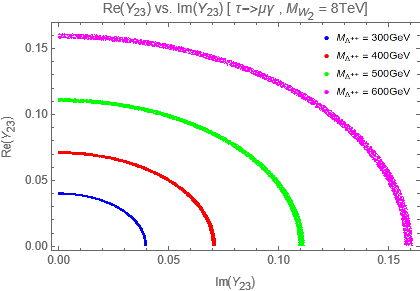}
         \caption{$Y_{23}$ coupling parameter space of $\tau \rightarrow \mu\gamma$ channel when $W_R = 8$ TeV}
         \label{taumug2}
     \end{subfigure}
     \hfill
     \begin{subfigure}[b]{0.3\textwidth}
         \centering
         \includegraphics[width=\textwidth]{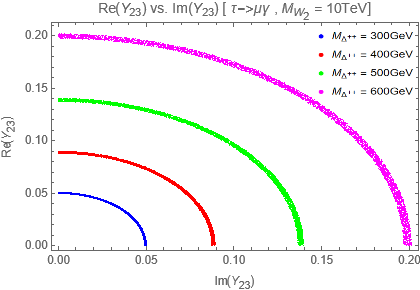}
         \caption{$Y_{23}$ coupling parameter space of $\tau \rightarrow \mu\gamma$ channel when $W_R = 10$ TeV}
         \label{taumug3}
     \end{subfigure}
     \caption{Off diagonal coupling parameter space from $l_i \rightarrow l_j\gamma$ channels}
     \label{llg}
\end{figure}

Here, we plotted $Re(Y_{ij})$ vs $Im(Y_{ij})$ for $W_R=6TeV$, $W_R=8TeV$ and $W_R=10TeV$ with doubly charged higgs masses:
\begin{itemize}
\item $300 GeV$, $400 GeV$, $500 GeV$ and $600 GeV$ for $\tau \rightarrow \mu \gamma$ process.  
\item $400 GeV$, $500 GeV$, $600 GeV$ and $700 GeV$ for $\tau \rightarrow e \gamma$ process.
\item $4 TeV$, $6 TeV$, $8 TeV$ and $10 TeV$ for $\mu \rightarrow e \gamma$ process.
\end{itemize}
The reason for choosing these specific masses is, that we focused on the phenomenologically interesting region discussed in Section \ref{Couplings vs Right handed Neutrino Mass} (off-diagonal couplings lie in the region $0 < Y_{ij} \leq 0.1$).
\newline
\newline
\textbf{Results:} Off-diagonal Couplings $Y_{12}$, $Y_{23}$ and $Y_{13}$ are calculated from the plots of the processes $\mu \rightarrow e \gamma$, $\tau \rightarrow \mu \gamma$ and $\tau \rightarrow e \gamma$ respectively and summarized in the table below -
\begin{table}[H]
        \centering
      \begin{tabular}{ | c | c | c | c |} 
      \hline
      & $W_2 = 6 TeV$ & $W_2 = 8 TeV$ & $W_2 = 10 TeV$\\
\hline 
$M_{\Delta^{++}} = 4 TeV$ & $Y_{12} = 0.01504$ & $Y_{12} = 0.02057$ & $Y_{12} = 0.02623$\\
\hline
$M_{\Delta^{++}} = 6 TeV$ & $Y_{12} = 0.03152$ & $Y_{12} = 0.04475$ & $Y_{12} = 0.05771$\\
\hline
$M_{\Delta^{++}} = 8 TeV$ & $Y_{12} = 0.04943$ & $Y_{12} = 0.07402$ & $Y_{12} = 0.09859$\\
\hline
$M_{\Delta^{++}} = 10 TeV$ & $Y_{12} = 0.06719$ & $Y_{12} = 0.10594$ & $Y_{12} = 0.14423$\\
\hline
\end{tabular}
\caption{$Y_{12}$ coupling values from $\mu \rightarrow e \gamma$ process}
\label{y12meg}
\end{table}
\begin{table}[H]
        \centering
      \begin{tabular}{ | c | c | c | c |} 
      \hline
      & $W_2 = 6 TeV$ & $W_2 = 8 TeV$ & $W_2 = 10 TeV$\\
\hline 
$M_{\Delta^{++}} = 400 GeV$ & $Y_{13} = 0.04522$ & $Y_{13} = 0.06012$ & $Y_{13} = 0.07569$\\
\hline
$M_{\Delta^{++}} = 500 GeV$ & $Y_{13} = 0.07147$ & $Y_{13} = 0.09499$ & $Y_{13} = 0.11848$\\
\hline
$M_{\Delta^{++}} = 600 GeV$ & $Y_{13} = 0.10237$ & $Y_{13} = 0.13796$ & $Y_{13} = 0.17255$\\
\hline
$M_{\Delta^{++}} = 700 GeV$ & $Y_{13} = 0.14009$ & $Y_{13} = 0.1885$ & $Y_{13} = 0.23387$\\
\hline
\end{tabular}
\caption{$Y_{13}$ coupling values from $\tau \rightarrow e \gamma$ process}
\label{y13teg}
\end{table}
\begin{table}[H]
        \centering
      \begin{tabular}{ | c | c | c | c |} 
      \hline
      & $W_2 = 6 TeV$ & $W_2 = 8 TeV$ & $W_2 = 10 TeV$\\
\hline 
$M_{\Delta^{++}} = 300 GeV$ & $Y_{23} = 0.02936$ & $Y_{23} = 0.04001$ & $Y_{23} = 0.04873$\\
\hline
$M_{\Delta^{++}} = 400 GeV$ & $Y_{23} = 0.05249$ & $Y_{23} = 0.07093$ & $Y_{23} = 0.08772$\\
\hline
$M_{\Delta^{++}} = 500 GeV$ & $Y_{23} = 0.0826$ & $Y_{23} = 0.11104$ & $Y_{23} = 0.13784$\\
\hline
$M_{\Delta^{++}} = 600 GeV$ & $Y_{23} = 0.11888$ & $Y_{23} = 0.15941$ & $Y_{23} = 0.19994$\\
\hline
\end{tabular}
\caption{$Y_{23}$ coupling values from $\tau \rightarrow \mu \gamma$ process}
\label{y23tmg}
\end{table}

\subsection{Estimation of AMM and EDM of Charged leptons}
\subsubsection{Formulas for EDM and MDM of Charged Leptons}
\label{Formulas for EDM and MDM of Charged Leptons}
In the Appendix \ref{Calculation of EDM and MDM Loop}, we systematically computed the one-loop expressions corresponding to relevant diagrams, as illustrated in Figure \ref{alldiagm}, concerning the Electric Dipole Moment (EDM) and Magnetic Dipole Moment (MDM) of charged leptons. Additionally, we have engaged in a comprehensive discussion regarding fundamental quantum field theoretic structures pertinent to the calculation (such as form factors, Gordon identities, etc.). Furthermore, we addressed Various critical issues (divergences, loop-related challenges, etc.) are addressed in detail in Appendix \ref{QFT Structure of Electric and Magnetic Dipole Moment}. Consequently, the relevant dipole expressions associated with BSM particles, serving as mediators in the loop, for our calculation are as follows with diagrams:
\newline
\newline
\textbf{For $W_2$ loop:}
\begin{align*}
\vcenter{\hbox{\includegraphics[width=40mm]{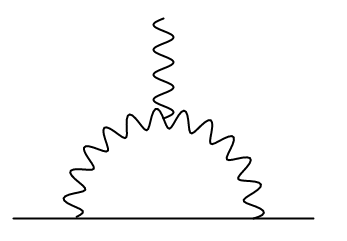}}}
\qquad
\begin{aligned}
d_f(W_2) &= \frac{1}{2} \frac{m_N Im(G^{W*}_R G^W_L)}{16\pi^2 M^2_{W_2}} \int_0^1 dx \frac{(1-x)[3(1-x)-sx^2]}{1-x+rx-sx(1-x)}; 
\\
r &=\frac{m_N^2}{M^2_{W_2}} \ \& \ s =\frac{m_l^2}{M^2_{W_2}}
\end{aligned}
\end{align*}
\textbf{For $Z_2$ loop:}
\begin{align*}
\vcenter{\hbox{\includegraphics[width=40mm]{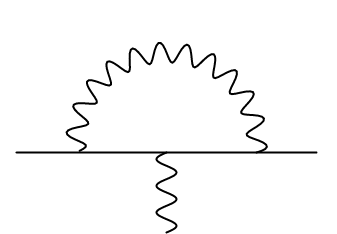}}}
\qquad
\begin{aligned}
d_f(Z_2) &= \frac{m_\tau Im(G^{W*}_R G^W_L)}{16\pi^2 M^2_{Z_2}} \int_0^1 dx \frac{x[4(1-x)+x(r-s)]}{1-x+rx-sx(1-x)}; 
\\
r &= \frac{m_\tau^2}{M^2_{Z_2}} \ \& \ s =\frac{m_l^2}{M^2_{Z_2}}
\end{aligned}
\end{align*}
\textbf{For $H^+$ loop:}
\begin{align*}
\vcenter{\hbox{\includegraphics[width=40mm]{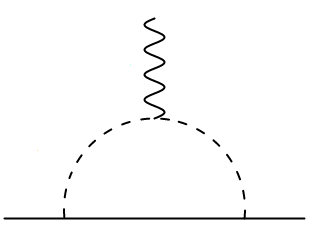}}}
\qquad
\begin{aligned}
d_f(H^+) &= -\frac{m_N Im(G^{H*}_R G^H_L)}{16\pi^2 M^2_{H^+}} \int_0^1 dx \frac{x(1-x)}{1-x+rx-sx(1-x)}; 
\\
r &=\frac{m_N^2}{M^2_{H^+}} \ \& \ s =\frac{m_l^2}{M^2_{H^+}}
\end{aligned}
\end{align*}
\textbf{For $H^0$ loop:}
\begin{align*}
\vcenter{\hbox{\includegraphics[width=40mm]{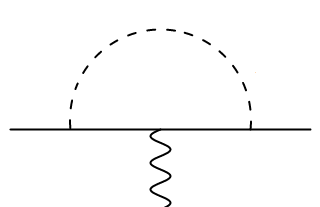}}}
\qquad
\begin{aligned}
d_f(H^0) &= -\frac{m_\tau Im(G^{H*}_R G^H_L)}{16\pi^2 M^2_{H^0}} \int_0^1 dx \frac{x^2}{1-x+rx-sx(1-x)}; 
\\
r &=\frac{m_\tau^2}{M^2_{H^0}} \ \& \ s=\frac{m_l^2}{M^2_{H^0}}
\end{aligned}
\end{align*}
\textbf{For $H^{++}$ loop:}
\begin{align*}
\vcenter{\hbox{\includegraphics[width=55mm]{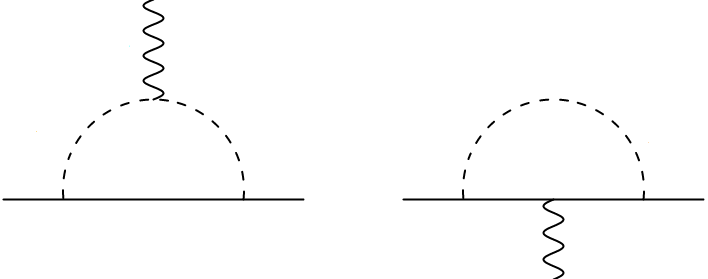}}}
\qquad
\begin{aligned}
d_f(H^{++}) &= -\frac{m_\tau Im(G^{H*}_R G^H_L)}{16\pi^2 M^2_{H^{++}}} \int_0^1 dx \frac{2x(1-x)+x^2}{1-x+rx-sx(1-x)}; 
\\
r &=\frac{m_\tau^2}{M^2_{H^{++}}} \& \ s=\frac{m_l^2}{M^2_{H^{++}}}
\end{aligned}
\end{align*}
Note that, we only mentioned the expressions for EDM. The MDM expressions can be found by
\begin{itemize}
\item replacing $d_f \rightarrow \Delta a_f$ and $Im(G_R^* G_L) \rightarrow Re(G_R^* G_L)$
\item multiplying each expressions with an extra factor $2m_l$ (where $l =$ the charged lepton).
\end{itemize}
In this calculation, we focused on dipole functions contribution within the framework of Type-II Seesaw compatible Minimal Left-Right Symmetric Model (LRSM). These dipole functions exhibit a direct dependence on a combination of couplings and are proportionally related to internal fermion masses while being inversely proportional to boson masses within the loop i.e.
$$Dipole \ function \propto (G_R^* G_L) \times \frac{m_0}{M_{Boson}^2}$$

\subsubsection{Results \& Discussion}

To test the model, we have to calculate the total contribution in AMM and EDM of charged leptons by adding all the expressions mentioned previously i.e.
$$d_f = d_f(W_2) + d_f(Z_2) + d_f(H^0) + d_f(H^+) + d_f(H^{++})$$
$$\Delta a_f = \Delta a_f (W_2) + \Delta a_f (Z_2) + \Delta a_f (H^0) + \Delta a_f (H^+) + \Delta a_f (H^{++})$$ 
using Wolfram Mathematica where we consider off-diagonal couplings values evaluated in the previous section \ref{Parameter Space of Couplings}  with corresponding mass scales of bosons. From the CLFV data set in Table \ref{y13t3e}, \ref{y23t3m}, \ref{y13teg}, \ref{y23tmg}; only the phenomenologically interesting regions discussed in Section \ref{Couplings vs Right handed Neutrino Mass} (i.e. off-diagonal coupling $> 0.1$ are excluded ) are considered in the computation of dipole functions. Hence, we see the model's predictions of EDM and AMM, summarized below:\begin{itemize}
\item \textbf{For $l_i \rightarrow 3 l_j$ Process}\begin{itemize}

\item \textbf{Electron EDM and AMM from $\tau \rightarrow 3e$ :}
\begin{itemize}
\item $Y_{11} = 0.1$, $M_{\Delta^{++}}=600 GeV$ and $Y_{13}= 0.09206$ 
$$d_e = 2166.34 TeV^{-1} \ \& \ \Delta a_e =-0.00442799$$

\item $Y_{11} = 0.01$, $M_{\Delta^{++}}=200 GeV$ and $Y_{13}= 0.10213$ 
$$d_e = 0.0000434273 TeV^{-1} \ \& \ \Delta a_e = -8.87654 \times 10^{-11}$$
\end{itemize}

\item \textbf{Muon EDM and AMM from $\tau \rightarrow 3\mu$ :}
\begin{itemize}
\item $Y_{22} = 0.1$, $M_{\Delta^{++}}=600 GeV$ and $Y_{23}= 0.10041$ 
$$d_\mu = -0.00199494 TeV^{-1} \ \& \ \Delta a_\mu = 1.11 \times 10^{-6}$$

\item $Y_{22} = 0.01$, $M_{\Delta^{++}}=200 GeV$ and $Y_{23}= 0.11154$ 
$$d_\mu = -0.00261085 TeV^{-1} \ \& \ \Delta a_\mu =1.107 \times 10^{-6}$$
\end{itemize}
\end{itemize}

\item \textbf{For $l_i \rightarrow l_j \gamma$ Process}\begin{itemize}

\item \textbf{Electron EDM and AMM from $\tau \rightarrow e \gamma$ :}
\begin{itemize}
\item $W_R = 6 TeV$, $M_{\Delta^{++}}=600 GeV$ and $Y_{13} = 0.10237$ 
$$d_e = 2.47567 \times 10^8 TeV^{-1} \ \& \ \Delta a_e = -506.026$$

\item $W_R = 8 TeV$, $M_{\Delta^{++}}=500 GeV$ and $Y_{13} = 0.09499$ 
$$d_e = 5.25942 \times 10^{-6} TeV^{-1} \ \& \ \Delta a_e = -1.07503 \times 10^{-11}$$

\item $W_R = 10 TeV$, $M_{\Delta^{++}}=400 GeV$ and $Y_{13} = 0.07569$ 
$$d_e = 536.564 TeV^{-1} \ \& \ \Delta a_e = -0.00109674$$ 
\end{itemize}

\item \textbf{Muon EDM and AMM from $\tau \rightarrow \mu \gamma$ :}
\begin{itemize}
\item $W_R = 6 TeV$, $M_{\Delta^{++}}=500 GeV$ and $Y_{23} = 0.0826$ 
$$d_\mu = -0.271753 TeV^{-1} \ \& \ \Delta a_\mu = 0.000115223$$

\item $W_R = 8 TeV$, $M_{\Delta^{++}}=400 GeV$ and $Y_{23}= 0.07093$ 
$$d_\mu = -0.189033 TeV^{-1} \ \& \ \Delta a_\mu = 0.0000801498$$

\item $W_R = 10 TeV$, $M_{\Delta^{++}}=400 GeV$ and $Y_{23} = 0.08772$ 
$$d_\mu = 1.05401 TeV^{-1} \ \& \ \Delta a_\mu = -0.000446899$$ 
\end{itemize}
\end{itemize}

\end{itemize}
Hence, we see that no dataset was identified within the confines of the current experimental values of Table \ref{exptvalue}. The computed values for Electric Dipole Moment (EDM) and Anomalous Magnetic Moment (AMM) significantly exceed the recent experimental limits. This discrepancy indicates an imperative requirement for additional Beyond Standard Model (BSM) particles and an expanded gauge structure, necessitating appropriate extensions to the Higgs sector. Such extensions would entail a non-minimal modification of the model, thereby rescuing the Left-Right Symmetric Model from exclusion. 
\newline
\newline
In summary, our comprehensive analysis of Electric Dipole Moment (EDM) and Magnetic Dipole Moment (MDM) within the framework of the Parity manifest minimal Left-Right Symmetric Model (LRSM) has provided significant insights. It is evident from the results that our minimal LRSM fails to simultaneously resolve the EDM/MDM and CLFV puzzles associated with charged leptons.

As such, it becomes apparent that additional experimental data and constraints are imperative. These constraints are needed not only within the Higgs sector but also in the W boson sector. Further experimentation and data collection are essential to refine our understanding of the couplings and interactions in the LRSM framework, ultimately advancing our pursuit of a more comprehensive and coherent explanation for the behavior of charged leptons.

\section{Conclusion and Future Perspectives}
\label{Conclusion}
Our primary objective in this study was to identify potential candidates for Physics Beyond the Standard Model by investigating phenomena such as neutrino physics, anomalous magnetic moment, CP violation, and charged lepton flavor violation. In this study, we have identified the Seesaw mechanism as a viable explanation for the smallness of neutrino masses and their mixing. Consequently, we endeavored to construct an ultraviolet (UV) theory that extends beyond the Standard Model by incorporating additional components such as right-handed neutrinos and to achieve a natural decoupling of right-handed neutrinos in the infrared (IR) limit (Standard Model) and indicated the necessity of a triplet scalar field. Hence, this extension accommodates both Seesaw type-I and type-II mechanisms, aligning perfectly with neutrino observations. Hence, the Left-Right Symmetric Model emerges as the most natural extension of the Standard Model.
\newline
\newline
Within this framework, we evaluated off-diagonal couplings from experimental bounds of CLFV channels and then conducted calculations for Electric Dipole Moments (EDM) and Magnetic Dipole Moments (MDM). Our results indicated exceeding values from recent experiment values due to mass insertions, highlighting the implication of this model. Notably, we observed that EDM, AMM, and CLFV all manifest at the same energy scale and are contingent on the same coupling parameters ($G_L$ and $G_R$).

Concerning the Higgs sector, we isolated the physical modes contributing to the Higgs loop. These contributions are dependent on the mixing of Higgs fields $\Delta_L$ and $\Delta_R$ (with no other possible chiral flipping mechanism). Through their vacuum expectation values (vev), they give rise to real couplings that influence EDM, AMM, and CLFV channels determined by the Higgs mixing coefficients, denoted as $\beta_i$ (see section \ref{Higgs Potential}). Our analysis of manifest and pseudo-manifest Left-Right Symmetric Models revealed that due to the Higgs structure and associated phenomenology ($v_L \ll k, k' \ll v_R$), the coefficients of $\beta_i$ are inherently zero. This characteristic has been extensively studied, and no symmetry has been identified that would constrain $\beta_i = 0$ within the model \cite{Senjanovic:1975rk} \cite{deshpande1991left}. We hope that the minimal Left-Right Symmetric Model can be embedded into higher ultraviolet (UV) theories such as Grand Unified Theories (GUTs) and Supersymmetry (SUSY), where symmetries exist to enforce $\beta_i = 0$. This Higgs scale phenomenology emerges from the interplay between neutrino physics and gauge boson mass observations in detectors, leading to significant suppression of the mixing which is why the model can't resolve the EDM/MDM and CLFV puzzles associated with charged leptons simultaneously. Henceforth, the Parity Manifest Minimal Left-Right Symmetric Model falls short in substantiating conclusive evidence as a Beyond Standard Model (BSM) candidate.
\newline
\newline 
To date, there has been no experimental evidence confirming the Minimal Left-Right Symmetric Model, although it has not been definitively ruled out either. Hence, there is potential for extensions or modifications to rescue the model. Recent studies have explored possible extensions based on multiplets accommodated within the Left-Right framework and the origin of CP violation, offering promising avenues for further investigation. Such possible extensions are - 
\begin{itemize}
\item \textbf{2 Scalar bi-doublet in Manifest LRSM}: This is like 2 HDM i.e. considering a second SM like Higgs doublet of standard model and couple to particles similarly. Then, demand the observed results are contributions of both these doublets.
 
In LRSM, we can extend to 2 bi-Higgs doublet models. Then couple both bi-doublets to theory like $\phi$ and calculate contributions from both fields to match the observation. In these extensions, we get more scalar particles that can be studied and given bounds to match observed CLFV.
\item \textbf{Extra Triplet Scalar in Manifest LRSM} : Considering a second triplet Scalar, exactly same as $\Delta_L = (\Delta^{++},\Delta^{+},\Delta^0)_L : (3,1,2)$ and $\Delta_R = (\Delta^{++},\Delta^{+},\Delta^0)_R : (1,3,2)$ to match the observation.
\item \textbf{Triplet Fermion in Manifest LRSM} \cite{Ma}: We can have type-III Seesaw compatible LRSM where
femionic triplet - $\Sigma_L = (\Sigma^{+},\Sigma^{0},\Sigma^-)_L : (3,1,0)$ and $\Sigma_R = (\Sigma^{+},\Sigma^{0},\Sigma^-)_R : (1,3,0)$ are introduced. In such models, scalar doublets ($\phi$) get vev which automatically generates Majorana DoF. No triplet scalar is needed. 
\item \textbf{Combination of Manifest, Pseudo Manifest, and Exact LRSM}: The Manifest LRSM provides more minimization conditions than the number of vev's leading to a fine-tuning problem. On the other hand, the Pseudo-manifest LRSM leads to a light triplet Higgs which is excluded by observation. With these results in mind, for minimal LRSM to be realistic and natural, manifest+pseudo-manifest+explicit LRSM must be taken into account at a time and then use Dark matter, AMM constraints, and embedding in Higher UV models like GUT.
\end{itemize}
Note that, we did not include the 2 Scalar doublet model, Singlet scalar and Singlet fermion($S : (1, 1, 0)$) as extensions since 2 Scalar doublet is excluded (by observed FCNC) due to large tree-level FCNC contribution, Singlet scalar is excluded as it's not seesaw compatible and lastly Singlet fermion($S : (1, 1, 0)$) contributes to Dark matter \cite{Ma}.
\newline
\newline
{\textbf{Acknowledgments}} 

Both Rafid Buksh and Samim Ul Islam gratefully acknowledge Rashik Buksh Rafsan for generously dedicating his time to assist with high-performance computing resources and demonstrating exceptional programming skills during the computational aspects of this research.
\newpage
\appendix
\section{Effective Field Theory}
\label{Effective Field Theory}
 \subsection{Introduction}

Effective field theory (EFT) represents a powerful approach in the realm of quantum physics, particularly when dealing with systems operating at different scales compared to an underlying or ``full theory". Often, to extract quantitative information from experiments, it is unnecessary to delve into the intricacies of the complete underlying theory. Instead, EFT allows us to harness the predictive power of an asymptotic limit within the full theory, thereby making calculations feasible with a different set of parameters. EFT, in itself, constitutes a fully-fledged quantum theory. It enables us to compute S-matrix elements through its experimentally measurable Lagrangian, with the added benefit of finite error in such computations. Formally, EFT results emerge as an expansion concerning a parameter denoted as $\delta$. The expansion is typically truncated at $\delta^n$, and any terms containing powers higher than $(n+1)$ of $\delta$ contribute to the error component of the theory. This expansion approach allows us to express interaction terms in terms of this parameter.

Our theory imposes certain constraints contingent on the expansion coefficients and parameters, necessitating that experimental results adhere to these conditions for meaningful comparisons. Therefore, the foundational principle behind constructing an EFT lies in situations where a quantum field theory encompasses multiple energy or length scales. In such cases, a simplified theory can be systematically developed by expanding the ratio of these scales. To illustrate this concept, let's consider an example:
\newline
\newline
The Standard Model (SM) of particle physics, while highly successful, leaves some fundamental questions unanswered, such as neutrino mixing and charged lepton flavor-violating (CLFV) processes. It is plausible that beyond the SM (BSM) physics introduces new, heavy particles with masses significantly greater than the electroweak symmetry breaking scale, denoted as $M \gg v$. While a complete Lagrangian for the high-energy UV theory remains elusive, we can construct a low-energy effective theory, known as the Standard Model Effective Field Theory (SMEFT). This low-energy limit represents a remnant of the UV theory and extends the familiar SM Lagrangian with higher-dimensional local operators comprising Standard Model fields.
\newline
\newline
Within the EFT framework, interaction terms in the Lagrangian are written as an expansion of its degrees of freedom and some couplings and a term $\Lambda$. This $\Lambda$ is called the cut-off of our theory. The reason $\Lambda$ is called cut-off will be clear in the next section. Now effective field theories are the appropriate theoretical tool to describe low energy physics, where low is defined concerning energy scale $\Lambda$. Now in this regime, we will be considering only relevant degrees of freedom. Although effective field theories contain an infinite number of terms, renormalizability is not an issue since at a given order in the energy expansion, the low-energy theory is specified by a finite number of couplings; this allows for an order by-order renormalization.
\newline
\newline
The theoretical basis of effective field theory (EFT) can be formulated as a theorem \cite{Pich}:
\newline
\newline
``For a given set of asymptotic states, perturbation theory with the most general Lagrangian containing all terms allowed by the assumed symmetries will yield the most general S-matrix elements consistent with analyticity, perturbative unitarity, cluster decomposition, and the assumed symmetries."

\subsection{Construction of Weinberg Operator}
To construct a $\mathcal{L}_{EFT}$ , we have to specify the following 3 requirements:
\begin{enumerate}
\item \textbf{Degrees of freedom (DOF)}: The first step when constructing an EFT is to figure out what are the degrees of freedom that are ``relevant" to describe the physical system one is interested in. These are the variables that will appear in the effective action. For example: In our thesis, we want to construct SMEFT so DoF will be the same as SM field contents like Higgs doublet, no $\nu_R$, etc.

The choice of degrees of freedom is an independent input, and it's a matter of art as much as science.

\item \textbf{Symmetries}: The second step in constructing an EFT consists of identifying the symmetries that constrain the form of the effective action, and therefore the dynamics of the system. Symmetries can come in many different flavors: they can be global, gauged, accidental, spontaneously broken, anomalous, approximate, contracted, etc. For our thesis for example, since we want to construct SMEFT. Therefore, underlying theory respects SM gauge group $G_{SM} = SU(3)_C \times SU(2)_L \times U(1)_Y$.

\item \textbf{Expansion Parameter}: The key to handling an action with an infinite number of terms lies in the fact that all EFTs feature one or more expansion parameters. These are small quantities controlling the impact that the physics we choose to neglect could potentially have on the degrees of freedom we choose to keep. For example, in particle physics, these expansion parameters are often ratios of energy scales $\frac{E}{\Lambda}$, where E is the characteristic
energy scale of the process one is interested in, and $\Lambda$ is the typical energy scale of the UV physics one is neglecting. Observable quantities are calculated in
perturbation theory as series in these small parameters. For this strategy to work, it is crucial to have an explicit power counting scheme, meaning that we should be able to assign a definite order in the expansion parameter to each term in the effective action. This ensures that only a finite number of terms contribute at any given order in
perturbation theory and that we can decide upfront which terms to keep in the action based on the desired level of accuracy.
\end{enumerate}
According to the above requirements, we will now construct all possible Dim 5 Effective Operators in detail (that is, we will illustrate the method and interesting physical result) :
\begin{enumerate}
\item \textbf{Ansatz} - 

First naively consider that operators may consist of only fermions or only scalars or only vectors. Since in dim 4 space-time $$[H] = 1 , \ [\psi] = \frac{3}{2}, \ [A] = 1$$
Therefore, it is not possible to construct Dim 5 operators considering only fermions or only scalars, or vectors.

\item \textbf{Ansatz} - 

Consider operators consist of 2 fermions and 2 scalars. \begin{enumerate}
\item If scalars are $H$ and $H^*$ then total hypercharge (Y) of fermions $\psi_1$  and $\psi_2$ is $0$ and we need a multiplet and its charge-conjugate but cannot make nonvanishing Lorentz scalar of dim 3 ($\psi \overline{\psi}$)
\item So, we are left with building blocks $H$, $H$, $\psi_1$  and $\psi_2$. Now, forming $SU(2)_L$ invariants : $H^T i\sigma_2 H = 0$ which implies $\psi_1$  and $\psi_2$ must be doublets so we are left with lepton ($L_L$) or quark ($Q_L$) doublet.

Hence, we form $SU(2)_L$ and $U(1)_Y$ invariant terms:$L_L^T C^{-1}i\sigma_2 H$ and $H^T i\sigma_2 L_L$. To make Lorentz scalar, we need to connect them $$ \mathcal{O}^{(5)} = L_{Li}^T C^{-1} i\sigma_2 HH^T i\sigma_2 L_{Lj} = (\overline{L^c_{Li}} \cdot \widetilde{H^*}) (\widetilde{H^\dagger} \cdot L_{Lj})$$

\end{enumerate}

\end{enumerate}

Therefore, Dim 5 Effective Lagrangian $$ \mathcal{L}_5 = c^{ij}\frac{( \overline{L^c_{Li}} \cdot \widetilde{H^*}) (\widetilde{H^\dagger} \cdot L_{Lj} )}{\Lambda} + H.c.$$
where, $L_{Li} = \begin{pmatrix}
\nu \\ e
\end{pmatrix}_{Li}$ are usual the lepton doublets, $H = \begin{pmatrix}
H^{+} \\
H^{0}
\end{pmatrix}$ is the Higgs doublet, $\widetilde{H}=i\tau_2 H^*$, $ C =i\gamma_2 \gamma_0 $ and $c^{ij}$ is the coupling strength known as Wilson coefficient. $\Lambda$ is the model-dependent energy scale suppressed by the scale of new physics. This is the only Dim 5 operator first mentioned by Weinberg \cite{weinberg1979baryon} which violates the lepton number and generates Majorana mass for $\nu_L$.

\section{Discussions on Essential QFT Structures}
\label{QFT Structure of Electric and Magnetic Dipole Moment}
\subsection{EDM \& MDM Form Factors}

The amplitude for a particle($p\rightarrow p'$) scattering from a heavy target($k \rightarrow k'$) is given by \cite{Peskin:1995ev} 
$$ iM=ie^2 (\bar{u}(p')\Gamma^{\mu}(p,p')u(p) \frac{1}{q^2} \bar{u}(k')\gamma^{\mu}u(k)  ) $$
where $e$ is the coupling constant, $q=p'-p=k-k'$ is the momentum exchange, $\Gamma^{\mu}$ is the quantum corrected vertex function. The $\Gamma^{\mu}$ function appears in the S-matrix element for scattering of an electron from an external electromagnetic field. The interacting Hamiltonian is given by 
$$ \Delta H_{int} = e \int d^3x  A_\mu^{cl}(x) j^\mu(x)  $$
Thus, with the vertex correction, the amplitude is given by 
$$ -ie \bar{u}(p')\Gamma^{\mu}(p,p')u(p) A_\mu^{cl}(p'-p) $$
where $A_\mu^{cl}(p'-p)$ is the Fourier transform of $A_\mu^{cl}(x)$. In the lowest order $\Gamma^{\mu}=\gamma^{\mu}$. In general, $\Gamma^{\mu}$ depends on $p,p',\gamma^{\mu}$. For a parity-violating theory, it can also depend on $\gamma^5$. But the form of $\Gamma^{\mu}$ is constrained. This object $\Gamma^{\mu}$ transforms like a vector under Lorentz transformation. Thus, we can write, 
$$  \Gamma^{\mu} = A \gamma^{\mu} + B (p'+p)^\mu + C (p'-p)^\mu$$

There could be dependency on $\slashed{p}$ or $\slashed{p'}$. As $\slashed{p}u(p)=mu(p)$ and $\bar{u}(p')\slashed{p'}=\bar{u}(p')m$, there is no dependency on $\slashed{p}$ or $\slashed{p'}$. Since the only non-trivial scalar is $q^2$, the coefficients must be a function of $q^2$ only. Now, due to Ward identity (i.e. Gauge invariance), $q_\mu \Gamma^\mu = 0$. Thus, $C=0$. The Gordon identity is given by 
$$ \bar{u}(p')\gamma^{\mu}(p,p')u(p) = \bar{u}(p')[\frac{p'^\mu+p^\mu}{2m}+\frac{i\sigma^{\mu \nu}q_\nu}{2m}]u(p) $$
Thus, we can write using the Gordon identity, 
$$ \Gamma^\mu(p',p)=\gamma^\mu F_1(q^2)+\frac{i\sigma^{\mu \nu}q_\nu}{2m}F_2(q^2)$$
We also need to take $\gamma^5$ terms since we are dealing with parity-violating theory. Therefore, the Gordon identity becomes 
$$ \bar{u}(p') \frac{i\sigma^{\mu \nu}q_\nu \gamma^5}{2m} u(p) = - \bar{u}(p') \frac{p'^\mu+p^\mu}{2m}{2m}u(p)  $$
Thus, we can write the $\Gamma^\mu$ function as (using this Gordon identity)
$$ \Gamma^\mu(p',p)=\gamma^\mu F_1(q^2)+\frac{i\sigma^{\mu \nu}q_\nu}{2m}F_2(q^2) + \frac{i\sigma^{\mu \nu}q_\nu \gamma^5}{2m}F_3(q^2)$$
These functions of $q^2$ are called form factors. $F_1(q^2)$ is the electromagnetic charge form factor, $F_2(q^2)$ is the magnetic dipole moment form factor and $F_3(q^2)$ is the electric dipole moment form factor. 

\subsection{Gordon Identity for EDM and MDM}

The general form of $\Gamma^\mu$ is, including the $\gamma^5$ term, 
$$ \Gamma^\mu = A\gamma^\mu + B (p'+p)^\mu + D (p'+p)^\mu \gamma^5 $$
Using $\sigma^{\mu \nu}= \frac{i}{2}[\gamma^\mu,\gamma^\nu]$ and Clifford algebra $\{\gamma^\mu,\gamma^\nu\} = 2\eta^{\mu \nu}$
\begin{enumerate}
\item \textbf{For MDM}
$$\bar{u}(p')[\frac{p'^\mu+p^\mu}{2m}+\frac{i\sigma^{\mu \nu}q_\nu}{2m}]u(p)$$
$$= \frac{1}{4m}\bar{u}(p')(2\eta^{\mu \nu}p'_{\nu}+2\eta^{\mu \nu}p_{\nu}-[\gamma^\mu,\gamma^\nu]p'_{\nu}+(\gamma^\mu+\gamma^\nu)p_{\nu})u(p)$$
$$= \frac{1}{4m}\bar{u}(p')((\{\gamma^\mu,\gamma^\nu\}-[\gamma^\mu,\gamma^\nu])p'_{\nu}+(\{\gamma^\mu,\gamma^\nu\}+[\gamma^\mu,\gamma^\nu])p_{\nu})u(p)$$
$$= \frac{1}{2m}\bar{u}(p')(\slashed{p'}\gamma^\mu+\gamma^\mu\slashed{p})u(p)$$
$$= \bar{u}(p')\gamma^{\mu}u(p)$$
Here, $ \slashed{p}u(p)=mu(p)$ and $\bar{u}(p')\slashed{p'}=\bar{u}(p')m $

\item \textbf{For EDM}
$$\bar{u}(p') \frac{i\sigma^{\mu \nu}q _\nu \gamma^5}{2m} u(p)$$ 
$$= \frac{i}{2m} \frac{i}{2} \bar{u}(p')(\gamma^\mu \gamma^\nu-\gamma^\nu \gamma^\mu)(p'-p)_\nu \gamma^5 u(p)$$
$$= \frac{i}{2m} \frac{i}{2} \bar{u}(p')[(2\eta^{\mu \nu}-2\gamma^\nu\gamma^\mu)p'_\nu -(2\gamma^\mu\gamma^\nu-2\eta^{\mu \nu})p_\nu] \gamma^5 u(p)$$
$$= \frac{i}{2m} \frac{i}{2} \bar{u}(p')2[p'^\mu - \slashed{p'}\gamma^\mu - \gamma^\mu \slashed{p}+p^\mu] \gamma^5 u(p)$$
$$= - \bar{u}(p') \frac{p'^\mu+p^\mu}{2m} \gamma^5 u(p)$$
Here, $\{\gamma^5,\gamma^\mu \} = 0$ , $ \slashed{p}u(p)=mu(p)$ and $\bar{u}(p')\slashed{p'}=\bar{u}(p')m$

\end{enumerate}

\subsection{Non-relativistic Limit of EDM \& MDM Form Factor}

The Lagrangian for a fermionic particle, the Dirac Lagrangian is given by,
$$ \mathcal{L} = \overline{\psi}(i\slashed{\partial} -m)\psi $$
where $\overline{\psi} = \psi^{\dagger}\gamma^{0}$ is the conjugate field. The $\gamma^{\mu}$ matrices in 4-dim denotes a collection of four $4 \times 4$ matrices satisfying the conditions. 
\begin{center}
$\{\gamma^{\mu},\gamma^{\nu}\} = 2\eta^{\mu\nu}$
\end{center}
\begin{center}
$\gamma^{0}\gamma^{\mu}\gamma^{0} = (\gamma^{\mu})^{\dagger}$
\end{center}
The first condition implies energy-momentum conservation and the second condition implies hermitian Hamiltonian. These are the defining relations for gamma matrices. Gamma matrices are the solutions to these equations. Using this we get the definition $\sigma^{\mu\nu} = \frac{i}{2}[\gamma^{\mu},\gamma^{nu}]$ where,
$$\sigma^{0i} = i
\begin{bmatrix}
\sigma^{i} & 0\\
0 & -\sigma^{i}
\end{bmatrix}
, \sigma^{ij} = - \epsilon^{ijk} \begin{bmatrix}
\sigma^{k} & 0 \\
0 & \sigma^{k}
\end{bmatrix}$$
These matrices correspond to the generators of boost and rotations for fermions(spin $\frac{1}{2}$ particles). From
Dirac theory the explicit form of the 4 spinors are
\begin{center}
$u(p) =  \begin{bmatrix}
\frac{\sqrt{E+m}}{2m}\phi(0) \\ \frac{\bm{\sigma}.\bm{p}}{2m(E+m)}\phi(0)
\end{bmatrix}\approx
\begin{bmatrix}
\phi(0) \\ \frac{\bm{\sigma}.\bm{p}}{2m}\phi(0)
\end{bmatrix}$ (in non-relativistic limit $E \approx m$)
\end{center}
where, $\phi(0) = \begin{bmatrix}
1 \\ 0
\end{bmatrix}$ for spin $\frac{1}{2}$ particles and $\begin{bmatrix}
0 \\ 1
\end{bmatrix}$ for spin $-\frac{1}{2}$ particles. From Maxwell's electrodynamics, the magnetic field is given by $\vec{B} = \vec{\nabla}\times\vec{A}$ where $A^{\mu} = (V,\vec{A})$ is relativistic 4-vector potential and considering the convention, the metric tensor is $\eta^{\mu\nu} = diag(1,-1,-1,-1)$
\begin{enumerate}
\item \textbf{For MDM}

We will consider the Dirac basis, but the results of calculations don't depend on the choice of basis. It will result in the same on a Weyl or Majorana basis. Depending on the problem, we select a basis that makes the calculation simple. Gamma matrices in the Dirac basis are given by
\begin{center}
$ \gamma^{0} = 
\begin{bmatrix}
\mathds{1} & 0\\
0 & -\mathds{1}
\end{bmatrix}
, \gamma^{i} = \begin{bmatrix}
0 & \sigma^{i} \\
-\sigma^{i} & 0
\end{bmatrix}$
\end{center}
For magnetic dipole moment, we compute the amplitude for elastic Coulomb scattering of a non-relativistic electron from a region of nonzero electrostatic vector potential i.e. setting ${A_\mu}^{cl}(x) = (0, A^{cl}(x))$. Then the amplitude for scattering from this field is 
$$i\mathcal{M} = -ie\overline{u}({p}')[\gamma^{i} F_1(0)+\frac{i\sigma^{i\nu}q_\nu}{2m}F_2(0)]u(p){A_i}^{cl}(q) $$
$$ = -ieu^{\dagger}({p}')[\gamma^{0}\gamma^{i}A_i F_1(0)+\frac{i}{2m}\gamma^{0}\sigma^{ij}q_j A_i F_2(0)]u(p)$$
\begin{center}
$= -ieu^{\dagger}({p}')
\begin{bmatrix}
\mathds{1} & 0\\
0 & -\mathds{1}
\end{bmatrix}
\begin{bmatrix}
0 & \sigma^{i} \\
-\sigma^{i} & 0
\end{bmatrix}
u(p)A_i F_1(0)-ie\frac{i}{2m}u^{\dagger}({p}')\begin{bmatrix}
\mathds{1} & 0\\
0 & -\mathds{1}
\end{bmatrix} 
\epsilon^{ijk}
\begin{bmatrix}
-\sigma^{k} & 0 \\
0 & -\sigma^{k}
\end{bmatrix}u(p)q_j A_i F_2(0)$
\end{center} 
\begin{center}
$= -ie
\begin{bmatrix}
{\phi^{\dagger}}'(0) & \frac{\bm{\sigma}.{\bm{p}}'}{2m}{\phi^{\dagger}}'(0)
\end{bmatrix}
\begin{bmatrix}
0 & \sigma^{i} \\
\sigma^{i} & 0
\end{bmatrix}
\begin{bmatrix}
\phi(0) \\ \frac{\bm{\sigma}.\bm{p}}{2m}\phi(0)
\end{bmatrix}
A_i F_1(0)$
\end{center}
\begin{center}
$-ie\frac{(-1)}{2m}
\begin{bmatrix}
{\phi^{\dagger}}'(0) & \frac{\bm{\sigma}.{\bm{p}}'}{2m}{\phi^{\dagger}}'(0)
\end{bmatrix} 
\begin{bmatrix}
\sigma^{k} & 0 \\
0 & -\sigma^{k}
\end{bmatrix}
\begin{bmatrix}
\phi(0) \\ \frac{\bm{\sigma}.\bm{p}}{2m}\phi(0)
\end{bmatrix}(-i\epsilon^{ijk}q_j A_i) F_2(0)$
\end{center}
\begin{center}
$= -ie
\begin{bmatrix}
\frac{\bm{\sigma}.{\bm{p}}'}{2m}{\phi^{\dagger}}'(0)\sigma^{i} & {\phi^{\dagger}}'(0)\sigma^{i}
\end{bmatrix}
\begin{bmatrix}
\phi(0) \\ \frac{\bm{\sigma}.\bm{p}}{2m}\phi(0)
\end{bmatrix}
A_i F_1(0)$
\end{center}
\begin{center}
$-ie\frac{(-1)}{2m}
\begin{bmatrix}
{\phi^{\dagger}}'(0)\sigma^{k} & -\frac{\bm{\sigma}.\bm{p}}{2m}{\phi^{\dagger}}'(0)\sigma^{k}
\end{bmatrix} 
\begin{bmatrix}
\phi(0) \\ \frac{\bm{\sigma}.\bm{p}}{2m}\phi(0)
\end{bmatrix}B_k F_2(0)$
\end{center}
$$= -i\frac{e}{2m}{\phi^{\dagger}}'(0)[(\bm{\sigma}.{\bm{p}}'\sigma^{i}+\sigma^{i}\bm{\sigma}.\bm{p})A_i F_1(0)-(\sigma^{k}-\frac{(\bm{\sigma}.{\bm{p}}')(\bm{\sigma}.\bm{p})}{4m^{2}})B_k F_2(0)]\phi(0)$$
$$ =-i\frac{e}{2m}{\phi^{\dagger}}'(0)[[(\delta_{ji}+i\epsilon_{jik}\sigma^{k}){p^{j}}'+(\delta_{ij}+i\epsilon_{ijk}\sigma^{k})p^{j}]A_i F_1(0)-(\sigma^{k}-\frac{(\bm{\sigma}.{\bm{p}}')(\bm{\sigma}.\bm{p})}{4m^{2}}\sigma^{k})B_k F_2(0)]\phi(0) $$
$$= -i\frac{e}{2m}{\phi^{\dagger}}'(0)[[(p^{j}+{p^{j}}')+i\epsilon_{ijk}\sigma^{k}(p^{j}-{p^{j}}')]A_i F_1(0)-(\sigma^{k}-\frac{(\bm{\sigma}.{\bm{p}}')(\bm{\sigma}.\bm{p})}{4m^{2}}\sigma^{k})B_k F_2(0)]\phi(0)$$
$$= -i\frac{e}{2m}{\phi^{\dagger}}'(0)[-\sigma^{k}(-i\epsilon^{ijk}q_j A_i) F_1(0)-\sigma^{k}B_k F_2(0)]\phi(0)$$
$$= -i\frac{e}{2m}{\phi^{\dagger}}'(0)[-\sigma^{k}B_k F_1(0)-\sigma^{k}B_k F_2(0)]\phi(0)$$
$$= -i(-[F_1 (0)+F_2 (0)]\frac{e}{m}\frac{\langle\sigma^{k}\rangle}{2} B_k)$$
Here,\begin{itemize}
\item $(p^{j}+{p^{j}}')$ term is neglected since it's spin independent term. 
\item $\frac{(\bm{\sigma}.{\bm{p}}')(\bm{\sigma}.\bm{p})}{4m^{2}}\sigma^{k}$ term is neglected. 
\end{itemize}
Now, we can interpret $\mathcal{M}$ as a Born approximation to the scattering from a potential well. The potential is just that of a magnetic moment interaction,
$$V(x)= -\vec{\mu}\cdot\vec{B}$$
where,
$$\vec{\mu} = -[F_1 (0)+F_2 (0)]\frac{e}{m}\frac{\langle\sigma^{k}\rangle}{2}$$
is magnetic moment can be rewritten in the standard form
$$\vec{\mu} = g \frac{e}{2m}\vec{S}$$
where, $g$ is Land\'e factor. Thus, 
$$g = 2[F_1 (0)+F_2 (0)] = 2+2F_2 (0)$$ 
$$=> a = \frac{g-2}{2} = F_2 (0)$$
Here, $a$ is a dimensionless parameter called Anomalous Magnetic Moment.
\item \textbf{For EDM}

We will consider the Weyl basis, but the results of calculations don't depend on the choice of basis. It will result in the same on a Dirac or Majorana basis. Depending on the problem, we select a basis that makes the calculation simple. As we are dealing with chirality explicitly, the Weyl basis is the better basis system. Gamma matrices in the Weyl basis are given by 
$$ \gamma^0 = \begin{bmatrix}
0 & \mathds{1} \\
\mathds{1} & 0
\end{bmatrix} , \gamma^i =  \begin{bmatrix}
0 & \sigma^i \\
-\sigma^i & 0
\end{bmatrix} , \gamma^5 =  \begin{bmatrix}
-\mathds{1} & 0 \\
0 & \mathds{1}
\end{bmatrix} $$

Now, consider only the electric dipole moment part in the amplitude 
$$ i\mathcal{M} = -ie \bar{u(p')} \frac{i\sigma^{\mu \nu}q_\nu \gamma^5}{2m} F_3(q^2) u(p) A_\mu(p'-p) $$
$$ = \frac{1}{2m} \bar{u(p')} [\sigma^{0 \nu}q_\nu \gamma^5 A_0(p'-p) + \sigma^{i \nu}q_\nu \gamma^5 A_i(p'-p)] F_3(0) u(p)  $$ 

$$ = \frac{e}{2m} \bar{u(p')} [\sigma^{0 i}q_i \gamma^5 A_0 + \sigma^{i 0}q_0 \gamma^5 A_i] F_3(0) u(p)  $$

$$ = \frac{e}{2m} \bar{u(p')} \sigma^{0 i} \gamma^5 [q_i A_0 - q_0 A_i] F_3(q^2) u(p)  $$

$$ = \frac{e}{2m} u^\dagger(p') \gamma^0 \sigma^{0 i} \gamma^5  u(p) [q_i A_0 - q_0 A_i] F_3(q^2)  $$

$$ = -i \frac{e}{2m} u^\dagger(p')   \begin{bmatrix}
0 & \sigma^i \\
\sigma^i & 0
\end{bmatrix}  u(p) [q_i A_0 - q_0 A_i] F_3(0)   $$

$$ = -i \frac{e}{2m} \begin{bmatrix}
\phi'^\dagger(0) & \frac{\vec{\sigma}\cdot \vec{p'}}{2m}\phi'^\dagger(0) 
\end{bmatrix}    \begin{bmatrix}
0 & \sigma^i \\
\sigma^i & 0
\end{bmatrix}  \begin{bmatrix}
\phi(0) \\
\frac{\vec{\sigma}\cdot \vec{p}}{2m}\phi(0) 
\end{bmatrix} [q_i A_0 - q_0 A_i] F_3(0)$$

$$= -i \frac{e}{2m} \frac{m}{2m} [\phi'^\dagger(0) \sigma^i \phi(0)+\phi'^\dagger(0) \sigma^i \phi(0)][q_i A_0 - q_0 A_i] F_3(0)$$

$$= -i \frac{1}{2m} \cdot 2 \langle \frac{\sigma^i}{2} \rangle[q_i A_0 - q_0 A_i] F_3(0)$$

$$ -i \frac{e}{2m}\langle\sigma^i \rangle (q_i A_0 - q_0 A_i) F_3(0) $$

$$= -i \frac{e}{2m}\langle\sigma^i \rangle (q_i V + q_0 A_i) F_3(0) $$

$$= -i \frac{e}{2m}\langle\sigma^i \rangle (i E_i) F_3(0) $$
Here, the electric field from Maxwell's theory,
$$\vec{E}(x) = -\partial_i V - \partial_0 A_i => i \vec{E}(x) = -i(\partial_i V + \partial_0 A_i)$$
Now, we go to momentum space by Fourier transformation ($-i\partial_\mu \rightarrow q_\mu$) where $q_\mu=p'_\mu-p_\mu$ is the momentum exchange and get, $ i\vec{E}(q) = q_i V +q_0 A_i $. Thus, the final expression becomes 
$$- [-F_3(0) \frac{e}{m} \frac{\langle \sigma^k \rangle}{2}]E_k $$
where,$$d_f = -\frac{F_3(0)}{2m} = -F_{EDM}(0)$$ is the Electric dipole moment leading to the non-relativistic  electric dipole Hamiltonian 
$$\mathcal{H}_{EDM}= -e d_f\langle \vec{\sigma} \rangle \cdot\vec{E}$$
The Dipole operator is not symmetric in the sense that it is described by a vector in the classical non-relativistic limit. It has a preferred direction, unlike the monopole term which has no preferred direction. That's the reason we get an Electric field along a specific direction, specified by spin vector expectation, defining this term as dipole and so the $d_f$ is the electric dipole moment (with direct analogy with classical limit), in our case, of the fundamental particle. Monopole term in Hamiltonian($qV$) is spin vector independent and the strength is called the electric charge of the fundamental particle.

\end{enumerate}

\subsection{Possible 1-Loop Diagrams for EDM \& MDM}

\begin{figure}[H]
\begin{center}
\includegraphics[width=100mm]{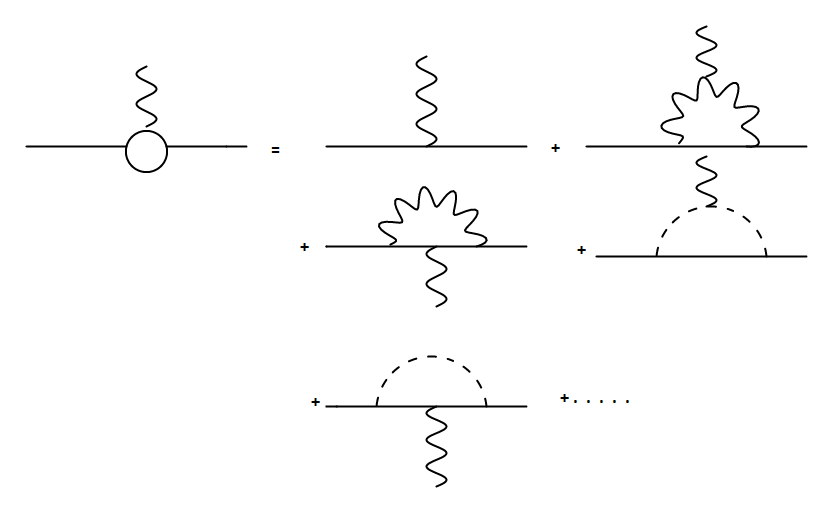}
\end{center}
\caption{All possible vertex.}
\label{possiblevertex}
\end{figure}

Fermion interaction with Gauge Boson is given by
$$\mathcal{L}^{UV}_{int} = \overline{\psi} \gamma^{\mu} \psi W^a_\mu T_a $$
At the IR limit, this interacting Lagrangian becomes
\begin{center}
$\mathcal{L}^{IR}_{int}= \overline{\psi} \gamma^{\mu} \psi\overline{\psi} \gamma_{\mu} \psi$
\end{center}
So, allowed vertex of this theory are $\gamma^{\mu}$ and $\gamma^{\mu}\gamma_5 $. With constructing effective vertex $\Gamma^{\mu}$ and using Gordon identity, we get vertices $\gamma^{\mu}$ occurring at tree level and $\sigma^{\mu\nu}q_\nu$ and $\sigma^{\mu\nu}q_\nu \gamma_5$ occurring at loop level, shown in Figure \ref{possiblevertex}
\graphicspath{{./possible vertex/}}
As we are only focusing on EDM and MDM, we need to calculate loop corrections of the relevant diagrams. The diagrams that contribute to the EDM and MDM form factor are the relevant diagrams. The first quantum correction is the 1-loop diagram contributions. Both EDM and MDM are electromagnetic properties. So, to calculate these, we need to couple fermions to photons. Fermion current couples to photons and other gauge bosons. Thus, relevant diagrams can be calculated using Feynman rules.

\begin{figure}[H]
\begin{center}
\includegraphics[width=80mm]{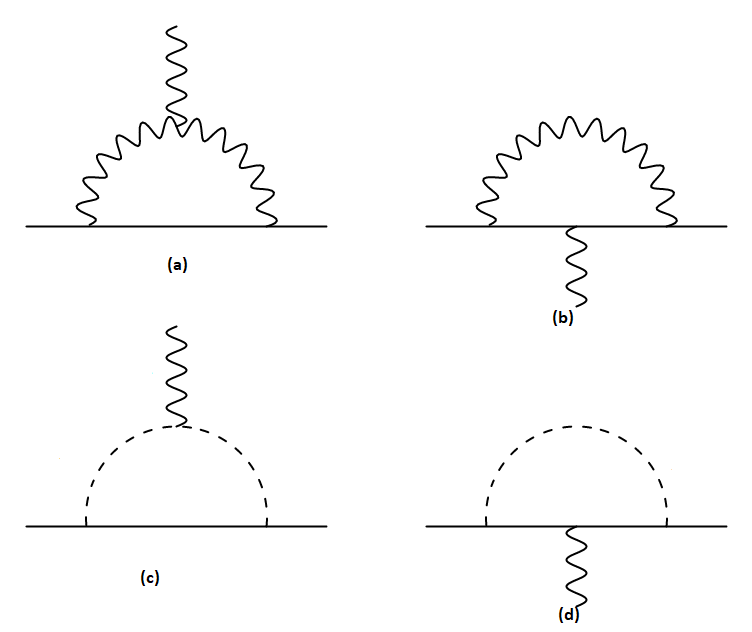}
\end{center}
\caption{Relevant Loop Diagrams for EDM and MDM Contribution}
\label{alldiagm}
\end{figure}

In LRSM, fermions couple to both $SU(2)_L$ and $SU(2)_R$ gauge bosons and also Higgs scalar fields. At one loop level, EDM and MDM of charged leptons receive contributions from all four diagrams of Figure \ref{alldiagm} for any gauge fixing (unitary, $R_\zeta$, t-Hooft, etc.). 

\begin{figure}[H]
\begin{center}
\includegraphics[width=80mm]{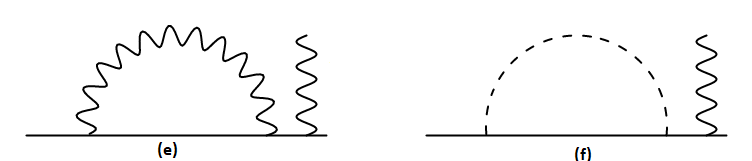}
\end{center}
\caption{W-boson and Higgs loops contributing to EM charge form factors (photon coupled to external fermion)}
\label{extleg}
\end{figure}

But diagrams where photon couple to internal fermions is irrelevant for CLFV because, in the case of charged leptons, the internal fermions are the corresponding neutrinos(as electron and neutrinos together form the $SU(2)$ doublets). The neutrinos have no electromagnetic charge i.e. neutrinos are EM charge neutral and so can not couple to photon. Thus, we have three relevant diagrams for CLFV. One is the charged W-bosons, couple to photon, the relevant diagram is  $(a)$ in Figure \ref{alldiagm}, and the other two are the physically charged scalar loops where singly charged scalars couple to photon and fermion couple to photon Figure \ref{alldiagm} $(c)$. But there is also a doubly charged scalar. Thus, for the loop contribution of doubly charged scalars, the internal fermion is also charged lepton and photon this time coupled to internal fermion Figure \ref{alldiagm}$(d)$.
\graphicspath{ {./alldiagm/} }
The photon could be coupled to the external fermion leg, still giving the three-point function like the above diagrams. These are given in diagrams (e) and (f) in Figure \ref{extleg}. These are not relevant because these diagrams don't contribute to the MDM and EDM form factor. These give a contribution to charge form factor $F_1(q^2)$. 
\graphicspath{ {./extleg/} }
There is one more possibility and it is half W-loop and half scalar loop. We will show in the calculation that this diagram has no contribution to the MDM and EDM form factor.  
\graphicspath{ {./half/} }
\begin{figure}[H]
\begin{center}
\includegraphics[width=40mm]{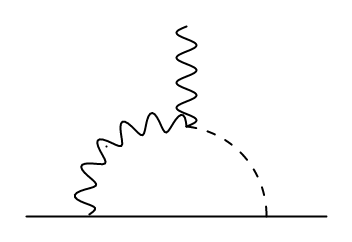}
\end{center}
\caption{W-boson and Higgs's loops contributing to EM charge form factors (Photon-W boson-Higgs scalar coupling)}
\label{half}
\end{figure}

\subsection{Feynman Rules}

For Feynman Rules, we will follow reference \cite{Nieves:1986uk} in our convention. The one-loop diagram of weak gauge bosons that contribute to electromagnetic form factors is shown in Figure \ref{Wloop}. If we choose to calculate form factors in the general $\zeta$ gauge, then additional diagrams appear where one or two W boson lines are replaced by nonphysical Higgs or Goldstone modes which are absorbed by W in the unitary gauge. We argued that the unitary gauge is the most economical gauge choice for this calculation. Then these extra diagrams are not needed for W-loops. 
\newline
\newline
As we want to study the extended Higgs EDM and MDM loop contributions, we need to calculate these diagrams using the physical Higgs. These are physical particles and can not to eliminated using any gauge choice. 

We can directly generalize the QED propagators and interaction vertex to nonabelian theory (for W bosons). As the bosons are nonabelian, they are electromagnetically charged and so they couple to a photon of QED. For fermion coupling to W bosons and Higgs H, we consider the interaction Lagrangian 
$$\mathcal{L}_{int}^W = \bar{\psi} \gamma_\mu [G_L^W (\frac{1-\gamma_5}{2})+G_R^W (\frac{1+\gamma_5}{2})]\psi W^\mu  $$
$$\mathcal{L}_{int}^H = \bar{\psi} [G_L^H (\frac{1-\gamma_5}{2})+G_R^H (\frac{1+\gamma_5}{2})]\psi H $$
where $G_{L,R}^W$ and $G_{L,R}^H$ are matrices in flavor space of all fermions. These $W^\mu$ and $H$ fields are mass eigenstates and so they are hermitian. From these interaction terms, we can calculate the couplings of bosons with fermions. Then using this, we can look at the scattering process $f^+ f^- \rightarrow W^+ W^-$, where $f$'s are charged leptons, we can extract the $W-W-\gamma$ coupling. Then consider the process $W_i + \nu_A^M \rightarrow \gamma +l_a$, where neutrino $\nu_A^M$ is rotated to the mass eigenstate as it is on the external leg, we can extract out the $W-H-\gamma$ coupling. Summarizing all these Feynman Rules, the results are given below.
\begin{figure}[H]
     \centering
     \begin{subfigure}[b]{0.4\textwidth}
         \centering
         \includegraphics[width= \textwidth]{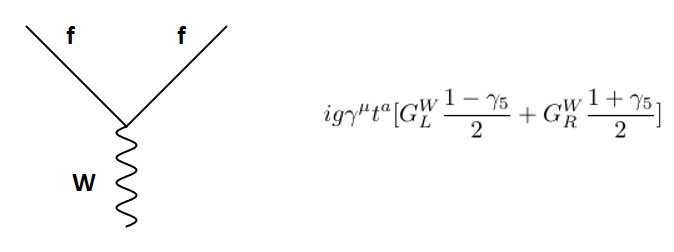}
         \caption{fermion-fermion-W vertex }
         \label{feyn1}
     \end{subfigure}
     \hfill
     \begin{subfigure}[b]{0.4\textwidth}
         \centering
         \includegraphics[width= \textwidth]{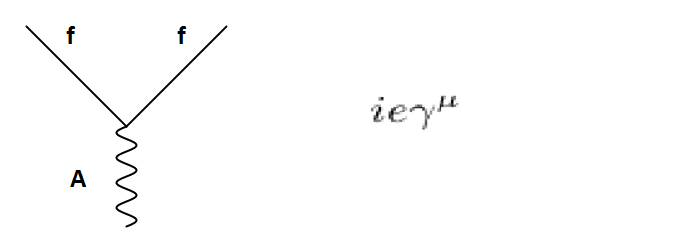}
         \caption{fermion-fermion-$\gamma$ vetex }
         \label{feyn2}
     \end{subfigure}
     \vfill
     \begin{subfigure}[b]{0.4\textwidth}
         \centering
         \includegraphics[width= \textwidth]{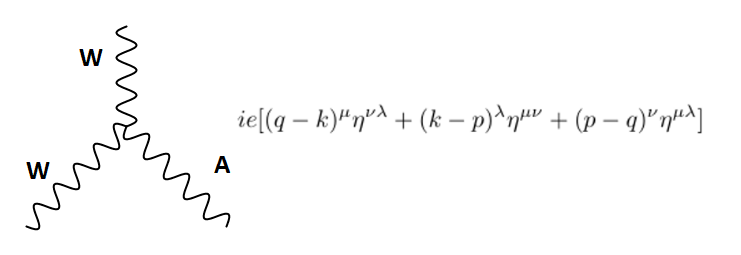}
         \caption{W-W-$\gamma$ vertex }
         \label{feyn4}
     \end{subfigure}
     \hfill
     \begin{subfigure}[b]{0.4\textwidth}
         \centering
         \includegraphics[width= \textwidth]{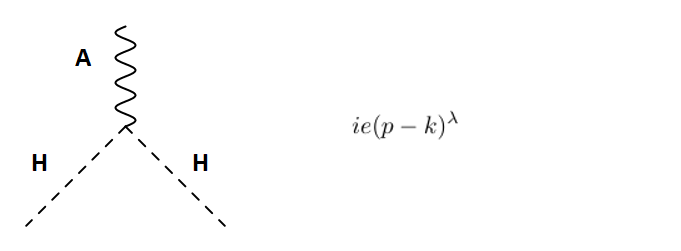}
         \caption{H-H-$\gamma$ vertex }
         \label{feyn5}
     \end{subfigure}
     \vfill
     \begin{subfigure}[b]{0.4\textwidth}
         \centering
         \includegraphics[width= \textwidth]{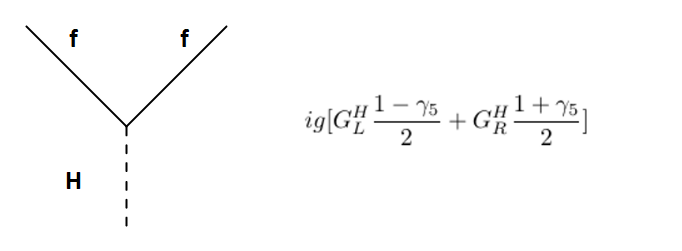}
         \caption{fermion-fermion-H vertex }
         \label{feyn3}
     \end{subfigure}
     \hfill
     \begin{subfigure}[b]{0.4\textwidth}
         \centering
         \includegraphics[width= \textwidth]{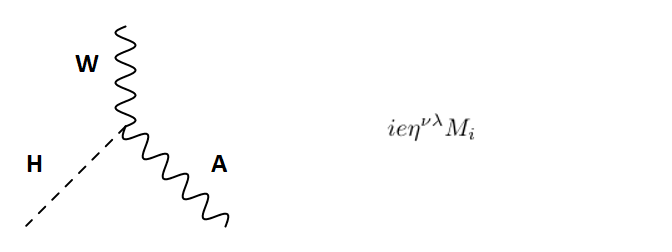}
         \caption{W-H-$\gamma$ vertex }
         \label{feyn6}
     \end{subfigure}
     
     \caption{Essential Feynman Rules.}
     \label{feyn}
\end{figure}

\subsection{Divergences}

Ultraviolet (UV) and Infrared divergences (IR) of any diagram are the divergence of the diagram contribution occurring at high and low momenta respectively. These divergences are not physical. These divergences are absorbed in the bare parameters and quantum corrections in such a way that the physical or observable quantities are finite. Whatever happens in the theory, eventually the scattering amplitude is finite because it is an observable quantity. 

To handle this kind of divergence, usually, we regulate the theory with a Pauli-Villars regulator or Dimensional regulator \cite{Peskin:1995ev}. Then, we define physics at reference energy or momenta i.e. observables are finite, and then determine physics at any momentum concerning that reference point. This is the Renormalization theory. By renormalization, we make the physics finite. But the defining objects are infinite i.e. diverging, but it is not a problem as it is not observable. 

To remove IR divergence, we can add a vertex diagram and external leg bremsstrahlungs and have a finite theory \cite{Peskin:1995ev}. There is a trick to solve it easily. IR divergence occurs due to off-shell bosons having zero mass. The trick is to add a very small mass to the off-shell boson, get the theory finite, and then take the mass limit to zero. Assuming a small mass for a massless boson, for instance, a photon which is not nonphysical since this small mass is way below the detector and so it is also unobservable. 
\newline
\newline
For our relevant diagrams in Figure \ref{alldiagm}, the off-shell bosons ($W, H$) are massive. So, there is no issue with IR divergence. From power counting of momentum, we can find UV divergence. Loop momentum runs from $-\infty$ to $\infty$. So, if the number of momentum (loop) power is greater in the numerator, the corresponding diagram may be UV divergent. But it is not necessarily divergent because symmetry like gauge invariance sometimes reduces the order of divergence.

In our case (Figure \ref{alldiagm}), we can calculate the degree of divergence. For a fermionic propagator in the loop, the momentum dependency for a large momentum follows as $\frac{1}{k}$ for $k$ loop momentum. For a nonabelian boson the propagator depends on momentum like $ \frac{-i(\eta_{\mu \nu}-\frac{k_\mu k_\nu}{M^2})}{k^2-M^2}$. For a scalar propagator, the momentum dependency is $  \frac{-i}{k^2-M^2}$. There is $\int d^4 k$ for one loop momentum integral. 

From power counting (Appendix \ref{Effective Field Theory}), scalar loops have no divergence. The first term in the nonabelian gauge boson propagator is finite i.e. divergence-free. But the second term, the nonabelian part is linear divergent. But we will see the calculation in Section \ref{Calculation of EDM and MDM Loop} that this cancels perfectly giving us a finite expression.

\subsection{Unitary Gauge vs Feynman -'t Hooft Gauge}

In gauge theories, each physical configuration of the system has an equivalent class of local field configuration. Any two configurations in the equivalent class give the same physics and are related by a gauge transformation. That's why we demand in gauge theories that under gauge transformations, the theory i.e. Lagrangian must be invariant.  Now, fixing a gauge choice is like taking one particular configuration from every equivalent class. Different gauge fixing are used for different purposes. But they all are equivalent i.e. same physical description by definition. 
\newline
\newline
A unitary gauge is a particular choice of gauge fixing in a gauge theory with SSB (Spontaneous Symmetry Breaking). In an SSB theory, the broken generators give Goldstone modes in the theory. These Goldstone modes are absorbed in the broken gauge boson direction giving those gauge bosons mass. Thus, on this basis, the Goldstone modes components are zero. It is sufficient to integrate the gauge boson loops and physical Higgs modes. There is no need to determine the contribution of Goldstone directions separately. There are three polarizations for each gauge boson instead of two. This extra degree of freedom comes from the absorbed Goldstone modes. In electroweak theory, the degrees of freedom in a unitarity gauge are the massive spin $1$ weak bosons with three polarizations each, the photon with two polarizations, and the scalar Higgs boson.
\newline
\newline
On the other hand, in $R_\zeta$ gauge, instead of fixing the gauge by constraining the gauge field, a gauge breaking term is added to the physical (gauge invariant) Lagrangian,
$$ \delta \mathcal{L} = - \frac{( \partial_\mu A^\mu )^2}{2 \zeta} $$
Fixing a gauge means fixing a number for $\zeta$. The Feynman -'t Hooft gauge is the one for which $\zeta = 1$. This gauge makes calculations simple. In this gauge, broken bosons don't absorb those Goldstone modes that correspond to each broken generator. Thus, in the W-loop contributions, we need to integrate the Goldstone modes also. This is a simpler gauge as gauge boson propagators are exactly similar to abelian theory (no quadratic momentum dependent term) and so there is no UV divergence. Thus, the calculation is simpler but we also need to add Goldstone contributions to the loop which being scalar, gives a simpler propagator. 
\newline
\newline
But as we said, in the explicit calculation, we will find that this divergence in the quadratic momentum-dependent term ($\frac{p^\mu p^\nu}{M^2}$) is canceled by symmetry. That's why we will do the calculation in the unitary gauge. If there were divergence, but the abelian part ($\eta^{\mu \nu}$) is finite, then 't Hooft gauge is the best option. 

In Section \ref{Calculation of EDM and MDM Loop}, we will use the unitary gauge to calculate the W-loop diagram (a) and only physical scalar loop diagram (c) and (d) from Figure \ref{alldiagm}. We don't need to include the Goldstone mode's ( as it is absorbed by weak bosons) loop contribution due to the unitary gauge.

\subsection{Feynman Parametrization}

The expression for the most general form of 1-loop contribution has denominators the products of the denominators of the propagators of the particle in the loop. Feynman parametrization \cite{Peskin:1995ev}, \cite{Srednicki:2007qs} is a trick to combine these denominators. To prove the general structure we will use the following identity: 
$$ \frac{1}{ab} = \int_0^1 \frac{dx}{[ax+b(1-x)]^2} $$
With the Delta function property, we can write this as 
$$ \frac{1}{ab} = \int_0^1 \int_0^1 dx dy \delta(x+y-1) \frac{1}{[ax+by]^2} $$
Taking the derivative with respect to a and b, we get 
$$  \frac{1}{a^n b^m} = \frac{\Gamma(m+n)}{\Gamma(m)\Gamma(n)} \int_0^1 \int_0^1 dx dy \delta(x+y-1) \frac{x^{n-1}y^{m-1}}{[ax+by]^{m+n}} $$
Using the induction (with $n,m,\cdots = 1$) we get, 
$$ \frac{1}{a_1a_2 \cdots a_n} = \Gamma(n) \int_0^1 dx_1 \int_0^1 dx_2 \cdots \int_0^1 dx_n \delta(x_1 +x_2 +\cdots x_n) \frac{1}{[a_1x_1+a_2x_2+\cdots a_n x_n]^n} $$
For our case, we need to combine three denominators. Thus taking $n=3$, we get=
$$ \frac{1}{a_1a_2 a_3} = \Gamma(3) \int_0^1 dx_1 \int_0^1 dx_2 \int_0^1 dx_3 \delta(x_1 +x_2 + x_3) \frac{1}{[a_1x_1+a_2x_2+a_3 x_3]^3} $$
There is an advantage of this trick. By plugging in the explicit denominator forms, we can arrange the denominator term $ a_1x_1+a_2x_2+a_3 x_3 $ as a general quadratic expression from where by shifting momentum variable we can remove the linear dependency in the denominator. Shift the momentum is valid since then $d^4k$ is unchanged. Thus, momentum redefinition makes the integral simpler because linear terms in the numerators can be ignored due to spherically symmetric integration with respect to momentum. These parameters $x,y,\cdots$ that make the loop integral simpler (now easy to evaluate) are called Feynman parameters. We will do this for the $n=3$ case in our calculation explicitly.

\subsection{Why chiral flipping is a must for EDM and MDM operators?}

\begin{figure}[H]
\begin{center}
\includegraphics[width=100mm]{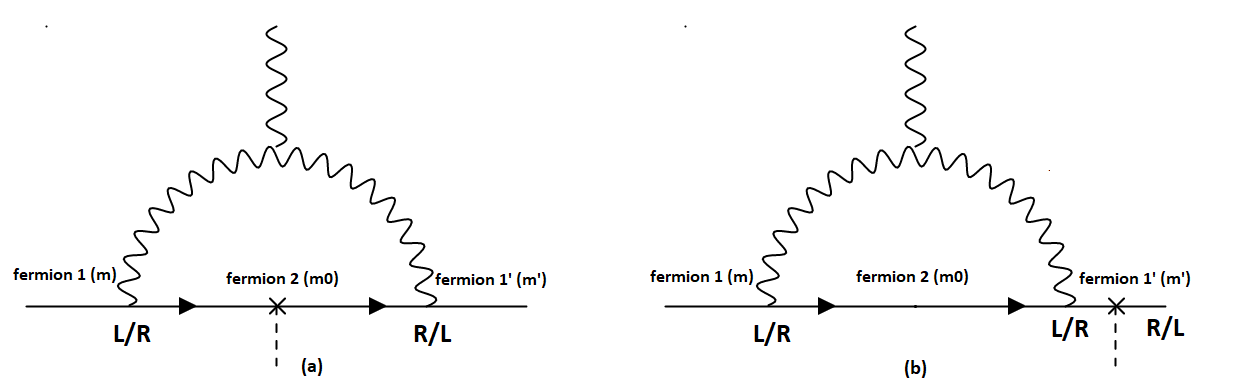}
\end{center}
\caption{Chiral flipping (mass insertion) (a) inside the loop (b) outside the loop}
\label{Chiralflipping}
\end{figure}

The projection operator is defined by 
$$P_{L/R} = \frac{1\mp\gamma^5}{2}$$
with the properties \begin{itemize}
\item $P_L P_R = \frac{1}{4}(1-\gamma^5)(1+\gamma^5)= \frac{1}{4}(1-(\gamma5)^2)=0$
\item $P_L P_L = \frac{1}{4}(1-\gamma^5)(1-\gamma^5)= \frac{1}{4}(1-\gamma5)^2= \frac{1}{2}(1-\gamma^5)=P_L$
\item $P_R P_R = \frac{1}{4}(1+\gamma^5)(1+\gamma^5)= \frac{1}{4}(1+\gamma5)^2= \frac{1}{2}(1+\gamma^5)=P_R$
\end{itemize}
In the chiral basis, we can write Dirac spinor as 
$$ \psi = \psi_L+\psi_R$$
where $\psi_L = P_L \psi$ and $\psi_R = P_R \psi$ are the chiral eigen modes. 
$$\gamma^5 \psi_{L/R} = \mp\psi_{L/R}$$
The Lagrangian of the nonrenormalization operator is given by 
$$\mathcal{L}_5 = \frac{\mu^M_{ij}}{2}\overline{\psi}_i \sigma^{\mu\nu} \psi_j F_{\mu\nu}+ \frac{\mu^E_{ij}}{2}\overline{\psi}_i \sigma^{\mu\nu}\gamma_5 \psi_j F_{\mu\nu}$$
where $F_{\mu \nu}$ is the field strength tensor of spin-1 field (e.g. Electromagnetism). \begin{itemize}
\item To study the MDM structure, consider $\bar{\psi} \sigma^{\mu \nu}\psi$. Using the conjugate definition $\bar{\psi} = \psi^\dagger \gamma^0$, we write
$$ \psi^\dagger \gamma^0 \sigma^{\mu \nu}\psi  $$
Now, this term in the chiral basis of Dirac fermion($\psi=\psi_L+\psi_R$), we get, 
$$ (\psi_L+\psi_R)^\dagger \gamma^0 \sigma^{\mu \nu}(\psi_L+\psi_R) $$
$$ (P_L \psi+P_R \psi)^\dagger \gamma^0 \sigma^{\mu \nu} (P_L\psi+P_R\psi) $$
If we do not consider the chiral flipping, then the term is 
$$(P_R \psi)^\dagger \gamma^0 \sigma^{\mu \nu} P_R\psi $$
Using $P_{L,R}^\dagger=P_{L,R}$, the hermiticity of $\gamma^5$, we get
$$ \psi^\dagger P_R \gamma^0 \sigma^{\mu \nu} P_R\psi $$
As $\gamma^5$ matrices anti-commutes with all other gamma matrices $\{\gamma^5,\gamma^\mu\}=0$, the projection operator, passing through three gamma matrices, will change. Thus, 
$$ \psi^\dagger  \gamma^0 \sigma^{\mu \nu} P_L P_R\psi $$
Since $P_L P_R =0 $, we get zero contribution to MDM if there is no chiral flipping.

\item Similarly, To study the EDM structure, consider $\bar{\psi} \sigma^{\mu \nu}\gamma^5 \psi$. Using the conjugate definition $\bar{\psi} = \psi^\dagger \gamma^0$, we write

$$ \psi^\dagger \gamma^0 \sigma^{\mu \nu}\gamma^5 \psi  $$

Now, this term in the chiral basis of Dirac fermion($\psi=\psi_L+\psi_R$), we get, 

$$ (\psi_L+\psi_R)^\dagger \gamma^0 \sigma^{\mu \nu}\gamma^5 (\psi_L+\psi_R) $$

$$ (P_L \psi+P_R \psi)^\dagger \gamma^0 \sigma^{\mu \nu}\gamma^5 (P_L\psi+P_R\psi) $$

If we do not consider the chiral flipping, then the term is 

$$(P_R \psi)^\dagger \gamma^0 \sigma^{\mu \nu}\gamma^5 P_R\psi $$

Using $P_{L,R}^\dagger=P_{L,R}$, the hermiticity of $\gamma^5$, we get

$$ \psi^\dagger P_R \gamma^0 \sigma^{\mu \nu}\gamma^5 P_R\psi $$

As $\gamma^5$ matrices anti-commutes with all other gamma matrices $\{\gamma^5,\gamma^\mu\}=0$, the projection operator, passing through three gamma matrices, will change. Thus, 

$$ \psi^\dagger \gamma^0 \sigma^{\mu \nu}\gamma^5 P_L P_R\psi $$

Since $P_L P_R =0 $, we get zero contribution to EDM if there is no chiral flipping.
\end{itemize}
Therefore, only the chiral flipping terms contribute to the EDM and MDM form factor since $P_{L, R}^2 =P_{L, R}$ is a non-zero contribution. This is not any kind of imposition. The QFT structure itself allows only chiral flipping terms to contribute. This does not mean in the loop, there must be chiral flipping. The loop contribution may give L-L no chiral flipping contribution and outside the loop we can insert mass to flip chirality and thus can get EDM and MDM contribution. Both diagrams (chiral flipping inside (a) and outside (b) the loop) are shown in Figure \ref{Chiralflipping}.
\graphicspath{{./Chiral flipping/}}
It seems like both kinds of diagrams will give similar contributions to  EDM and MDM, but actually, it's not. We can neglect the contribution of the diagram with chiral flipping outside the loop because the loop function is like enhancing the  EDM and MDM contribution at the quantum correction level. If the mass is inserted outside the loop (Figure \ref{Chiralflipping}(b)) then there is no enhancement of the contribution which leads to negligible contribution. But the diagram with chiral flipping inside (Figure \ref{Chiralflipping}(a)) will have mass insertion inside which leads to enhanced contribution. 

\subsection{Mass Insertions for Chiral Flipping}

Due to fermionic propagators and gauge boson or scalar coupling structure, we will get a tensor (to be specific a vector) structure numerator($N^\mu$). The diagrams from Figure \ref{alldiagm} contributing to the EDM and MDM have chiral flipping in the external fermionic leg like Figure \ref{Chiralflipping}(b). That's why there will be a factor $G_R^*G_L$ in the numerator of the loop function (see Section \ref{Calculation of EDM and MDM Loop}). This chiral flipping is allowed in a theory if the internal propagator fields have masses. Because if there is no mass introduced in the Lagrangian for the internal particles, there will be no other way for a left-handed to mix with a right-handed field. The mass term in the Lagrangian is the only source for a chiral field to change its chirality. Therefore, mass insertion is very important for this contribution. The QFT structure (i.e. also at the loop level) will exactly pick the mass inserted term. 

We will see this in the explicit calculation in Section \ref{Calculation of EDM and MDM Loop}. No kinetic or interaction term, in our theory, can mix between different handedness of fields (this is obvious from the theory because left and right-handedness correspond to different particles, and kinetic and interaction terms are connected to particle identities).

Fermions kinetic term in chiral basis is given by, 
$$ \mathcal{L}^{Fermion}_{kin} =  i \bar{\psi}_R \gamma^\mu \partial_\mu \psi_R +i \bar{\psi}_L \gamma^\mu \partial_\mu \psi_L $$
Fermion coupling to bosons ($\psi - A$ int) in chiral basis is given by, 
$$ \mathcal{L}^{Fermion-Boson}_{int} = g_R \bar{\psi}_R \gamma^\mu \psi_R W_{R\mu}^a t^a  + g_L \bar{\psi}_L \gamma^\mu \psi_L W_{L\mu}^a t^a$$
where $W_{L, R}$ corresponds to gauge bosons of the $SU(2)_{L, R}$ group and $\psi_{L, R}$ are the Dirac 4 spinors in chiral basis defined with the definition
$$ P_{+} \psi = \frac{1+\gamma^5}{2} \psi = \psi_R  $$
$$ P_{-} \psi = \frac{1-\gamma^5}{2} \psi = \psi_L  $$
But the mass term for fermions and gauge bosons in the Lagrangian is 
$$ \mathcal{L}^{Fermion}_{Dirac} = m_{D} \bar{\psi}_R \psi_L + m_{D} \bar{\psi}_L \psi_R $$
$$ \mathcal{L}^{Gauge}_{Dirac} = \frac{1}{2} M^2_D W^+_L W^-_R + h.c.  $$
\textbf{Remark} : These mass terms are from SSB of the $\Phi$ Higgs breaking terms. This breaking gives us Dirac-type mass(L-R). Here we don't consider Majorana-type mass caused by SSB of $\Delta$ Higgs scalar. Because in the Majorana mass term($L-L$ or $R-R$), the chiral flipping doesn't occur and that's why we don't consider the Majorana mass terms here. 
$$ \mathcal{L}^{Fermion}_{Majorana} = M_{m} \psi_L^T C^{-1} \psi_L + M_{m} \psi_R^T C^{-1} \psi_R  $$
$$ \mathcal{L}^{Gauge}_{Majorana} = \frac{1}{2} M_m^2 W^+_L W^-_L + \frac{1}{2} M_m^2 W^+_R W^-_R + h.c.  $$
In LRSM, only neutral leptons i.e. neutrinos are coupled to $\Delta$ in the fermionic sector. Charged leptons don't get majorana mass. Although neutrinos get majorana mass in the loop as the chiral flipping is enabled. Due to the Lagrangian, we can see that the Majorana mass term is not responsible for chiral flipping(since it doesn't couple two different chiral fermions). Similarly, only the Dirac mass term of gauge bosons contributes to chiral flipping in the loop function.
\newline
\newline
If the incoming charged lepton is left-handed, then in the loop, if there are no mass insertions, the outgoing charged lepton can not be right-handed. Because the W-boson or the neutrino(neutral lepton) in the loop function is left-handed coupling gauge group $SU(2)_L$ bosons and left-handed neutrino. They can not couple to right-handed fermions (charged leptons) as right-handed fermions couple to a different group in our LR theory $SU(2)_R$ and right-handed charged leptons are not connected with left-handed neutrinos in the loop. So, we need mass insertions so that left neutrinos in the loop can mix with right-handed neutrinos and also the same for bosons. Thus, we get the contribution of the loop function for EDM and MDM. This mass for the neutrino is the Dirac mass for neutrinos since chiral mixing is allowed in the Dirac mass term.

\subsection{Higgs-Fermion Coupling}

The relevant Lagrangian for Higgs-Fermion and Gauge boson-Fermion couplings are given by, 
$$ \mathcal{L}_{int}^{H-\psi}= \bar{\psi} [G_L^H (\frac{1-\gamma^5}{2})+G_R^H (\frac{1+\gamma^5}{2})]  \psi H$$
$$ \mathcal{L}_{int}^{W-\psi}= \bar{\psi} \gamma_\mu [G_L^W (\frac{1-\gamma^5}{2})+G_R^W (\frac{1+\gamma^5}{2})]  \psi W^\mu$$
where $G_{L, R}^W$ and $G_{L, R}^H$ are flavor space matrices of all fermions $f$ and the is one term like this for each mass eigenstate $W^\mu$ or $H$. Because in unitary gauge only physical modes are contributing. Physical modes have a definite eigenstate and real value. Thus, $W^\mu$ and $H$ are hermitian, 
$$G_L^{W\dagger}=G_L^W ,G_R^{W\dagger}=G_R^W ,G_L^{H\dagger}=G_R^H  $$
and that's why hermitian conjugate is omitted in the Lagrangian. 

Now, from this Lagrangian, we can determine the Fermion-Higgs vertex coupling. Delta $\Delta$ fields couple with fermions(LR model $\mathcal{L}_{Yukawa}(\Delta)$). The conjugate field of the fermion is opposite chirality to the actual field. We will prove this below. For this, the coupling for the conjugate field ($\bar{\psi}$) will be 
$$[G_L^H (\frac{1+\gamma^5}{2})+G_R^H (\frac{1-\gamma^5}{2})]$$ 
as it is opposite chiral to $\psi$. 
$$ P_L=\frac{1-\gamma^5}{2} \rightarrow P_R=\frac{1+\gamma^5}{2} $$ 
and 
$$ P_R=\frac{1+\gamma^5}{2} \rightarrow P_L=\frac{1-\gamma^5}{2} $$
Thus, 
$$[G_L^H P_L + G_R^H P_R] \rightarrow[G_L^H P_R + G_R^H P_L] $$
Now, let's prove that the conjugate field is opposite chiral. First, we have to define majorana conjugate. Majorana fermion are the real solutions of the Dirac equation. 
$$ (i\slashed{\partial}-m) \psi =0 $$
Thus, Majorana fields are those for which $\psi^*=\psi$(reality condition) i.e. real in the Majorana (real) basis. Thus, majorana conjugate is $\hat{\psi}_M = \psi^*_M$ in the majorana basis. For the Dirac or Weyl basis, we need a unitary transformation of the real representation gamma matrices and the real Majorana field. 
$$ \gamma^\mu = U \gamma^\mu_M U^\dagger $$
$$ \psi = U \psi_M $$
Thus, the reality condition in general basis ($\gamma^\mu$) is 
$$ U^\dagger \psi = (U^\dagger \psi)^* $$
which implies $\psi = UU^T \psi^*$

As $U$ is a unitary matrix, define another unitary matrix $C$ such that $UU^T=\gamma^0 C$. Thus, the Majorana conjugate $\hat{\psi}$ in general basis is 
$$ \hat{\psi}= \gamma^0 C \psi^* $$
Due to this unitary nature, the $C$ matrix satisfies a property which is,  
$$ C^{-1}\gamma_\mu C = -\gamma_\mu^T $$

As $\gamma_5$ is hermitian, $\gamma_5^*=\gamma_5^T$, we can write 
$$ C^{-1}\gamma_5 C = \gamma_5^T $$
or $C\gamma_5^T=\gamma_5 C$ that defines the action of passing $C$ matrix through $\gamma_5$.

We have defined the Majorana field and its conjugate. So, we can check the chirality explicitly. A left chiral (Weyl) field is defined by $P_L  \chi=0$. This field $\chi$ to be a Majorana field, it must satisfy $\hat{\chi}=\chi$ where $\hat{\chi}= \gamma^0 C \chi^*$. Using these defining relations, we can find the handedness of Majorana conjugate. 

The left chiral field is given by, 
$$ (1-\gamma_5) \chi =0 $$

Now, to get the Majorana conjugate, we need to take the complex conjugate and then multiply it by $\gamma_0 C$.

$$ \gamma_0 C (1-\gamma_5^*) \chi^* =0 $$

Using the property $C\gamma_5^* = C\gamma_5^T = \gamma_5 C$, we get 
$$ \gamma_0 (1-\gamma_5) C \chi^* =0 $$

Since, $\gamma_5$ anti-commutes with $\gamma_\mu$ that is $\{\gamma_5,\gamma_\mu\}=0$, we can get the majorana conjugate, $\hat{\chi}$, 
$$ (1+\gamma_5) \gamma_0 C \chi^* =0  $$
using the defining relation of majorana conjugate field $\hat{\chi}=\gamma_0 C \chi^*$, 
$$ (1+\gamma_5) \hat{\chi}=0 $$

Thus, the conjugate majorana field is right-handed if the majorana field is left-handed. This is the reason why we have to be careful with the Higgs coupling, the coupling of the conjugate field with Higgs in the loop function must be opposite chirality. The conjugation is a toggle operator($\hat{\hat{\psi}}=\psi$). Thus, our Majorana field is given by,
$$ \psi(x) = \chi(x)+\hat{\chi}(x) $$

In the case of the $\phi$ field, it is not hard to verify. Due to gamma structure, this chirality flipping is occurring. Consider the conjugate field. 

$$ \bar{\psi}_L = (\psi_L^\dagger \gamma^0) $$
$$  = (P_L\psi)^\dagger \gamma^0  $$
$$ =\psi^\dagger P_L \gamma^0 $$

Using gamma matrix property, $\{\gamma^\mu, \gamma^5\}=0$ we get, 

$$ \bar{\psi}_L = \psi^\dagger \gamma^0 P_R  $$

Higgs scalar coupling with fermion has $\gamma^5$ and coupling constants in the vertex factor(See Feynman Rules). This is the reason why chirality is flipped in the vertex factor for the conjugate field only for Higgs-scalar coupling.

$$(-i) [G_L^H (\frac{1-\gamma^5}{2})+G_R^H (\frac{1+\gamma^5}{2})] $$

For W-boson and fermion coupling, there is an extra $\gamma^\mu$ coming from the vertex of gauge boson-fermion coupling. 

$$ (-i\gamma^\mu)[G_L^W (\frac{1-\gamma^5}{2})+G_R^W (\frac{1+\gamma^5}{2})] $$

Thus, using the anti-commutation relation again, we get no flipping in the chirality of the conjugate field in the case of W-boson and fermion coupling. 

\subsection{Doubly Charged Higgs Vertex}

In the Higgs sector, we not only have singly charged scalars (like $\delta^+$) which give contribution via diagram (c) in Figure \ref{alldiagm} but also have neutral and doubly charged scalars ($\delta^0$ or $\delta^{++}$). They give contribution via diagram(d) in Figure \ref{alldiagm}. We can see this in two ways. We can see the couplings in the Lagrangian to find the vertex, or we can use $U(1)_{EM}$ invariance which implies that the vertex must be neutral. If a $\delta^{++}$ couples with a singly charged lepton and a neutrino, then it can not conserve electric charge at the vertex. It necessarily has to couple with two singly charged leptons. 
\graphicspath{{./delta/}}
\begin{figure}[H]
\begin{center}
\includegraphics[width=30mm]{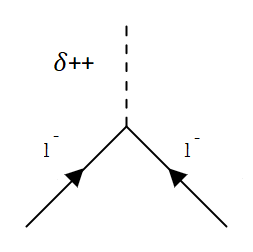}
\end{center}
\caption{$\delta^{++}$ coupling to 2 singly charged leptons}
\label{delta}
\end{figure}
The $\Delta$ coupling to the fermionic doublet term in the Lagrangian is 
$$ L_{Li}^T G_{Lij} C^{-1}i\tau_2 \Delta_L L_{Lj}+ (L \rightarrow R) + h.c. $$ 
where $$\Delta = \frac{1}{\sqrt{2}} \begin{pmatrix}
\delta^+ & \sqrt{2}\delta^{++} \\
\sqrt{2}\delta^0 & -\delta^+
\end{pmatrix} \ \& \ L = \begin{pmatrix}
 \nu \\
  e
\end{pmatrix} $$
With explicitly writing the terms we can find the $\delta^{++}$ coupling term, 
$$ Y \overline{L^c_L} i\tau_2 L_L \delta^{++} + Y^\dagger \overline{L_L} i\tau_2 L_L^c \delta^{++}  $$

Both ways, the doubly charged delta should couple to two leptons, and thus the internal fermionic line is charged under $U(1)_{EM}$. Thus, photon in this case can couple to loop's internal fermion, diagram(d) in Figure \ref{alldiagm}. A similar argument implies for $\delta^0$.

\section{Calculation of EDM and MDM Loop}
\label{Calculation of EDM and MDM Loop}

In this section, we will explicitly calculate: W-loop contribution (Figure \ref{alldiagm} (a) ) and Higgs loop (Figure \ref{alldiagm} (c), (d) ). Our main concern for BSM phenomenology is both Higgs loops in Figure \ref{alldiagm} (c) \& (d). Doing W-loop calculation is just for consistency with \cite{Ecker:1983dj}. The reason for not doing a calculation of Z-loop contribution (Figure \ref{alldiagm} (b) ) explicitly is contribution is smaller than W-loop due to heavy mass but will be included in the computations of EDM and MDM in Section \ref{Formulas for EDM and MDM of Charged Leptons}.
 
\subsection{Diagram 1: W-loop contributions}
\graphicspath{{./Wloop/}}
\begin{figure}[H]
\begin{center}
\includegraphics[width=80mm]{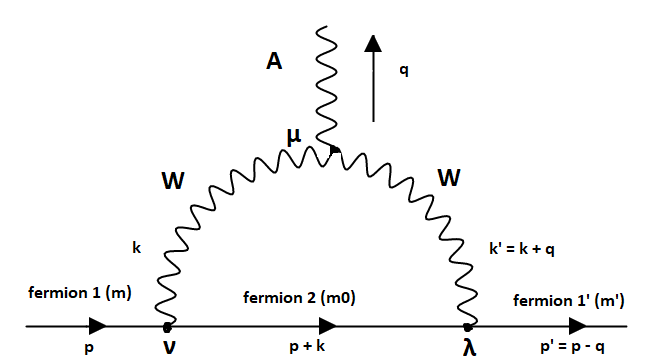}
\end{center}
\caption{W-loop contributing diagram}
\label{Wloop}
\end{figure}
 
Consider, the internal fermion's mass = $m_0$ and external fermion's mass = $m$. Here, fermions couple with W-boson($W$) which has the mass $M_W$. The interaction Lagrangian is given by 
$$ -\mathcal{L} = \bar{\psi} \gamma^\mu [G_L^{W}(\frac{1-\gamma _5}{2})+G_R^{W}(\frac{1+\gamma _5}{2})] \psi W^\mu $$
Thus, the loop contribution is (using the corresponding Feynman rules of Figure \ref{feyn1} and \ref{feyn4})
\begin{center}
$\delta \Gamma_1 ^{\mu }=\int \frac{d^{4}k}{(2\pi)^{4}}(i\gamma^{\lambda})[G_L^{W}\frac{1-\gamma _5}{2}+G_R^{W}\frac{1+\gamma _5}{2}]^{*}\frac{i(\slashed{p}+\slashed{k}+m_0)}{(p+k)^{2}-m_0^{2}+i\epsilon}(i\gamma^{\nu})[G_L^{W}\frac{1-\gamma _5}{2}+G_R^{W}\frac{1+\gamma _5}{2}]\frac{(-i)(\eta _{\nu{\nu}'}-\frac{k_{\nu}k_{\nu}'}{M_W^{2}})}{k^{2}-M_W^{2}+i\epsilon}(ie)[(k+q+k)^{{\mu}}\eta_{{\nu}'{\lambda}'}-(q+k+q)_{{\nu}'} \eta _{\mu{\lambda}'}-(k-q)_{{\lambda}'}\eta _{\mu{\nu}'}]\frac{(-i)(\eta _{\lambda{\lambda}'}-\frac{k_{\lambda}k_{\lambda}'}{M_W^{2}})}{{k}'^{2}-M_W^{2}+i\epsilon}tr[t^at^b]$
\end{center}
where $q=p-p'$ is the momentum provided by the photon. This implies using $p^2=m^2$ and ${p'}^2= {m'}^2= 0$ since $m >> m'$ hence $m'\approx 0$
$$ {p'}^2 = (p-q)^2 $$
$$=> {m'}^2 = m^2 - 2p\cdot q+q^2 $$
$$=> 2p\cdot q = m^2 + q^2 $$
This is the momentum conservation relation. For MDM $\langle\mu\rangle = -F_{MDM}(q^2=0)$, we will set $q^2=0$ and also due to momentum conservation relation $2p\cdot q = m^2$. This will simplify our calculation a lot. The trace of generators is due to non-abelian nature. From normalization $tr[t^at^b]=\frac{1}{2}\delta^{ab}$ which is known as the group factor.

Now, using Feynman parametrization we can combine these three denominators,
$$\frac{1}{[(p+k)^{2}-m_0^{2}+i\epsilon][k^{2}-M_W^{2}+i\epsilon][{k}'^{2}-M_W^{2}+i\epsilon]}=\int_{0}^{1}dx_1dx_2dx_3\frac{2\delta(x_1+x_2+x_3-1)}{D^{3}}$$
where, $$D = x_1[(p+k)^{2}-m_0^{2}+i\epsilon]+x_2[k^{2}-M_W^{2}+i\epsilon]+x_3[{k}'^{2}-M_W^{2}+i\epsilon]$$
$$=x_1[p^{2}+k^{2}+2p.k-m_0^{2}]+x_2[k^{2}-M_W^{2}]+x_3[k^{2}+2k.q+q^{2}-M_W^{2}]+i\epsilon $$
$$ =k^{2}+2k.(x_1p+x_3q)+x_1m^{2}+x_3q^{2}-x_1m_0^{2}-M_W^{2}(1-x_1) $$
We define, $$l = k+x_1p+x_3q $$
$$ => l^{2}=k^{2}+2k.(x_1p+x_3q)+x_3^{2}q^{2}+x_1^{2}p^{2}+2p.qx_1x_3 $$
\newline
With this trick, we can write our combined denominator in terms of $l^2$ and $l$ independent term. The advantage of this trick is that there is no linear $l$ term in the denominator. Thus, numerators with linear in $l$ can be dropped due to the integral limit from $-\infty$ to $+\infty$(all odd power in $l$ in the numerator vanishes due to symmetric integration limit). Thus, using the delta function, i.e. $ x_1+x_2+x_3=1$, we can write, 
$$l^{2}-D=M_W^{2}[1-x_1+\frac{m_0^{2}}{M_W^{2}} x_1+\frac{m^{2}}{M_W^{2}}(x_1-1)+x_1x_3]-\frac{q^{2}}{M_W^{2}}(x_{1}x_{3}+x_{2}x_{3})]-i\epsilon = \Delta-i\epsilon $$
Numerator, $ N^{\mu}=\frac{(-e)}{4}\gamma^{\lambda}[G_L^{W}(1-\gamma _5)+G_R^{W}(1+\gamma _5)]^{*}(\slashed{p}+\slashed{k}+m_0)\gamma^{\nu}[G_L^{W}(1-\gamma _5)+G_R^{W}(1+\gamma _5)](\eta _{\nu{\nu}'}-\frac{k_{\nu}k_{\nu}'}{M_W^{2}})[(2k+q)^{\mu}\eta_{{\nu}'{\lambda}'}-(k+2q)_{{\nu}'} \eta _{\mu{\lambda}'}-(k-q)_{{\lambda}'}](\eta _{\lambda{\lambda}'}-\frac{k_{\lambda}k_{\lambda}'}{M_W^{2}})$
\newline
\newline
$=\frac{(-e)}{4}[G_R^{W*}G_L^{W}\{\gamma^{\lambda}(\slashed{p}+\slashed{k}+m_0)\gamma^{\nu}-\gamma^{\lambda}(\slashed{p}+\slashed{k}-m_0)\gamma^{\nu}+\gamma^{\lambda}(\slashed{p}+\slashed{k}-m_0)\gamma^{\nu}\gamma _5-\gamma^{\lambda}(\slashed{p}+\slashed{k}+m_0)\gamma^{\nu}\gamma _5\}+G_L^{W*}G_R^{W}\{\gamma^{\lambda}(\slashed{p}+\slashed{k}+m_0)\gamma^{\nu}-\gamma^{\lambda}(\slashed{p}+\slashed{k}-m_0)\gamma^{\nu}-\gamma^{\lambda}(\slashed{p}+\slashed{k}-m_0)\gamma^{\nu}\gamma _5+\gamma^{\lambda}(\slashed{p}+\slashed{k}+m_0)\gamma^{\nu}\gamma _5\}] [(2k+q)^{\mu}\eta_{\nu{\lambda}'}-\frac{1}{M_W^{2}}(2k+q)^{\mu}k_{\nu}k_{\lambda}'-(k+2q)_{\nu} \eta _{\mu{\lambda}'}+\frac{1}{M_W^{2}}k.(k+2q)k_{\nu}\eta _{\mu{\lambda}'}-(k-q)_{{\lambda}'}\eta _{\mu\nu}+\frac{1}{M_W^{2}}(k-q)_{{\lambda}'}k_{\nu}k_{\mu}](\eta _{\lambda{\lambda}'}-\frac{k_{\lambda}k_{\lambda}'}{M_W^{2}})$
\newline
\newline
$=-em_0[Re(G_R^{W*}G_L^{W})\gamma^{\lambda}\gamma^{\nu}-iIm(G_R^{W*}G_L^{W})\gamma^{\lambda}\gamma^{\nu}\gamma _5][\{(2k+q)^{\mu}\eta _{\nu\lambda}-(k+2q)_{\nu} \eta _{\mu\lambda}-(k-q)_{\lambda}\eta _{\mu\nu}\}+\frac{1}{M_W^{2}}\{-(2k+q)^{\mu}k_{\nu}k_{\lambda}+k.(k+2q)k_{\nu}\eta _{\mu\lambda}+(k-q)_{\lambda}k_{\nu}k_{\mu}-(2k+q)^{\mu}{k_{\nu}}'{k_{\lambda}}'+(k+2q)_{\nu}{k_{\mu}}'{k_{\lambda}}'+{k}'.(k-q){k_{\lambda}}'\eta _{\mu\nu}\}+\frac{1}{M_W^{4}}\{(2k+q)^{\mu}k_{\nu}(k.{k}'){k_{\lambda}}'-k.(k+2q)k_{\nu}{k_{\mu}}'{k_{\lambda}}'-{k}'.(k-q)k_{\mu}k_{\nu}{k_{\lambda}}'\}]$
\newline
\newline
Here, we see that only the term linear in mass is contributing. This is the mass insertion we were expecting for the chiral flipping. We argued that this mass must be Dirac mass which indeed is. As we are calculating the chiral flipping diagram, the structure automatically takes the terms that have mass insertions. Also, we see the boson's mass coming in the denominator already, reflecting the fact that only mass terms in the Lagrangian can flip chirality.
\newline
\newline
$N^{\mu} = -em_0[Re(G_R^{W*}G_L^{W})-iIm(G_R^{W*}G_L^{W})\gamma _5]\gamma^{\lambda}\gamma^{\nu}[\{(2k+q)^{\mu}\eta _{\nu\lambda}-(k+2q)_{\nu} \eta _{\mu\lambda}-(k-q)_{\lambda}\eta _{\mu\nu}\}+\frac{1}{M_W^{2}}\{-(2k+q)^{\mu}k_{\nu}k_{\lambda}+k.(k+2q)k_{\nu}\eta _{\mu\lambda}+(k-q)_{\lambda}k_{\nu}k_{\mu}-(2k+q)^{\mu}{k_{\nu}}'{k_{\lambda}}'+(k+2q)_{\nu}{k_{\mu}}'{k_{\lambda}}'+{k}'.(k-q){k_{\lambda}}'\eta _{\mu\nu}\}+\frac{1}{M_W^{4}}\{(2k+q)^{\mu}k_{\nu}(k.{k}'){k_{\lambda}}'-k.(k+2q)k_{\nu}{k_{\mu}}'{k_{\lambda}}'-{k}'.(k-q)k_{\mu}k_{\nu}{k_{\lambda}}'\}]$
\newline
\newline
$=-em_0[Re(G_R^{W*}G_L^{W})-iIm(G_R^{W*}G_L^{W})\gamma _5][\{4(2k+q)^{\mu}-\gamma^{\mu}(\slashed{k}+2\slashed{q})-(\slashed{k}-\slashed{q})\gamma^{\mu}\}+\frac{1}{M_W^{2}}\{-(2k+q)^{\mu}k^{2}+(k^{2}+2k.q)\gamma^{\mu}\slashed{k}+(k^{2}-\slashed{q}\slashed{k})k^{\mu}-(2k+q)^{\mu}(k+q)^{2}+(k+q)^{\mu}(k^{2}+2\slashed{k}\slashed{q}+\slashed{q}\slashed{k})+(\slashed{k}+\slashed{q})\gamma^{\mu}k^{2}\}]$
\newline
\newline
$=-em_0[Re(G_R^{W*}G_L^{W})-iIm(G_R^{W*}G_L^{W})\gamma _5][\{8k^{\mu}+4q^{\mu}-\gamma^{\mu}\slashed{k}-\slashed{k}\gamma^{\mu}-2\gamma^{\mu}\slashed{q}+\slashed{q}\gamma^{\mu}\}+\frac{1}{M_W^{2}}\{-(2k+q)^{\mu}(2k^{2}+2k.q)+\gamma^{\mu}\slashed{k}(k^{2}+2k.q-k^{2})+2k^{\mu}\slashed{k}+\slashed{q}\gamma^{\mu}k^{2}+k^{\mu}(k^{2}-\slashed{q}\slashed{k}+k^{2}+2\slashed{k}\slashed{q}+\slashed{q}\slashed{k})+q^{\mu}(k^{2}+2\slashed{k}\slashed{q}+\slashed{q}\slashed{k})\}]$
\newline
\newline
$=-em_0[Re(G_R^{W*}G_L^{W})-iIm(G_R^{W*}G_L^{W})\gamma _5][\{8k^{\mu}+4q^{\mu}-2k^{\mu}-2\gamma^{\mu}\slashed{q}-\gamma^{\mu}\slashed{q}+2q^{\mu}\}+\frac{1}{M_W^{2}}\{k^{\mu}(-4k.q+2\slashed{k}\slashed{q})+q^{\mu}(-k^2+2\slashed{k}\slashed{q}+\slashed{q}\slashed{k}-2k.q)+\slashed{q}\gamma^{\mu}k^{2}+\gamma^{\mu}\slashed{k}2k.q\}]$
\newline
\newline
$=-em_0[Re(G_R^{W*}G_L^{W})-iIm(G_R^{W*}G_L^{W})\gamma _5][\{6k^{\mu}+6q^{\mu}-3\gamma^{\mu}\slashed{q}\}+\frac{1}{M_W^{2}}\{k^{\mu}(-4k.q+2\slashed{k}\slashed{q})+q^{\mu}(-k^2+\slashed{k}\slashed{q})+\slashed{q}\gamma^{\mu}k^{2}+\gamma^{\mu}\slashed{k}2k.q]$
\newline
\newline
Now, to evaluate $\gamma^\mu \slashed{q}$
$$\bar{u}(p')\gamma^\mu \slashed{q}u(p)$$
$$= \bar{u}(p')\gamma^\mu (\slashed{p}-\slashed{p}')u(p) $$
$$ =\bar{u}(p')(\gamma^\mu \slashed{p} - \gamma^\mu\slashed{p}')u(p) $$
$$=\bar{u}(p')(\gamma^\mu m + \slashed{p}' \gamma^\mu - 2p'^\mu )u(p)$$
$$=\bar{u}(p')(\gamma^\mu m -2(p-q)^\mu)u(p) $$
Now, substituting $\gamma^\mu \slashed{q} =\gamma^\mu m -2(p-q)^\mu$ and $k^\mu = l^\mu-x_1p^\mu-x_3q^\mu$, we get,
\newline
\newline
$=-em_0[Re(G_R^{W*}G_L^{W})-iIm(G_R^{W*}G_L^{W})\gamma _5][\{6(l-x_1p-x_3q)^{\mu}-3(-2p^{\mu}+2q^{\mu}+\gamma^{\mu}m)+6q^{\mu}\}+\frac{1}{M_W^{2}}\{(l-x_1p-x_3q)^{\mu}[-4(l.q-x_1p.q)+2(\slashed{l}\slashed{q}-x_1\slashed{p}\slashed{q})]+q^{\mu}[-(l-x_1p-x_3q)^{2}+(\slashed{l}-x_1\slashed{p}-x_3\slashed{q})\slashed{q}]+\slashed{q}\gamma^{\mu}(l-x_1p-x_3q)^{2}+\gamma^{\mu}(\slashed{l}-x_1\slashed{p}-x_3\slashed{q})2(l.q-x_1 p.q)\}]$
\newline
\newline
$=-em_0[Re(G_R^{W*}G_L^{W})-iIm(G_R^{W*}G_L^{W})\gamma _5][\{6l^{\mu}-6(x_1-1)p^{\mu}-6x_3q^{\mu}-3\gamma^{\mu}m\}+\frac{1}{M_W^{2}}\{l^{\mu}(-4l.q+2\slashed{l}\slashed{q})-(x_1 p+x_3 q)^{\mu}(4x_1 p.q-2x_1 \slashed{p}\slashed{q})+q^{\mu}[-l^{2}-(x_1 p+x_3 q)^{2}-x_1 \slashed{p}\slashed{q}+2x_1 x_3 \slashed{p}\slashed{q}]+\slashed{q}\gamma^{\mu}[l^{2}+(x_1 p+x_3 q)^{2}]+\frac{l^{2}}{2}\gamma^{\mu}\slashed{q}+\gamma^{\mu}(x_1 \slashed{p}+x_3 \slashed{q})2x_1 p.q\}]$
\newline
\newline
Using $l^a l^b \rightarrow \frac{l^2}{4} \eta^{a b}$ we need to find $l^\mu\slashed{l}\slashed{q}$, $\gamma^\mu\slashed{l} l.q$ and $l^\mu l.q$ 
$$\bar{u}(p')l^\mu\slashed{l}\slashed{q} u(p)$$
$$= \bar{u}(p')l^\mu l^\beta\gamma_\beta u(p)\slashed{q}u(p) $$
$$=\bar{u}(p')\frac{l^2}{4} \eta^{\mu \beta}\gamma_\beta\slashed{q} u(p) $$
$$=\bar{u}(p')\frac{l^2}{4}\gamma^\mu\slashed{q} u(p) $$
and 
$$\bar{u}(p')\gamma^\mu\slashed{l} l.q u(p)$$
$$=\bar{u}(p')\gamma^\mu\gamma_\alpha l^\beta q_\beta u(p)$$
$$=\bar{u}(p')\frac{l^2}{4}\gamma^\mu\slashed{q}u(p)$$
and
$$\bar{u}(p')l^\mu l.q u(p)$$
$$=\bar{u}(p')l^\mu l^\alpha q_\alpha u(p)$$
$$=\bar{u}(p')\frac{l^2}{4} \eta^{\mu \alpha}q_\alpha u(p)$$
$$=\bar{u}(p')\frac{l^2}{4}q^\mu u(p)$$
Now, substituting $l^\mu\slashed{l}\slashed{q} = \frac{l^2}{4}\gamma^\mu\slashed{q}$, $\gamma^\mu\slashed{l} l.q = \frac{l^2}{4}\gamma^\mu\slashed{q}$ and $l^\mu l.q = \frac{l^2}{4}q^\mu $ we get,
\newline
\newline
$=-em_0[Re(G_R^{W*}G_L^{W})-iIm(G_R^{W*}G_L^{W})\gamma _5][\{6l^{\mu}-6(x_1-1)p^{\mu}-6x_3q^{\mu}-3\gamma^{\mu}m\}+\frac{1}{M_W^{2}}\{\frac{l^{2}}{4}(-4q^{\mu}+2\gamma^{\mu}\slashed{q})-(x_1 p+x_3 q)^{\mu}(4x_1 p.q-2x_1 \slashed{p}\slashed{q})+q^{\mu}[-l^{2}-(x_1 p+x_3 q)^{2}-x_1 \slashed{p}\slashed{q}+2x_1 x_3 \slashed{p}\slashed{q}]+l^{2}\slashed{q}\gamma^{\mu}+\slashed{q}\gamma^{\mu}{x_1}^{2}p^{2}+\frac{l^{2}}{2}\gamma^{\mu}\slashed{q}+\gamma^{\mu}(x_1 \slashed{p}+x_3 \slashed{q})2x_1 p.q\}]$
\newline
\newline
$=-em_0[Re(G_R^{W*}G_L^{W})-iIm(G_R^{W*}G_L^{W})\gamma _5][\{6l^{\mu}-6(x_1-1)p^{\mu}-6x_3q^{\mu}-3\gamma^{\mu}m\}+\frac{1}{M_W^{2}}\{q^{\mu}[-(x_1 p+x_3 q)^{2}-x_1 \slashed{p}\slashed{q}+2x_1 x_3 \slashed{p}\slashed{q}]+\slashed{q}\gamma^{\mu}{x_1}^{2}m^{2}-x_1 p^{\mu}(4x_1 p.q-2x_1 \slashed{p}\slashed{q})+\gamma^{\mu}\slashed{p}2{x_1}^{2}p.q\}]$
\newline
\newline
Now, we need to find $\slashed{p}\slashed{q}$ 
$$\bar{u}(p')\slashed{p}\slashed{q}u(p)$$
$$=\bar{u}(p')p_\mu \gamma^\mu \slashed{q}u(p) $$
$$=\bar{u}(p')p_\mu(-2(p-q)^\mu)u(p) $$
$$=\bar{u}(p')(-2p^2 + 2p\cdot q)u(p) $$
$$=\bar{u}(p')(-m^2)u(p) $$
substituting $\slashed{p}\slashed{q} = -m^2 $ we get,
\newline
\newline
$=-em_0[Re(G_R^{W*}G_L^{W})-iIm(G_R^{W*}G_L^{W})\gamma _5][\{6l^{\mu}-6(x_1-1)p^{\mu}-6x_3q^{\mu}-3\gamma^{\mu}m\}+\frac{1}{M_W^{2}}\{q^{\mu}(-x_1^{2}m^{2}-3x_1 x_3 m^{2}+x_1 m^{2})+2{x_1}^{2}m^{2}p^{\mu}-4{x_1}^{2}m^{2}p^{\mu}\}]$
\newline
\newline
$=-em_0[Re(G_R^{W*}G_L^{W})-iIm(G_R^{W*}G_L^{W})\gamma _5][\{-3(x_1-1)2p^{\mu}-6x_3 q^{\mu}+6l^{\mu}-3\gamma^{\mu}m\}+\frac{m^{2}}{M_W^{2}}\{-2p^{\mu}{x_1}^{2}m^{2}-q^{\mu}x_1 (x_1 +3x_3 -1)\}]$
\newline
\newline
To apply Gordon identity to go to EDM and MDM form factor, we need to find $(p^\mu+p^{'\mu})$. We can write
$ 2p^\mu = p^\mu + p'^{\mu} + q^\mu $, then the expression becomes
\newline
\newline
$=-em_0[Re(G_R^{W*}G_L^{W})-iIm(G_R^{W*}G_L^{W})\gamma _5][\{-3(x_1-1)(p^{\mu}+{p}'^{\mu})+(-6x_3-3x_1+3)q^{\mu}+6l^{\mu}-3\gamma^{\mu}m\}+\frac{m^{2}}{M_W^{2}}\{-{x_1}^{2}(p^{\mu}+{p}'^{\mu})-q^{\mu}x_1 (2x_1 +3x_3 -1)\}]$
\newline
\newline
$=-em_0[Re(G_R^{W*}G_L^{W})-iIm(G_R^{W*}G_L^{W})\gamma _5][\{3(1-x_1)(-i\sigma^{\mu\nu}q_\nu)+3(x_2-x_3)q^{\mu}+6l^{\mu}-3\gamma^{\mu}m\}+\frac{m^{2}}{M_W^{2}}\{-{x_1}^{2}(-i\sigma^{\mu\nu}q_\nu)-x_1 (x_1+x_3)q^{\mu}+x_1 (x_2-x_3)q^{\mu}\}]$
\newline
\newline
Note that,\begin{itemize}
\item $(x_2-x_3)$ term is odd $x_2$ and $x_3$, but the denominator is even $x_2x_3$. That is, under $x_2 \rightarrow -x_2$ and $x_3 \rightarrow -x_3$, the numerator changes sign i.e. odd but the denominator doesn't i.e. even. Thus, it gives zero contribution to the integration.
\item $(x_1+x_3)$ term is even therefore, it does not vanish in $l$ integration. We can interpret this as photon($\gamma$) contributes in total $|\mathcal{M}|^{2}$ such that $\mathcal{M}= \alpha F(q^{2}=0) + (q^{\mu}$ term) i.e. $q^{\mu}$ term exists because in $l_i \rightarrow l_j \gamma$ process photon($\gamma$) must be on-shell i.e. physical($q^{2}= 0$).
\item $\gamma^{\mu}m$ term is ignored since it contributes to electric charge.  
\item The $\frac{1}{M_W^4}$ terms vanish. Due to its complexity, it is not done explicitly but it vanishes. Moreover, this term is highly suppressed by the weak boson mass. Thus, combining both $M_W^0$ and $M_W^2$ terms, we get the W-loop contribution. 
\item $l^{\mu}$ term  will not contribute due to momentum integral $\int \frac{d^{4}l}{(2\pi)^{4}}\frac{l^{\mu}}{l^{\mu}-\Delta}=0$.
\end{itemize}
Thus, combining everything, the loop contribution is
\newline
\newline
$$\delta \Gamma_1 ^{\mu }=\frac{1}{2}\int \frac{d^{4}l}{(2\pi)^{4}}\int_{0}^{1}dx_1dx_2dx_3\frac{2\delta(x_1+x_2+x_3-1)N^{\mu}}{D^{3}}$$
\newline
\newline
Now, we need to do Wick Rotation to do the integration($l^0 \rightarrow il^0_E$). We will ignore the subscript E(Euclidean). Then
$$\frac{1}{2} \int \frac{d^4 l}{(2\pi)^4} \frac{1}{(l^2-\Delta)^3} \rightarrow -i \int \frac{d^4 l}{(2\pi)^4} \frac{1}{(l^2+\Delta)^3} $$
The general formula for doing this kind of loop integral is given by 
$$ \int \frac{d^d l}{(2\pi)^d} \frac{(l^2)^a}{(l^2+\Delta)^b} = \frac{\Gamma(b-a-\frac{d}{2})\Gamma(a+\frac{d}{2})}{(4\pi)^\frac{d}{2}\Gamma(b) \Gamma(\frac{d}{2})} \Delta^{-(b-a-\frac{d}{2})}  $$
We can prove this by using $a=0$, then differentiating and using gamma matrix properties. This formula is directly taken from \cite{Srednicki:2007qs}. 

For our case, using $d=4,a=0,b=3$ and Wick rotating(extra $-i$ factor), we get 

$$ \int \frac{d^4 l}{(2\pi)^4} \frac{1}{(l^2-\Delta)^3} =  \frac{(-i)}{(4\pi)^2}\frac{1}{2\Delta} $$

Using this, we can do the $l$ integration. Thus, our loop contribution is
\begin{center}
$\delta \Gamma_1 ^{\mu }=-em_0m[Re(G_R^{W*}G_L^{W})-iIm(G_R^{W*}G_L^{W})\gamma _5]\frac{(-i)}{(4\pi)^{2}}\int_{0}^{1}dx_1dx_2dx_3\frac{2\delta(x_1+x_2+x_3-1)}{2\Delta}[3(1-x_1)-\frac{m^{2}}{M_W^{2}}{x_1}^{2}](-i\frac{\sigma^{\mu\nu}q_\nu}{2m})$
\end{center} 
Therefore,
$$\delta \Gamma_1 ^{\mu } = \frac{em_0m[Re(G_R^{W*}G_L^{W})-iIm(G_R^{W*}G_L^{W})\gamma _5]}{16\pi^{2}M_W^{2}}I_1(r,s)\frac{\sigma^{\mu\nu}q_\nu}{2m}$$
where, $$I_1(r,s)= \int_0^1  dx  \frac{(1-x)[3(1-x)-sx^2]}{1-x+rx-sx(1-x)}; r=\frac{m_0^{2}}{M_W^{2}} \& s=\frac{m^{2}}{M_W^{2}}$$
From the definition of EDM and MDM form factor, $\delta\Gamma^\mu = F_{2}(q^2)\frac{\sigma^{\mu\nu}q_\nu}{2m} +F_{EDM}(q^2)\sigma^{\mu\nu}q_\nu\gamma^5 $, and AMM definition $a = F_{2}(q^2=0)$, EDM definition $d_f = -F_{EDM}(q^2=0)$. So we get the anomalous magnetic moment \& electric dipole moment :
$$a = mRe(G_R^{W*}G_L^{W})\frac{em_0}{16\pi^{2}M_W^{2}}\int_0^1  dx  \frac{(1-x)[3(1-x)-sx^2]}{1-x+rx-sx(1-x)}$$
$$d_f = \frac{Im(G_R^{W*}G_L^{W})}{2}\frac{em_0}{16\pi^{2}M_W^{2}}\int_0^1  dx  \frac{(1-x)[3(1-x)-sx^2]}{1-x+rx-sx(1-x)}$$
where $ r=\frac{m_0^{2}}{M_W^{2}}$ \& $s=\frac{m^{2}}{M_W^{2}}$
\newline
\newline
Note that, The expression of EDM ($d_f$) exactly matches with the expression in \cite{Ecker:1983dj}. 

\subsection{Diagram 2: H-loop contributions}

\graphicspath{ {./Hloop1/} }
\begin{figure}[H]
\begin{center}
\includegraphics[width=80mm]{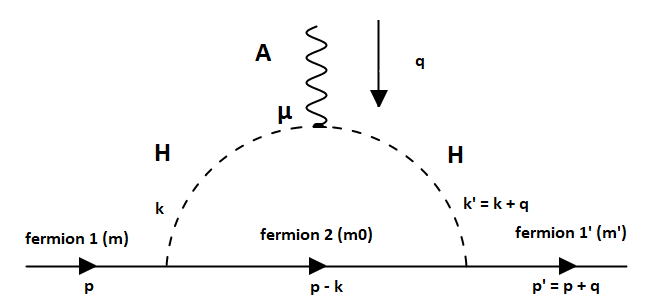}
\end{center}
\caption{H-loop (1) contributing diagram}
\label{Hloop1}
\end{figure}
The interaction (fermion-scalar) Lagrangian is given by  
$$-\mathcal{L} = \bar{\psi} [G_L^{H}(\frac{1-\gamma _5}{2})+G_R^{H}(\frac{1+\gamma _5}{2})] \psi H $$
Consider, the internal fermion's mass = $m_0$ and external fermion's mass = $m$. Here, fermions couple with scalar field  ($H$) which has the mass $M_H$. Recall, here for the conjugate field we have to use $[G_L^H P_R+G_R^H P_L] = [G_L^{H}(\frac{1+\gamma _5}{2})+G_R^{H}(\frac{1-\gamma _5}{2})]$ because the conjugate field of the Majorana fermion is opposite chiral. Using the fermion scalar vertex from Feynman rule of Figure \ref{feyn3} and \ref{feyn5}, we can write the H-loop contribution:
\begin{center}
$\delta \Gamma_2 ^{\mu }=\int \frac{d^{4}k}{(2\pi)^{4}}[G_L^{H}\frac{1+\gamma _5}{2}+G_R^{H}\frac{1-\gamma _5}{2}]^{*}\frac{i(\slashed{p}-\slashed{k}+m_0)}{(p-k)^{2}-m_0^{2}+i\epsilon}[G_L^{H}\frac{1-\gamma _5}{2}+G_R^{H}\frac{1+\gamma _5}{2}]\frac{(-i)}{k^{2}-M_H^{2}+i\epsilon}(ie)[(2k+q)^{\mu}] \frac{(-i)}{{k}'^{2}-M_H^{2}+i\epsilon}$
\end{center}
\begin{center}
$=\int\frac{d^{4}k}{(2\pi)^{4}}\frac{\frac{e}{4}[G_L^{H}(1+\gamma _5)+G_R^{H}(1-\gamma _5)]^{*}(\slashed{p}-\slashed{k}+m_0)[G_L^{H}(1-\gamma _5)+G_R^{H}(1+\gamma _5)][(2k+q)^{\mu}]}{[(p-k)^{2}-m_0^{2}+i\epsilon][k^{2}-M_H^{2}+i\epsilon][{k}'^{2}-M_H^{2}+i\epsilon]}$
\end{center}
Now, using Feynman parametrization we can combine these three denominators,
$$\frac{1}{[(p+k)^{2}-m_0^{2}+i\epsilon][k^{2}-M_H^{2}+i\epsilon][{k}'^{2}-M_H^{2}+i\epsilon]}=\int_{0}^{1}dx_1dx_2dx_3\frac{2\delta(x_1+x_2+x_3-1)}{D^{3}}$$
where, $$D = x_1[(p+k)^{2}-m_0^{2}+i\epsilon]+x_2[k^{2}-M_H^{2}+i\epsilon]+x_3[{k}'^{2}-M_H^{2}+i\epsilon]$$
$$=x_1[p^{2}+k^{2}+2p.k-m_0^{2}]+x_2[k^{2}-M_H^{2}]+x_3[k^{2}+2k.q+q^{2}-M_H^{2}]+i\epsilon $$
$$ =k^{2}+2k.(x_1p+x_3q)+x_1m^{2}+x_3q^{2}-x_1m_0^{2}-M_H^{2}(1-x_1) $$
We define, $$l = k-x_1p+x_3q $$
$$ => l^{2}=k^{2}+2k.(x_1p-x_3q)+x_3^{2}q^{2}+x_1^{2}p^{2}-2p.qx_1x_3 $$
Hence, $$ l^{2}-D=M_H^{2}[1-x_1+\frac{m_0^{2}}{M_H^{2}} x_1+\frac{m^{2}}{M_H^{2}}\{x_1(x_1-1)+x_1x_3\}]-i\epsilon = \Delta-i\epsilon $$
Numerator,$ N^{\mu} = \frac{e}{4}[G_R^{H*}G_L^{H}(1-\gamma _5)(\slashed{p}-\slashed{k}+m_0)(1-\gamma _5)+G_L^{H*}G_R^{H}(1+\gamma _5)(\slashed{p}-\slashed{k}+m_0)(1+\gamma _5)](2k+q)^{\mu}$
\newline
\newline
$=\frac{e(2k+q)^{\mu}}{4}[G_R^{H*}G_L^{H}\{(\slashed{p}-\slashed{k}+m_0)+\gamma _5(\slashed{p}-\slashed{k}+m_0)\gamma _5-(\slashed{p}-\slashed{k}+m_0)\gamma _5-\gamma _5(\slashed{p}-\slashed{k}+m_0)\}+G_L^{H*}G_R^{H}\{(\slashed{p}-\slashed{k}+m_0)+\gamma _5(\slashed{p}-\slashed{k}+m_0)\gamma _5+(\slashed{p}-\slashed{k}+m_0)\gamma _5+\gamma _5(\slashed{p}-\slashed{k}+m_0)\}]$
\newline
\newline
$=\frac{e(2k+q)^{\mu}}{4}[G_R^{H*}G_L^{H}\{(\slashed{p}-\slashed{k}+m_0)-(\slashed{p}-\slashed{k}-m_0)-(\slashed{p}-\slashed{k}+m_0)\gamma _5+(\slashed{p}-\slashed{k}-m_0)\gamma _5\}+G_L^{H*}G_R^{H}\{(\slashed{p}-\slashed{k}+m_0)-(\slashed{p}-\slashed{k}-m_0)+(\slashed{p}-\slashed{k}+m_0)\gamma _5-(\slashed{p}-\slashed{k}-m_0)\gamma _5\}]$
\newline
\newline
$=\frac{e(2k+q)^{\mu}}{4}[G_R^{H*}G_L^{H}\{2m_0-2m_0\gamma_5\}+G_L^{H*}G_R^{H}\{2m_0+2m_0\gamma_5\}]$
\newline
\newline
$=em_0(2k+q)^{\mu}[Re(G_R^{H*}G_L^{H})-iIm(G_R^{H*}G_L^{H})\gamma _5]$
\newline
\newline
Now, substituting $k^\mu = l^\mu+x_1p^\mu-x_3q^\mu$ we get,
\newline
\newline
$= em_0[Re(G_R^{H*}G_L^{H})-iIm(G_R^{H*}G_L^{H})\gamma _5][2l+2x_1p+(1-2x_3q)]^{\mu}$
\newline
\newline
To apply Gordon identity to have EDM and MDM form factor, we need to find $(p^\mu+p^{'\mu})$. We can write $2p^\mu = p^\mu + p'^{\mu} - q^\mu $, then the expression becomes
\newline
\newline
$= em_0[Re(G_R^{H*}G_L^{H})-iIm(G_R^{H*}G_L^{H})\gamma _5][x_1(p^{\mu}+{p}'^{\mu})+(1-x_1-2x_3)q^{\mu}+2l^{\mu}]$
\newline
\newline
$= em_0[Re(G_R^{H*}G_L^{H})-iIm(G_R^{H*}G_L^{H})\gamma _5][x_1(-i\sigma^{\mu\nu}q_\nu)+(x_2-x_3)q^{\mu}+2l^{\mu}]$
\newline
\newline
Note that,\begin{itemize}
\item $(x_2-x_3)$ term is odd $x_2$ and $x_3$, but the denominator is even $x_2x_3$. That is, under $x_2 \rightarrow -x_2$ and $x_3 \rightarrow -x_3$, the numerator changes sign i.e. odd but the denominator doesn't i.e. even. Thus, it gives zero contribution to the integration.
\item $l^{\mu}$ term  will not contribute due to momentum integral $\int \frac{d^{4}l}{(2\pi)^{4}}\frac{l^{\mu}}{l^{\mu}-\Delta}=0$.
\end{itemize}
Thus, combining everything, the loop contribution is
$$\delta \Gamma_2 ^{\mu }= \int \frac{d^{4}l}{(2\pi)^{4}}\int_{0}^{1}dx_1dx_2dx_3\frac{2\delta(x_1+x_2+x_3-1)N^{\mu}}{D^{3}}$$
Now, we need to do Wick Rotation to do the integration($l^0 \rightarrow il^0_E$). We will ignore the subscript E(Euclidean). Then
$$\frac{1}{2} \int \frac{d^4 l}{(2\pi)^4} \frac{1}{(l^2-\Delta)^3} \rightarrow -i \int \frac{d^4 l}{(2\pi)^4} \frac{1}{(l^2+\Delta)^3} $$
The general formula for doing this kind of loop integral is given by 
$$ \int \frac{d^d l}{(2\pi)^d} \frac{(l^2)^a}{(l^2+\Delta)^b} = \frac{\Gamma(b-a-\frac{d}{2})\Gamma(a+\frac{d}{2})}{(4\pi)^\frac{d}{2}\Gamma(b) \Gamma(\frac{d}{2})} \Delta^{-(b-a-\frac{d}{2})}  $$
We can prove this by using $a=0$, then differentiating and using gamma matrix properties. This formula is directly taken from \cite{Srednicki:2007qs}.

For our case, using $d=4,a=0,b=3$ and Wick rotating(extra $-i$ factor), we get 
$$ \int \frac{d^4 l}{(2\pi)^4} \frac{1}{(l^2-\Delta)^3} =  \frac{(-i)}{(4\pi)^2}\frac{1}{2\Delta} $$
Using this, we can do the $l$-integration. Thus, our loop contribution is 
\begin{center}
$\delta \Gamma_2 ^{\mu } = em_02m[Re(G_R^{H*}G_L^{H})-iIm(G_R^{H*}G_L^{H})\gamma _5]\frac{(-i)}{(4\pi)^{2}}\int_{0}^{1}dx_1dx_2dx_3\frac{2\delta(x_1+x_2+x_3-1)}{2\Delta}x_1(-i\frac{\sigma^{\mu\nu}q_\nu}{2m})$
\end{center}
Therefore,
$$\delta \Gamma_2 ^{\mu } = -\frac{em_02m[Re(G_R^{H*}G_L^{H})-iIm(G_R^{H*}G_L^{H})\gamma _5]}{16\pi^{2}M_H^{2}}I_2(r,s)\frac{\sigma^{\mu\nu}q_\nu}{2m}$$
where, $$I_2(r,s)= \int_0^1  dx \frac{x(1-x)}{1-x+rx-sx(1-x)}; r=\frac{m_0^{2}}{M_H^{2}} \& s=\frac{m^{2}}{M_H^{2}}$$
From the definition of EDM and MDM form factor, $\delta\Gamma^\mu = F_{2}(q^2)\frac{\sigma^{\mu\nu}q_\nu}{2m} +F_{EDM}(q^2)\sigma^{\mu\nu}q_\nu\gamma^5 $, and AMM definition $a = F_{2}(q^2=0)$, EDM definition $d_f = -F_{EDM}(q^2=0)$. So we get the anomalous magnetic moment \& electric dipole moment :
$$a = -2mRe(G_R^{H*}G_L^{H})\frac{em_0}{16\pi^{2}M_H^{2}}\int_0^1  dx \frac{x(1-x)}{1-x+rx-sx(1-x)}$$
$$d_f = -Im(G_R^{H*}G_L^{H})\frac{em_0}{16\pi^{2}M_H^{2}}\int_0^1  dx \frac{x(1-x)}{1-x+rx-sx(1-x)}$$
where $ r=\frac{m_0^{2}}{M_H^{2}}$ \& $ s=\frac{m^{2}}{M_H^{2}}$
\newline
\newline
Note that, The expression of EDM ($d_f$) exactly matches with the expression in \cite{Ecker:1983dj}.

\subsection{Diagram 3: H-loop contributions}

\graphicspath{ {./Hloop2/} }
\begin{figure}[H]
\begin{center}
\includegraphics[width=80mm]{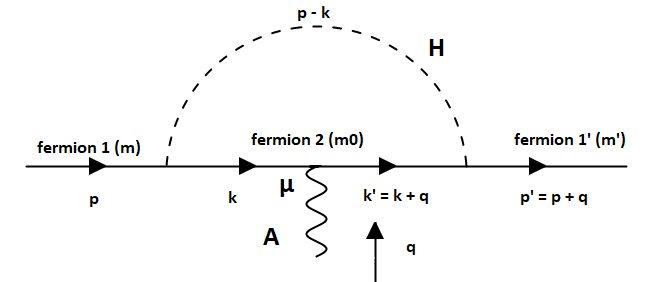}
\end{center}
\caption{H-loop (2) contributing diagram}
\label{Hloop2}
\end{figure}
 The interaction (fermion-scalar) Lagrangian is given by  
$$-\mathcal{L} = \bar{\psi} [G_L^{H}(\frac{1-\gamma _5}{2})+G_R^{H}(\frac{1+\gamma _5}{2})] \psi H $$
The structure is the same, but now the photon is coupled with the internal fermion. The loop contribution using the Feynman rules Figure Figure \ref{feyn3} and \ref{feyn2} is given by
\begin{center}
$\delta \Gamma_3 ^{\mu }=\int \frac{d^{4}k}{(2\pi)^{4}}[G_L^{H}\frac{1+\gamma _5}{2}+G_R^{H}\frac{1-\gamma _5}{2}]^{*}\frac{i({\slashed{k}}'+m_0)}{{k}'^{2}-m_0^{2}+i\epsilon}(ie\gamma^{\mu})\frac{i(\slashed{k}+m_0)}{k^{2}-m_0^{2}+i\epsilon}[G_L^{H}\frac{1-\gamma _5}{2}+G_R^{H}\frac{1+\gamma _5}{2}] \frac{(-i)}{(p-k)^{2}-M_H^{2}+i\epsilon}$
\end{center}
\begin{center}
$=\int\frac{d^{4}k}{(2\pi)^{4}}\frac{\frac{e}{4}[G_L^{H}(1+\gamma _5)+G_R^{H}(1-\gamma _5)]^{*}({\slashed{k}}'+m_0)\gamma^{\mu}(\slashed{k}+m_0)[G_L^{H}(1-\gamma _5)+G_R^{H}(1+\gamma _5)]}{[(p-k)^{2}-M_H^{2}+i\epsilon][k^{2}-m_0^{2}+i\epsilon][{k}'^{2}-m_0^{2}+i\epsilon]}$
\end{center}
Now, using Feynman parametrization we can combine these three denominators,
$$\frac{1}{[(p-k)^{2}-M_H^{2}+i\epsilon][k^{2}-m_0^{2}+i\epsilon][{k}'^{2}-m_0^{2}+i\epsilon]}=\int_{0}^{1}dx_1dx_2dx_3\frac{2\delta(x_1+x_2+x_3-1)}{D^{3}}$$
where, $$D = x_1[(p-k)^{2}-M_H^{2}+i\epsilon]+x_2[k^{2}-m_0^{2}+i\epsilon]+x_3[{k}'^{2}-m_0^{2}+i\epsilon]$$
$$=x_1[p^{2}+k^{2}-2p.k-M_H^{2}]+x_2[k^{2}-m_0^{2}]+x_3[k^{2}+2k.q+q^{2}-m_0^{2}]+i\epsilon $$
$$ =k^{2}+2k.(x_3q-x_1p)+x_1m^{2}+x_3q^{2}-x_1M_H^{2}-m_0^{2}(1-x_1) $$
We define, $$l = k-x_1p+x_3q $$
$$ => l^{2}=k^{2}+2k.(x_3q-x_1p)+x_3^{2}q^{2}+x_1^{2}m^{2}-2p.qx_1x_3 $$
Hence, $$ l^{2}-D=M_H^{2}[x_1+\frac{m_0^{2}}{M_H^{2}}(1-x_1)+\frac{m^{2}}{M_H^{2}}[x_1(x_1-1)+x_1x_3]-i\epsilon = \Delta-i\epsilon $$
Numerator,$N^{\mu}=\frac{e}{4}[G_R^{H*}G_L^{H}(1-\gamma _5)({\slashed{k}}'+m_0)\gamma^{\mu}(\slashed{k}+m_0)(1-\gamma _5)+G_L^{H*}G_R^{H}(1+\gamma _5)({\slashed{k}}'+m_0)\gamma^{\mu}(\slashed{k}+m_0)(1+\gamma _5)]$
\newline
\newline
$=\frac{e}{4}[G_R^{H*}G_L^{H}\{({\slashed{k}}'+m_0)\gamma^{\mu}(\slashed{k}+m_0)+\gamma _5({\slashed{k}}'+m_0)\gamma^{\mu}(\slashed{k}+m_0)\gamma _5-({\slashed{k}}'+m_0)\gamma^{\mu}(\slashed{k}+m_0)\gamma _5-\gamma _5({\slashed{k}}'+m_0)\gamma^{\mu}(\slashed{k}+m_0)\}+G_L^{H*}G_R^{H}\{({\slashed{k}}'+m_0)\gamma^{\mu}(\slashed{k}+m_0)+\gamma _5({\slashed{k}}'+m_0)\gamma^{\mu}(\slashed{k}+m_0)\gamma _5+({\slashed{k}}'+m_0)\gamma^{\mu}(\slashed{k}+m_0)\gamma _5+\gamma _5({\slashed{k}}'+m_0)\gamma^{\mu}(\slashed{k}+m_0)\}]$
\newline
\newline
$=\frac{e}{4}[G_R^{H*}G_L^{H}\{({\slashed{k}}'+m_0)\gamma^{\mu}(\slashed{k}+m_0)-({\slashed{k}}'-m_0)\gamma^{\mu}(\slashed{k}-m_0)+({\slashed{k}}'-m_0)\gamma^{\mu}(\slashed{k}-m_0)\gamma _5-({\slashed{k}}'+m_0)\gamma^{\mu}(\slashed{k}+m_0)\gamma _5\}+G_L^{H*}G_R^{H}\{({\slashed{k}}'+m_0)\gamma^{\mu}(\slashed{k}+m_0)-({\slashed{k}}'-m_0)\gamma^{\mu}(\slashed{k}-m_0)-
({\slashed{k}}'-m_0)\gamma^{\mu}(\slashed{k}-m_0)\gamma _5+({\slashed{k}}'+m_0)\gamma^{\mu}(\slashed{k}+m_0)\gamma _5\}]$
\newline
\newline
$=\frac{e}{2}[G_R^{H*}G_L^{H}\{m_0({\slashed{k}}'\gamma^{\mu}+\gamma^{\mu}\slashed{k})\}+G_L^{H*}G_R^{H}\{m_0({\slashed{k}}'\gamma^{\mu}+\gamma^{\mu}\slashed{k})\}]$
\newline
\newline
$=em_0[Re(G_R^{H*}G_L^{H})({\slashed{k}}'\gamma^{\mu}+\gamma^{\mu}\slashed{k})-iIm(G_R^{H*}G_L^{H})({\slashed{k}}'\gamma^{\mu}+\gamma^{\mu}\slashed{k})\gamma _5]$
\newline
\newline
$=em_0[Re(G_R^{H*}G_L^{H})-iIm(G_R^{H*}G_L^{H})\gamma _5]({\slashed{k}}'\gamma^{\mu}+\gamma^{\mu}\slashed{k})$
\newline
\newline
$=em_0[Re(G_R^{H*}G_L^{H})-iIm(G_R^{H*}G_L^{H})\gamma _5](\slashed{k}\gamma^{\mu}+\gamma^{\mu}\slashed{k}+\slashed{q}\gamma^{\mu})$
\newline
\newline
$= em_0[Re(G_R^{H*}G_L^{H})-iIm(G_R^{H*}G_L^{H})\gamma _5](2k^{\mu}+\slashed{q}\gamma^{\mu})$
\newline
\newline
Now, substituting $k^\mu = l^\mu+x_1p^\mu-x_3q^\mu$ we get,
\newline
\newline
$= em_0[Re(G_R^{H*}G_L^{H})-iIm(G_R^{H*}G_L^{H})\gamma _5][2(l+x_1p-x_3q)^{\mu}-\gamma^{\mu}\slashed{q}+2q^{\mu}]$
\newline
\newline
$= em_0[Re(G_R^{H*}G_L^{H})-iIm(G_R^{H*}G_L^{H})\gamma _5][2(l+x_1p-x_3q)^{\mu}-((-{\slashed{p}}'\gamma^{\mu}+2{p}'^{\mu}-\gamma^{\mu}m))+2q^{\mu}]$
\newline
\newline
$= em_0[Re(G_R^{H*}G_L^{H})-iIm(G_R^{H*}G_L^{H})\gamma _5][2l^{\mu}+2(x_1-1)p^{\mu}-2x_3q^{\mu}-\gamma^{\mu}m]$
\newline
\newline
To apply Gordon identity to have EDM and MDM form factor, we need to find $(p^\mu+p^{'\mu})$. We can write
$ 2p^\mu = p^\mu + p'^{\mu} - q^\mu $, then the expression becomes
\newline
\newline 
$= em_0[Re(G_R^{H*}G_L^{H})-iIm(G_R^{H*}G_L^{H})\gamma _5][(x_1-1)(-i\sigma^{\mu\nu}q_\nu)+(x_2-x_3)q^{\mu}+2l^{\mu}-\gamma^{\mu}m]$
\newline
\newline
Note that,\begin{itemize}
\item $\gamma^{\mu}m$ term is ignored since it contributes to electric charge. 
\item $(x_2-x_3)$ term is odd $x_2$ and $x_3$, but the denominator is even $x_2x_3$. That is, under $x_2 \rightarrow -x_2$ and $x_3 \rightarrow -x_3$, the numerator changes sign i.e. odd but the denominator doesn't i.e. even. Thus, it gives zero contribution to the integration.
\item $l^{\mu}$ term  will not contribute due to momentum integral $\int \frac{d^{4}l}{(2\pi)^{4}}\frac{l^{\mu}}{l^{\mu}-\Delta}=0$.
\end{itemize}
Thus, combining everything, the loop contribution is
$$\delta \Gamma_3 ^{\mu }= \int \frac{d^{4}l}{(2\pi)^{4}}\int_{0}^{1}dx_1dx_2dx_3\frac{2\delta(x_1+x_2+x_3-1)N^{\mu}}{D^{3}}$$
Now, we need to do Wick Rotation to do the integration($l^0 \rightarrow il^0_E$). We will ignore the subscript E(Euclidean). Then
$$\frac{1}{2} \int \frac{d^4 l}{(2\pi)^4} \frac{1}{(l^2-\Delta)^3} \rightarrow -i \int \frac{d^4 l}{(2\pi)^4} \frac{1}{(l^2+\Delta)^3} $$
The general formula for doing this kind of loop integral is given by 
$$ \int \frac{d^d l}{(2\pi)^d} \frac{(l^2)^a}{(l^2+\Delta)^b} = \frac{\Gamma(b-a-\frac{d}{2})\Gamma(a+\frac{d}{2})}{(4\pi)^\frac{d}{2}\Gamma(b) \Gamma(\frac{d}{2})} \Delta^{-(b-a-\frac{d}{2})}  $$
We can prove this by using $a=0$, then differentiating and using gamma matrix properties. This formula is directly taken from \cite{Srednicki:2007qs}. 

For our case, using $d=4,a=0,b=3$ and Wick rotating(extra $-i$ factor), we get 
$$ \int \frac{d^4 l}{(2\pi)^4} \frac{1}{(l^2-\Delta)^3} =  \frac{(-i)}{(4\pi)^2}\frac{1}{2\Delta} $$
Using this, we can do the $l$-integration. Thus, our loop contribution is
\begin{center}
$\delta \Gamma_3 ^{\mu } = em_02m[Re(G_R^{H*}G_L^{H})-iIm(G_R^{H*}G_L^{H})\gamma _5]\frac{(-i)}{(4\pi)^{2}}\int_{0}^{1}dx_1dx_2dx_3\frac{2\delta(x_1+x_2+x_3-1)}{2\Delta}(x_1-1)(-i\frac{\sigma^{\mu\nu}q_\nu}{2m})$
\end{center} 
Therefore,
$$\delta \Gamma_3 ^{\mu }= -\frac{em_02m[Re(G_R^{H*}G_L^{H})-iIm(G_R^{H*}G_L^{H})\gamma _5]}{16\pi^{2}M_H^{2}}I_3(r,s)\frac{\sigma^{\mu\nu}q_\nu}{2m}$$
where, $$I_3(r,s)= \int_0^1 dx  \frac{x^2}{1-x+rx-sx(1-x)}; r=\frac{m_0^{2}}{M_H^{2}}  \&  s=\frac{m^{2}}{M_H^{2}}$$
From the definition of EDM and MDM form factor, $\delta\Gamma^\mu = F_{2}(q^2)\frac{\sigma^{\mu\nu}q_\nu}{2m} +F_{EDM}(q^2)\sigma^{\mu\nu}q_\nu\gamma^5 $, and AMM definition $a = F_{2}(q^2=0)$, EDM definition $d_f = -F_{EDM}(q^2=0)$. So we get the anomalous magnetic moment \& electric dipole moment :  
$$a = -2mRe(G_R^{H*}G_L^{H})\frac{em_0}{16\pi^{2}M_H^{2}}\int_0^1 dx  \frac{x^2}{1-x+rx-sx(1-x)}$$

$$d_f = -Im(G_R^{H*}G_L^{H})\frac{em_0}{16\pi^{2}M_H^{2}}\int_0^1 dx  \frac{x^2}{1-x+rx-sx(1-x)}$$  
where $r=\frac{m_0^{2}}{M_H^{2}}$  and  $s=\frac{m^{2}}{M_H^{2}}$
\newline
\newline
Note that, The expression of EDM ($d_f$) exactly matches with the expression in \cite{Ecker:1983dj}.

\subsection{Diagram 4: H-loop contribution}

\graphicspath{ {./Hloop3/} }
\begin{figure}[H]
\begin{center}
\includegraphics[width=80mm]{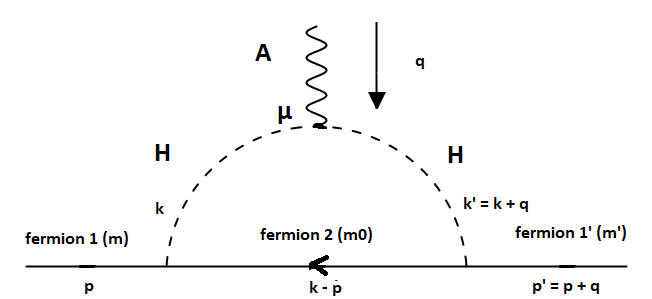}
\end{center}
\caption{H-loop (3) contributing diagram}
\label{Hloop3}
\end{figure}
This type of diagram is only for $\delta^{++}$ Higgs loops (see Section \ref{QFT Structure of Electric and Magnetic Dipole Moment}) which is exceptional. Because, in the Standard model, there is no doubly charged particle. As this doubly charged particle couples electrons with electrons (not electron and neutrino), the internal fermion is charged lepton. We have to sum over all possible internal propagators. This time we have to include all possible charged leptons for a specific i.e. electrons EDM and MDM contribution for $\delta^{++}$ loop. The difference from diagram 2 is that now, internal fermium is charged lepton and to conserve $U(1)_{EM}$ electric charge at the vertex, the internal fermion momentum direction is opposite. Similarly, as $\delta$ is coupling with fermions, here also conjugate field must be opposite chiral. Here specifically, $H=\delta^{++}$in our model. Thus, the loop contribution is, using the Feynman rules of Figure \ref{feyn3} and \ref{feyn5}
\begin{center}
$\delta \Gamma^\mu = \int \frac{d^{4}k}{(2\pi)^{4}}(-i)[G_L^{H}(\frac{1+\gamma _5}{2})+G_R^{H}(\frac{1-\gamma _5}{2})]^{*} \frac{i(\slashed{k}-\slashed{p}+m_0)}{(p-k)^2-m_0^2+i\epsilon} (-i) [G_L^{H}(\frac{1-\gamma _5}{2})+G_R^{H}(\frac{1+\gamma _5}{2})] \frac{-i}{k^2-M_H^2+i\epsilon} (-ie(2k+q)^\mu) \frac{-i}{(k+q)^2-M_H^2+i\epsilon}$
\end{center}
The denominator is the same as the denominator of Diagram 2. Thus, call denominator from diagram 2, $$D=l^2-\Delta+i\epsilon$$ where, $$\Delta = M_H^{2}[1-x_1+\frac{m_0^{2}}{M_H^{2}} x_1+\frac{m^{2}}{M_H^{2}}\{x_1(x_1-1)+x_1x_3\}]$$
Now, consider only the fermionic part from the numerator
$$[G_L^{H}(\frac{1+\gamma _5}{2})+G_R^{H}(\frac{1-\gamma _5}{2})]^{*} (\slashed{k}-\slashed{p}+m_0) [G_L^{H}(\frac{1-\gamma _5}{2})+G_R^{H}(\frac{1+\gamma _5}{2})]$$
$$= \frac{1}{4} [G_R^{H*}G_L^H (1-\gamma^5)(\slashed{k}-\slashed{p}+m_0)(1-\gamma^5)+G_L^{H*}G_R^H (1+\gamma^5)(\slashed{k}-\slashed{p}+m_0)(1+\gamma^5)]$$
$$= m_0[Re(G_R^{H*}G_L^{H})-iIm(G_R^{H*}G_L^{H})\gamma _5]$$
The total numerator is (there is an overall $-1$ occur from $-1$ and $i$ factors in the loop function), $$N^\mu = e m_0 [Re(G_R^{H*}G_L^{H})-iIm(G_R^{H*}G_L^{H})\gamma _5](2k+q)^\mu$$
This is the same expression as diagram 2. Both give the same structure because we can only measure physical quantities. So, a physical particle is the one which is in the mass eigen basis. Thus, this massive nature enables chiral flipping. This chiral flipping demands mass insertions in the theory. That's why from the whole (momentum + mass) structure, the EDM and MDM contribution extracts only the mass terms, as mass insertion is needed. This diagram differs from diagram 2 only in the momentum part i.e. opposite momentum. As EDM and MDM contribution only deals with or demands mass insertions, these two diagrams are the same because they are the same in the mass terms (differs in momentum slash part in the fermionic propagator). Thus, this diagram's contribution structure is the same as diagram 2.

\subsection{Sum Over All Possible Internal Fermion State}

In all calculations, we did not consider the mixing of the internal fermion. To fix this, we need to add diagrams with all possible internal fermions. For example, consider the W-loop contribution to EDM and MDM, Figure \ref{Wloop}. The internal fermion is the neutrino, we need to sum over all possible neutrino states in the loop which can not happen in the Standard model. Because in the minimal LR symmetric model, the neutrino has both Dirac and Majorana mass (seesaw compatible) and so neutrino mixes between flavors. 
\newline
Consider, charged lepton Yukawa coupling $Y^l$. By rotating with a unitary matrix $V_l$, we go to a new basis such that it is diagonal, 
$$ \hat{Y^l} = V_l Y^lV_l^\dagger = diag(y_e,y_\mu,y_\tau) $$

Similarly, for neutrino, we have to rotate by $V_\nu$ such that it diagonalizes the neutrino mass matrix. So, either we can go to charged lepton's mass basis(then neutrinos are in superposition) or go to neutrinos mass basis (where charged leptons are in superposition). In a minimal LR symmetric model, the neutrino has mass and so nontrivial mixing matrix $U_{PMNS} $  of the lepton sector exists 
$$ U_{PMNS}  = V_\nu^\dagger V_l$$ 
Going to charged leptons basis, we will get neutrinos in the loop as a superposition of all mass eigen basis. Thus, the minimal LR model enables mixing in the vertex of this theory as $U_{PMNS}$ is a non-trivial mixing matrix.


\begin{thebibliography}{}
\bibitem{sakharov}
Andrei D Sakharov, ``Violation of CP-invariance, C-asymmetry, and baryon asymmetry of the Universe", Arzamas-16 84--87(1998), World Scientific. doi: 10.1142/9789812815941$\_$0013

\bibitem{abazajian2011}
K.N. Abazajian et al. ``Cosmological and astrophysical neutrino mass measurements'', Astropart. Phys. 35 (2011) 177

\bibitem{weinberg1979baryon}
S. Weinberg, ``Baryon-and lepton-nonconserving processes'', Phys. Rev. Lett. 43, 1566 (1979)

\bibitem{Minkowski}
P.Minkowski, Phys. Lett. B 67 (1977) 421; R. Mohapatra, G. Senjanovi´c, Phys. Rev. Lett. 44 (1980) 912

\bibitem{Yanagida}
Yanagida, Workshop on unified theories and baryon number in the universe, ed. A. Sawada, A. Sugamoto (KEK, Tsukuba, 1979); S. Glashow, Quarks and leptons, Carg`ese 1979, ed. M. L´evy (Plenum, NY, 1980); M. Gell-Mann et al., Supergravity Stony Brook workshop, New York, 1979, ed. P. Van Niewenhuizen, D. Freeman (North Holland, Amsterdam, 1980).

\bibitem{Magg}
Magg, Wetterich, Phys. Lett. B 94 (1980) 61; G. Lazarides et al., Nucl. Phys.B 181 (1981) 287; Mohapatra, Senjanovi´c, Ref.[27]

\bibitem{Foot}
Foot, Lew, X.G.He and Joshi, Z. Phys. C 44, 441 (1989).

\bibitem{Forero}
D.V. Forero, M. Tortola, and J.W.F. Valle, Neutrino oscillations refitted, Phys.Rev.D 90 (2014) 9, 093006[arXiv:1405.7540 [hep-ph]] 

\bibitem{Casas:2001sr}
J. A. Casas and A. Ibarra, ``Oscillating neutrinos and $\mu \to e, \gamma$",Nucl. Phys. B 618 171--204 (2001) doi:0.1016/S0550-3213(01)00475-8 [arXiv:0103065 [hep-ph]].

\bibitem{Mohapatra:1974gc}
Rabindra N. Mohapatra and Jogesh C. Pati, ``A Natural Left-Right Symmetry", CCNY-HEP-74-2, Phys. Rev. D 11 2558 (1975),
doi:10.1103/PhysRevD.11.2558

\bibitem{Senjanovic:1975rk}
G.~Senjanovic and R.~N.~Mohapatra, ``Exact Left-Right Symmetry and Spontaneous Violation of Parity,'' Phys. Rev. D \textbf{12}, 1502 (1975) doi:10.1103/PhysRevD.12.1502

\bibitem{Grimus:1993fx}
W.~Grimus, ``Introduction to left-right symmetric models,''
UWTHPH-1993-10.

\bibitem{Corrigan:2015kfu}
E.~Corrigan, ``LEFT-RIGHT-SYMMETRIC MODEL BUILDING,'' LU-TP-15-30.

\bibitem{Liu:1985zp}
J.~Liu, ``Electric Dipole Moment of Electron in the Left-right Models'', Nucl. Phys. B \textbf{271}, 531-539 (1986)
doi:10.1016/S0550-3213(86)80024-4

\bibitem{Mohapatra:1979ia}
Rabindra N. Mohapatra and G.~Senjanovi\'c, ``Neutrino Mass and Spontaneous Parity Nonconservation", Phys. Rev. Lett. 44 912 (1980) doi: 10.1103/PhysRevLett.44.912
    
\bibitem{Senjanovic:2016bya}
Goran Senjanovi\'c, ``Is left\textendash{}right symmetry the key?", Mod. Phys. Lett. A 32 04 1730004 (2017) doi:10.1142/S021773231730004X [arXiv:1610.04209 [hep-ph]].

\bibitem{deshpande1991left}
NG Deshpande, JF Gunion, Boris Kayser and Fredrick Olness; ``Left-right-symmetric electroweak models with triplet Higgs field", Phys. Rev. D 44 3 837 (1991) doi: 10.1103/PhysRevD.44.837

\bibitem{chengandli}
Tai-Pei Cheng and Ling-Fong Li ``Gauge theory of elementary particle physics'', Oxford University Press, 1984. ISBN: 0-19-85961-3

\bibitem{edfSM}
M. Pospelov and A. Ritz, Electric dipole moments as probes of new physics, Annals Phys. 318 (2005) 119–169 [hep-ph/0504231]

\bibitem{Babu},
K. S. Babu, Bhaskar Dutta, and R. N. Mohapatra, Enhanced electric dipole moment of the muon in the presence of large neutrino mixing, FERMILAB-PUB-00-147-T, OSU-HEP-00-03, CTP-TAMU-20-00, UMD-PP-00-084, Phys. Rev. Lett. 85 5064--5067 (2000), [arXiv:hep-ph/0006329]

\bibitem{Muondf}
Muon (g-2) collaboration, G. W. Bennett et al., An Improved Limit on the Muon Electric Dipole Moment, Phys. Rev. D 80 052008 (2009), doi: 10.1103/PhysRevD.80.052008, [arXiv:0811.1207[hep-ex]]

\bibitem{edf}
Tanya S. Roussy et al., An improved bound on the electron\textquoteright{}s electric dipole moment, Science 381 6653 adg4084 (2023), doi: 10.1126/science.adg4084, [arXiv:2212.11841[physics. atom-ph]]

\bibitem{MDMEDM}
M. Abe et al., A New Approach for Measuring the Muon Anomalous Magnetic Moment and Electric Dipole Moment, PTEP 2019 (2019) 053C02 [arXiv:1901.03047]

\bibitem{frozenspin}
A. Adelmann et al., Search for a muon EDM using the frozen-spin technique,(2021) [arXiv:2102.08838]

\bibitem{Schwinger}
J. S. Schwinger, ``On quantum electrodynamics and the
magnetic moment of the electron", Phys. Rev. 73, 416 (1948)

\bibitem{Fermilab}
Muon g-2 collaboration, D. P. Aguillard et al., Measurement of the Positive Muon Anomalous Magnetic Moment to 0.20~ppm", FERMILAB-PUB-23-385-AD-CSAID-PPD, Phys. Rev. Lett. 131 16 161802 (2023), doi: 10.1103/PhysRevLett.131.161802, [arXiv:2308.06230[hep-ex]]

\bibitem{BNLg2}
Bennett GW, Bousquet B, Brown HN, Bunce G, Carey RM, Cushman P, et al., ``Final report of the E821 muon anomalous magnetic moment measurement at BNL", Phys Rev D (2006) 73:072003. doi:10.1103/PhysRevD.73.072003

\bibitem{Muong2SM}
T. Aoyama, N. Asmussen, M. Benayoun, J. Bijnens, T. Blum, M. Bruno, et al., ``The anomalous magnetic moment of the muon in the Standard Model. Phys Rep (2020) 887:1–166. doi:10.1016 j.physrep.2020.07.006 , [arXiv:2006.04822 [hep-ph]].

\bibitem{Fan:2022eto}
X.Fan, T. G. Myers, B. A. D.Sukra and G.Gabrielse, ``Measurement of the Electron Magnetic Moment, Phys. Rev. Lett. 130 7 071801 (2023) doi: 10.1103/PhysRevLett.130.071801, arXiv: 2209.13084 [physics.atom-ph]

\bibitem{Electrong2}
T. Aoyama, T. Kinoshita, and M. Nio, “Revised and Improved Value of the QED Tenth-Order Electron Anomalous Magnetic Moment,” Phys. Rev. D97 no.3, (2018) 036001, [arXiv:1712.06060 [hep-ph]]

\bibitem{Morel}
L. Morel, Z. Yao, P. Clad´e, and S. Guellati-Kh´elifa, “Determination of the fine-structure constant with an accuracy of 81 parts per trillion,” Nature 588 no. 7836, (2020) 61–65.

\bibitem{Parker}
R. H. Parker, C. Yu, W. Zhong, B. Estey, and H. M¨uller, “Measurement of the fine-structure constant as a test of the Standard Model,” Science 360 (2018) 191, arXiv:1812.04130[physics. atom-ph]

\bibitem{Cirigliano}
V. Cirigliano, A. Kurylov, M. J. Ramsey-Musolf and P. Vogel, Phys. Rev. D70, 075007 (2004), arXiv: 0404233 [hep-ph].

\bibitem{SINDRUM}
SINDRUM Collaboration, U. Bellgardt et al., Nucl. Phys. B299, 1 (1988).

\bibitem{Belle}
Belle Collaboration, K. Hayasaka, K. Inami, Y. Miyazaki, K. Arinstein, V. Aulchenko, T. Aushev, A. M. Bakich and A. Bay et al., Phys. Lett. B 687 (2010) 139 [arXiv:1001.3221[hep-ex]]

\bibitem{MEG:2016leq}
MEG Collaboration, A. M. Baldini, et al., ``Search for the lepton flavor violating decay $\mu^+ \rightarrow e^+ \gamma $ with the full dataset of the MEG experiment", Eur. Phys. J. C 76 8 434 (2016), doi: 10.1140/epjc/s10052-016-4271-x [arXiv:1605.05081[hep-ex]]

\bibitem{BaBar}
BaBar collaboration, Searches for Lepton Flavor Violation in the Decays $\tau^\pm \rightarrow e^\pm\gamma $ and $\tau^\pm \rightarrow \mu^\pm\gamma$, Phys. Rev. Lett. 104 (2010) 021802 [arXiv:0908.2381]

\bibitem{Aguila}
F. del Aguila, J. de Blas, and M. Perez-Victoria, Electroweak Limits on General New Vector Bosons, UG-FT-272-10, CAFPE-142-10, JHEP 09 033 (2010), doi: 10.1007/JHEP09(2010)033, [arXiv:1005.3998[hep-ph]]

\bibitem{ATLAS:2018gfm}
ATLAS collaboration, Morad Aaboud et al., Search for charged Higgs bosons decaying via $H^{\pm} \to \tau^{\pm}\nu_{\tau}$ in the $\tau$+jets and $\tau$+lepton final states with 36 fb$^{-1}$ of $pp$ collision data recorded at $\sqrt{s} = 13$ TeV with the ATLAS experiment, CERN-EP-2018-148, JHEP 09 139 (2018), doi: 10.1007/JHEP09(2018)139, [arXiv:1807.07915[hep-ex]]

\bibitem{CMS:2018rmh}
CMS collaboration, Albert M Sirunyan et al., Search for additional neutral MSSM Higgs bosons in the $\tau\tau$ final state in proton-proton collisions at $\sqrt{s}=$ 13 TeV, CMS-HIG-17-020, CERN-EP-2018-026, JHEP 09 007 (2018), doi: 10.1007/JHEP09(2018)007, [arXiv:1803.06553[hep-ex]]

\bibitem{Ma}
E. Ma, LHEP 01, 01 (2018)

\bibitem{Pich}
A. Pich, Effective field theory: Course, arXiv:9806303 [hep-ph]

\bibitem{Peskin:1995ev}
M.~E.~Peskin and D.~V.~Schroeder, ``An Introduction to Quantum Field Theory,''

\bibitem{Nieves:1986uk}
J.~F.~Nieves, D.~Chang and P.~B.~Pal, ``Electric Dipole Moment of the Electron in Left-right Symmetric Theories,'' Phys. Rev. D \textbf{33}, 3324-3328 (1986) doi:10.1103/PhysRevD.33.3324

\bibitem{Srednicki:2007qs}
M.~Srednicki, ``Quantum field theory,''

\bibitem{Ecker:1983dj}
G.~Ecker, W.~Grimus and H.~Neufeld, ``The Neutron Electric Dipole Moment in Left-right Symmetric Gauge Models'', Nucl. Phys. B \textbf{229}, 421-444 (1983) doi 10.1016/0550-3213(83)90341-3
\end{thebibliography}
\end{document}